\documentclass[preprint, sort&compress]{elsarticle}
\usepackage{graphicx,graphics}
\usepackage{epstopdf}
\usepackage{dcolumn}
\usepackage{amsmath,amssymb,amsfonts}
\usepackage{latexsym,verbatim}
\usepackage{bm}
\usepackage{textcomp} 
\usepackage{mathtools}
\usepackage{braket}
\usepackage{soul}
\usepackage{bbold}
\usepackage{color}
\usepackage{ulem}
\usepackage[breaklinks=false,colorlinks,citecolor=blue,linkcolor=blue,urlcolor=blue]{hyperref}
\usepackage[abs]{overpic}
\usepackage{tikz, bm}
\usepackage[compat=1.1.0]{tikz-feynman}
\begin{document}
\title{Phonon-mediated superconductivity in strongly correlated electron systems: a Luttinger-Ward functional approach}
\author[IT]{Andrea Secchi}
\ead{andrea.secchi@nano.cnr.it}
\author[P1,P2,IT]{Marco Polini}
\author[RU]{Mikhail I. Katsnelson}
\address[IT]{Istituto Italiano di Tecnologia, Graphene Labs, Via Morego 30, I-16163 Genova, Italy}
\address[P1]{Dipartimento di Fisica dell'Universit\`a di Pisa, Largo Bruno Pontecorvo 3, I-56127 Pisa,~Italy}
\address[P2]{School of Physics \& Astronomy, University of Manchester, Oxford Road, Manchester M13 9PL,~United Kingdom}
\address[RU]{Institute for Molecules and Materials, Radboud University,
Heyendaalseweg 135, 6525 AJ, Nijmegen, The Netherlands}

\date{\today}

\date{\today}

\begin{abstract}
{We use a Luttinger-Ward functional approach to study the problem of phonon-mediated superconductivity in electron systems with strong electron-electron interactions (EEIs). Our derivation does not rely on an expansion in skeleton diagrams for the EEI and the resulting theory is therefore nonperturbative in the strength of the latter. We show that one of the building blocks of the theory is the irreducible six-leg vertex related to EEIs. Diagrammatically, this implies five contributions (one of the Fock and four of the Hartree type) to the electronic self-energy, which, to the best of our knowledge, have never been discussed in the literature. Our approach is applicable to (and in fact designed to tackle superconductivity in) strongly correlated electron systems described by generic lattice models, as long as the glue for electron pairing is provided by phonons.}
\end{abstract}


\maketitle

\section{Introduction}
\label{sec:Introduction}

Our microscopic knowledge of superconductivity relies on the monumental theory laid down between the end of the Fifties and the beginning of the Sixties by Migdal and Eliashberg~\cite{Migdal58,Eliashberg60,Eliashberg61,Abrikosov,Scalapino66,ScalapinoParks,Marsiglio-Carbotte}, in a period where significant progress in the theoretical understanding of electron-phonon interactions (EPIs)~\cite{Nakajima54, Bardeen55, Baym61, Keating68, Gillis70, Maksimov76, Vogl76, recent} was made. 

Interest in the Migdal-Eliashberg theory, which represents one of the first applications of quantum field theory to condensed matter physics, arose out of the need to transcend the Bardeen-Cooper-Schrieffer (BCS) mean-field theory~\cite{BCS}. In brief, BCS theory relies on a quasiparticle reduced Hamiltonian approach, which is inadequate to describe at least three situations~\cite{ScalapinoParks}: 1) when the quasiparticle damping rate due to EPIs becomes comparable with the excitation energy; 2) when the BCS assumption of an effective two-body instantaneous attractive interaction between quasiparticles does not provide an adequate representation of the retarded nature of electron-phonon interactions; or 3) when normal and pairing correlations must be treated on equal footing.

From the Sixties to the present days, many researchers have invested a truly considerable energy in studying superconductivity in an extremely large number of situations and materials. In what follows we provide a largely incomplete list of past and contemporary research subfields in the realm of superconductivity:
\begin{itemize}
\item[i)] {\it Unconventional superconductors}. Huge efforts have been devoted to investigate {\it unconventional} superconductivity~\cite{Dagotto94,Orenstein00,Sachdev03,Yanase03,Capone09,Scalapino12,Anderson13,Fradkin15,Hussey18}, i.e.~superconductivity unrelated to EPIs but arising from other pairing glues, originating microscopically solely from repulsive electron-electron interactions (EEIs). These studies were motivated by the discovery of high-temperature superconductivity in the cuprates~\cite{Bednorz88}, which still pose tremendous challenges to theoretical condensed matter physics.

\item[ii)] {\it Non-adiabatic effects in the theory of superconductivity}. 
Since the theory by Eliashberg~\cite{Eliashberg60,Eliashberg61} relies on the Migdal theorem~\cite{Migdal58}, many researchers have studied the importance of {\it non-adiabatic} corrections~\cite{Engelsberg63,Pietronero92,Ikeda,Grimaldi95,Maksimov95,Cappelluti03,Cappelluti06,Gorkov16,Prakash17,Chubukov19}, which become important when the Migdal parameter $\hbar\omega_{\rm D}/E_{\rm F}$ ceases to be $\ll 1$. Here, $\omega_{\rm D}$ is the Debye frequency and $E_{\rm F}$ is the Fermi energy.

\item[iii)] {\it The Migdal-Eliashberg theory for real materials}. For understanding phonon-mediated superconductivity in real materials, the Eliashberg equations need to be supplemented by first principles calculations of the electronic band structure and EPIs. For a recent review of this vast research field, we invite the reader to consult Ref.~\cite{Giustino17}. In passing, we note that an {\it ab initio} theory of phonon-mediated superconductivity (SCDFT)---constructed by generalizing the Hohenberg-Kohn theorem~\cite{HK} of density functional theory to include the order parameter as an additional ``density"---has been recently laid down and applied to a number of materials~\cite{SCDFT}.

\item[iv)] {\it High-temperature superconductivity in hydrates}. The discovery of high-temperature superconductivity at high pressures in hydrogen-rich compounds (at $203~{\rm K}$ in ${\rm H}_3{\rm S}$~\cite{drozdov_nature_2015} and at $\sim 250~{\rm K}$ in ${\rm LaH}_{10}$~\cite{drozdov_nature_2019,somayazulu_prl_2019}) has re-attracted a great deal of attention to conventional phonon-mediated superconductors. Despite the underlying pairing mechanism in these compounds is conventional, several unconventional processes~\cite{errea_prl_2015,
errea_nature_2016,errea_arxiv_2019,quan_prb_2016,sano_prb_2016,liu_prb_2019,ghosh_prb_2019} compete in them, such as high phonon frequencies, quantum zero-point motion of H, strong anharmonic effects, and van Hove singularities. In this latter respect, for example, ${\rm H}_3{\rm S}$ displays~\cite{quan_prb_2016,sano_prb_2016,liu_prb_2019,ghosh_prb_2019} a pair of closely spaced van Hove singularities at $E_{\rm F}$. As a result, in a significant portion of the Brillouin zone, the quasiparticle velocity vanishes, technically invalidating the applicability of the Migdal theorem---see point ii) above---and, at the same time, greatly enhancing the role of EEIs.

\item[v)] {\it Superconductivity in magic-angle twisted bilayer graphene}. The discovery of superconductivity and correlated insulating states in magic-angle twisted bilayer graphene (MATBG)~\cite{cao_naturea_2018,cao_natureb_2018} has proven this system to be a rich playground for exploring strong correlations~\cite{yankowitz_science_2019,lu_nature_2019,kerelsky_nature_2019,xie_nature_2019,jiang_nature_2019,cao_arxiv_2019,hesp_arxiv_2019}. While the mechanism of superconductivity is still debated, several theoretical and experimental works point to the importance of electron-phonon interactions~\cite{polshyn_naturephys_2019,wu_prl_2018,lian_prl_2019,yudhistira_prb_2019,angeli_prx_2019}. On top of this, the electronic structure of MATBG displays nearly flat bands near the charge neutrality point~\cite{bistritzer_pnas_2011,sanjose_prl_2012,tarnopolsky_prl_2019,koshino_prx_2018,carr_prr_2019}. Similarly to the case of the hydrates, these undermine, at least in principle, the applicability of the Migdal-Eliashberg theory and are believed to be responsible for strong EEIs.
\end{itemize}
This work is mainly motivated by the experimental systems mentioned at points iv) and v) above. We pose the following simple question: What is the fate of the Migdal-Eliashberg theory in materials where EEIs play a dominant role? More precisely: Is it possible to construct a theory of conventional, phonon-mediated superconductivity in materials where EEIs require a nonperturbative treatment? In this work we answer this question affirmatively and present such theory. Our main result is that, in order to produce a consistent theory of EEIs to all orders, a building block which has so far been overlooked needs to be taken into account: the irreducible six-leg vertex related to EEIs. Under certain approximations discussed below, we derive a set of extended Eliashberg equations.

\section{Outline, approximations, and applicability}
\label{sec:Outline}

We here summarize the key features of our theory, which will be detailed in the remainder of this work. 

We consider an electron-phonon Hamiltonian (written for an arbitrary set of basis states) where EEIs are included in all generality, while EPIs are assumed to be linear in the ion displacements from their equilibrium positions, as in the Fr\"ohlich EPI Hamiltonian (Sect.~\ref{sec: Hamiltonian}). By using Matsubara path integrals, in Sect.~\ref{sect:LWF} we derive the general form of the Luttinger-Ward functional (LWF)~\cite{LuttingerWard,StefanuccivanLeeuwen} corresponding to this Hamiltonian, using an exact procedure that is nonperturbative in the EEI. We then make the first approximation, which consists in expanding the LWF in powers of the EPI matrix elements, up to the second order (Sect.~\ref{sec: Expansion}). As in any other LWF, the coefficients of such expansion are {\it fully interacting} many-body quantities, namely Green's functions (GFs) and interaction vertices. From this approximate LWF, in Sect.~\ref{sect:self-energy} (Sect.~\ref{sect:phononic-self-energy}) we derive the electronic (phononic) self-energy. In particular, the electronic self-energy consists of the EEI self-energy functional, plus two EPI self-energy functionals (Hartree and Fock contributions). Up to this point, our treatment is very general and can be applied to study the combined effects of EEIs and EPIs in any solid-state system. No approximations are made, except for the truncation of the LWF (such a truncation is necessary to develop any practical theory). Our main general results can be found in Eqs.~\eqref{Sigma Fock simplified}--\eqref{Sigma Hartree simplified}, \eqref{Lambda Fock}, and~\eqref{Lambda Hartree}.

Sect.~\ref{sec: Eliashberg} is specific to the problem of phonon-mediated superconductivity. Here, we apply the previously developed general formalism to derive a set of {\it extended Eliashberg equations}. Our aim is to provide a way to compute the anomalous components of the electronic self-energy, accounting as much as possible for the EEI effects derived in Sections~\ref{sec: Hamiltonian}-\ref{sect:self-energy}. This task requires some approximations. Most importantly, we need to adopt a tractable, explicit expression for the electronic self-energy functional; this cannot be done exactly, because the functional corresponding to the EEI self-energy is not known analytically (and even if it was, it would be overwhelmingly complicated). In Sect.~\ref{sec: Eliashberg} we therefore: 1)  neglect the EEI vertices appearing in the EPI self-energy functionals and 2) confine ourselves to the regime of temperatures close to the critical temperature, where the expressions can be linearized in the anomalous self-energy. We then assume that the EEI self-energy functional {\it in the normal state} can be obtained via other theoretical/computational means~\cite{DMFT, beyondDMFT, dual}, and plugged into the resulting extended Eliashberg equations, \eqref{Eliashberg I final}--\eqref{Eliashberg VI final}. These equations include the vertex corrections arising from the EEI self-energy functional, which are not captured by the Tolmachev-Morel-Anderson pseudopotential method~\cite{Tolmachev, MorelAnderson} and the McMillan formula~\cite{McMillan}.

A brief set of conclusions and future perspectives are reported in Sect.~\ref{sect:conclusions}. Numerous technical details are reported in \ref{app: Derivation Hamiltonian}--\ref{app:Nambu}.

\section{Model Hamiltonian}
\label{sec: Hamiltonian}
We consider a system of electrons and phonons, in the presence of EEIs and EPIs. 
The Hamiltonian has the following general form
\begin{align}
\hat{\cal H} & = \hat{\cal H}({\bm t}, {\bm U}, {\bm f}, {\bm M})   =  \hat{\cal H}_{\rm IE}({\bm t}) + \hat{\cal H}_{\rm EEI}({\bm U}) + \hat{\cal H}_{\rm IP}({\bm f}) + \hat{\cal H}_{\rm EPI}({\bm M}, {\bm f})~,
\label{H}
\end{align}
where the independent-electron (IE) term is
\begin{align}
\hat{\cal H}_{\rm IE}({\bm t}) = \sum_{\alpha, \beta} t_{\alpha, \beta} \hat{c}^{\dagger}_{\alpha} \hat{c}_{\beta}~,
\label{H_IE}
\end{align}
the electron-electron interaction (EEI) is
\begin{align}
\hat{\cal H}_{\rm EEI}({\bm U}) = \frac{1}{2} \sum_{\alpha, \beta, \gamma, \delta} U_{\alpha, \beta, \delta, \gamma} \, \hat{c}^{\dagger}_{\alpha} \hat{c}^{\dagger}_{\beta} \hat{c}_{\gamma} \hat{c}_{\delta}~,
\label{H_EEI}
\end{align}
the independent-phonon (IP) term is
\begin{align}
\hat{\cal H}_{\rm IP}({\bm f}) = \sum_{\kappa, \lambda} f_{\kappa, \lambda} \hat{b}^{\dagger}_{\kappa} \hat{b}_{\lambda}~,
\label{H_IP}
\end{align}
and the electron-phonon interaction (EPI) is
\begin{align}
\hat{\cal H}_{\rm EPI}({\bm M}, {\bm f}) = \sum_{\alpha, \beta, \lambda} 
 M_{\alpha, \beta}^{(\lambda)}  \hat{Q}_{\lambda} \hat{c}_{\alpha}^{\dagger} \hat{c}_{\beta}~.
\label{H_EPI}
\end{align}
In the above equations, we have denoted electron and phonon creation operators by $\hat{c}^{\dagger}_{\alpha}$ and $\hat{b}^{\dagger}_{\kappa}$, respectively. Greek indices denote sets of quantum numbers for both electron and phonon single-particle basis functions. For the sake of concreteness, electronic indices are given explicitly by $\alpha = (i_{\alpha}, \sigma_{\alpha})$ in a lattice representation, where $i_{\alpha}$ is the lattice site and $\sigma_{\alpha}$ is the spin-orbital index. The latter accounts for the component of the spin along a given direction, as well as for an orbital index in the case where two or more orbitals reside on a given lattice site (this also includes the case where the unit cell contains two or more atoms). For phonons, we have $\kappa = (i_{\kappa}, s_{\kappa})$, where $i_{\kappa}$ is, again, a lattice site, while $s_{\kappa}$ is a branch index. The full derivation of the EPI Hamiltonian, including the microscopic definition of the coefficients, is given in~\ref{app: Derivation Hamiltonian}. Without loss of generality, we assume that $M^{(\lambda)}_{\alpha, \beta} = M^{(\lambda)}_{\beta, \alpha}$. This condition, together with $t_{\alpha, \beta} = t^*_{\beta, \alpha}$, $f_{\kappa, \lambda} = f^*_{\lambda, \kappa}$, and $U_{\alpha, \beta, \delta, \gamma} = U^*_{\delta, \gamma, \alpha, \beta}$, guarantees that the Hamiltonian is Hermitian.

Equations \eqref{H}, \eqref{H_IE}, \eqref{H_EEI}, \eqref{H_IP}, and \eqref{H_EPI} explicitly display their dependence on the sets of parameters ${\bm t} \equiv \lbrace t_{\alpha, \beta} \rbrace$, ${\bm U} \equiv \lbrace U_{\alpha, \beta, \gamma, \delta} \rbrace$, ${\bm f} \equiv \lbrace f_{\kappa, \lambda} \rbrace$, and ${\bm M} \equiv \lbrace M^{(\lambda)}_{\alpha, \beta} \rbrace$. The phonon displacement operator $\hat{Q}_{\lambda}$ in \eqref{H_EPI} is defined as
\begin{align}
\hat{Q}_{\lambda}  
& =      \sum_{ \kappa}   f_{ \lambda, -\kappa }^{-1/2}  \, \frac{1}{\sqrt{2}} \left( \hat{b}^{\dagger}_{\kappa} + \hat{b}_{\kappa} \right)~, 
\label{Q operator}
\end{align}
where $- \kappa$ is the composite index that results after applying spatial inversion to the position index $i_{\kappa}$ of the composite index $\kappa$ (see~\ref{app: Q}).

The parameters ${\bm t}$, ${\bm U}$, and ${\bm f}$ have dimensions of energy, $[E]$. The field operators $\hat{c}$ and $\hat{b}$ are dimensionless. The operators $\hat{Q}$ have dimensions of $[E^{-1/2}]$, and therefore the ${\bm M}$ parameters have dimensions of $[E^{3/2}]$.

Although Eq.~\eqref{H_EPI} makes use of the widely used definition of the ${\bm M}$ parameters, for the present derivation of the LWF it is convenient to rewrite it as
\begin{align}
\hat{\cal H}_{\rm EPI}({\bm I}) = \sum_{\alpha, \beta, \kappa } I^{(\kappa)}_{\alpha, \beta}  \, \hat{q}_{\kappa} \hat{c}_{\alpha}^{\dagger} \hat{c}_{\beta}~,
\label{H_EPI for LW}
\end{align}
where we have introduced a dimensionless displacement operator
\begin{align}
\hat{q}_{\kappa} \equiv \frac{1}{\sqrt{2}} \left( \hat{b}^{\dagger}_{\kappa} + \hat{b}_{\kappa} \right)~,
\label{dimensionless q}
\end{align}
as well as the renormalized EPI parameters ${\bm I} \equiv \lbrace  I^{(\kappa)}_{\alpha, \beta} \rbrace $ given by
\begin{align}
 I^{(\kappa)}_{\alpha, \beta}  \equiv \sum_{\lambda} M_{\alpha, \beta}^{(\lambda)}     f_{ \lambda, -\kappa }^{-1/2}~,
\label{renormalized I}
\end{align}
with dimensions of energy, $[E]$. In this way, the operators have no hidden dependence on the parameters $\bm f$. We will revert to the use of the $\bm M$ parameters starting from Section~\ref{sec: second-order selfenergy}.

We carry out our formal derivations for a general basis of single-particle states. The position (lattice) representation may be particularly useful, e.g., to study the Hubbard model, or any strongly correlated system. However, in some cases it is useful to switch to the Bloch state representation, which diagonalizes the independent-particle Hamiltonians. We then denote the lattice coordinates by $\boldsymbol{R}_i$ (while the equilibrium position of a nucleus is, in general, $\boldsymbol{R}_{i, n} \equiv \boldsymbol{R}_i + \boldsymbol{B}_n$, where $\boldsymbol{B}_n$ is a vector of the unit cell basis), and we introduce the Fourier transform matrix elements,
\begin{align}
F_{\boldsymbol{k}, i} = \frac{1}{\sqrt{N}} e^{i \boldsymbol{k} \cdot \boldsymbol{R}_i}~,
\label{Fourier matrix}
\end{align}
where $N$ is the number of lattice sites; then
\begin{align}
& \sum_{i } F_{\boldsymbol{k}, i}   \, F^*_{ \boldsymbol{k}', i }   = \delta_{\boldsymbol{k}, \boldsymbol{k}'}~, \quad \sum_{\boldsymbol{k}}     F^*_{ \boldsymbol{k}, i } \, F_{\boldsymbol{k}, j}  = \delta_{i,j}~.
\end{align}
We assume that the matrices $\bm t$ and $\bm f$ are diagonal in the spin-orbital or branch indices, respectively:
\begin{align}
t_{\alpha, \beta} = \delta_{\sigma_{\alpha}, \sigma_{\beta}} t_{i_{\alpha}, i_{\beta}}(\sigma_{\alpha})~, \quad f_{\kappa, \lambda} = \delta_{s_{\kappa}, s_{\lambda}} f_{i_{\kappa}, i_{\lambda}}(s_{\kappa})~.
\end{align}
They are diagonalized with respect to the lattice index by the Fourier transform matrix,
\begin{align}
&  \sum_{i,j} F^*_{\boldsymbol{k}, i} \, t_{i, j}(\sigma) \, F_{ \boldsymbol{k}' , j}  = \delta_{\boldsymbol{k}, \boldsymbol{k}'} \, \epsilon_{\boldsymbol{k}, \sigma} \quad \Leftrightarrow \quad t_{i, j}(\sigma)   =   \sum_{\boldsymbol{k}} F_{\boldsymbol{k} , i}   \, \epsilon_{\boldsymbol{k}, \sigma} \, F^*_{\boldsymbol{k}, j}~, \nonumber \\
&  \sum_{i,j} F^*_{\boldsymbol{k}, i} \, f_{i, j}(s) \, F_{ \boldsymbol{k}' , j}  = \delta_{\boldsymbol{k}, \boldsymbol{k}'} \, \omega_{\boldsymbol{k}, s} \quad \Leftrightarrow \quad f_{i, j}(s)   =   \sum_{\boldsymbol{k}} F_{\boldsymbol{k} , i}   \, \omega_{\boldsymbol{k}, s} \, F^*_{\boldsymbol{k}, j}~, 
\label{Fourier diagonalization}
\end{align}
where $\epsilon_{\boldsymbol{k}, \sigma}$ is the electronic spectrum and $\omega_{\boldsymbol{k}, s}$ is the phonon dispersion. The operators are then transformed according to  
\begin{align}
& \hat{c}^{\dagger}_{i, \sigma} = \sum_{\boldsymbol{k}} F^*_{\boldsymbol{k}, i} \hat{c}^{\dagger}_{\boldsymbol{k}, \sigma}~, \quad \hat{c}_{i, \sigma} = \sum_{\boldsymbol{k}} F_{\boldsymbol{k}, i} \hat{c}_{\boldsymbol{k}, \sigma}~, \nonumber \\
& \hat{b}^{\dagger}_{i, s} = \sum_{\boldsymbol{k}} F^*_{\boldsymbol{k}, i} \hat{b}^{\dagger}_{\boldsymbol{k}, s}~, \quad \hat{b}_{i, s} = \sum_{\boldsymbol{k}} F_{\boldsymbol{k}, i} \hat{b}_{\boldsymbol{k}, s}~.
\end{align}

Because of the complete generality of the notation introduced in Eqs.~\eqref{H}, \eqref{H_IE}, \eqref{H_EEI}, \eqref{H_IP}, and \eqref{H_EPI}, the present formalism can be directly applied to both normal and superconducting systems; in the latter case, it is convenient to use the Gor'kov-Nambu representation of the electronic basis set~\cite{Anokhin96}, as we will see in Section~\ref{sec: Eliashberg}. 

\section{Derivation of the LWF}
\label{sect:LWF}

The LWF for a purely electronic system was derived in Ref.~\cite{Potthoff06} in a way that is {\it nonperturbative} in the EEI. Hence, it is suitable for applications to strongly correlated electronic systems, where the skeleton-diagram expansion~\cite{LuttingerWard, StefanuccivanLeeuwen}, which is typically used in the weakly-interacting regime, may fail to converge. Here, we extend the derivation of Ref.~\cite{Potthoff06} by including the phonon terms $\hat{\cal H}_{\rm IP}({\bm f})$ and $\hat{\cal H}_{\rm EPI}({\bm I})$ in the Hamiltonian. In our notation, Ref.~\cite{Potthoff06} only treats the electronic subsystem described by $\hat{\cal H}_{\rm IE}({\bm t})$ and $\hat{\cal H}_{\rm EEI}({\bm U})$.

\subsection{Definitions}
We start by introducing a number of useful quantities, and the corresponding notations. The grand potential of the system at temperature $T$ is
\begin{align}
\Omega_{{\bm t}, {\bm U}, {\bm f}, {\bm I}} = - T \ln({\cal Z}_{{\bm t}, {\bm U}, {\bm f}, {\bm I}})~,
\end{align}
where the partition function is
\begin{align}
{\cal Z}_{{\bm t}, {\bm U}, {\bm f}, {\bm I}} = {\rm tr} \exp\left[- \frac{\hat{\cal H}({\bm t}, {\bm U}, {\bm f}, {\bm I}) - \mu \hat{N}}{T} \right]~,
\end{align}
$\mu$ being the electron chemical potential and $\hat{N}$ the total electron number operator; the symbol ``tr'' denotes the trace over the Fock space associated with the Hamiltonian $\hat{\cal H}({\bm t}, {\bm U}, {\bm f}, {\bm I})$. In the framework of a path-integral formulation, we introduce the action
\begin{align}
  A^{(c, \overline{c}, b, b^*)}_{{\bm t}, {\bm U}, {\bm f}, {\bm I}}  
& \equiv \frac{1}{T} \sum_n \sum_{\alpha, \beta} \overline{c}_{\alpha}(i \omega_n) \left[ (i \omega_n + \mu) \delta_{\alpha, \beta} - t_{\alpha, \beta} \right] c_{\beta}(i \omega_n) \nonumber \\
& \quad + \frac{1}{T}\sum_n \sum_{\kappa, \lambda}  b^*_{\kappa}(i \Omega_n) \left(  i \Omega_n   \delta_{\kappa, \lambda} - f_{\kappa, \lambda} \right) b_{\lambda}(i \Omega_n) \nonumber \\
& \quad - \frac{1}{2} \sum_{\alpha, \beta, \gamma, \delta} U_{\alpha, \beta, \delta, \gamma} \int_0^{1/T} d\tau \, \overline{c}_{\alpha}(\tau) \overline{c}_{\beta}(\tau) c_{\gamma}(\tau) c_{\delta}(\tau)  \nonumber \\
& \quad - \sum_{\alpha, \beta, \kappa} I_{\alpha, \beta}^{(\kappa)} \int_0^{1/T} d\tau \, q_{\kappa}(\tau) \overline{c}_{\alpha}(\tau)  c_{\beta}(\tau)~,
\end{align}
where the fermionic and bosonic field operators appearing in the Hamiltonian $\hat{\cal H}({\bm t}, {\bm U}, {\bm f}, {\bm I})$ have been replaced by the corresponding Grassmann numbers ($\overline{c}, c$) and complex numbers ($b^*, b, q$), respectively. Moreover, $\omega_n = (2n+1) \pi T$ is a fermionic Matsubara frequency, $\Omega_n = 2 n \pi T$ is a bosonic Matsubara frequency, and~\cite{note1}
\begin{align}
& c_{\alpha}(i \omega_n) = T  \int_0^{1/T} d\tau \, e^{i \omega_n \tau}  c_{\alpha}(\tau)~,  \quad \quad \overline{c}_{\alpha}(i \omega_n) = T  \int_0^{1/T} d\tau \, e^{-i \omega_n \tau}  \overline{c}_{\alpha}(\tau)~, \nonumber \\
& b_{\kappa}(i \Omega_n) = T  \int_0^{1/T} d\tau \, e^{i \Omega_n \tau}  b_{\kappa}(\tau)~,  \quad \quad b^*_{\kappa}(i \Omega_n) = T  \int_0^{1/T} d\tau \, e^{-i \Omega_n \tau}  b^*_{\alpha}(\tau)~.
\label{Matsubara transform}
\end{align}
Finally,
\begin{align}
q_{\kappa}(\tau)  
  =         \frac{1}{\sqrt{2}} \left[ b^{*}_{\kappa}(\tau) + b_{\kappa}(\tau) \right]  =         \sum_{i \Omega_n}  \frac{e^{i \Omega_n \tau}}{\sqrt{2}} \left[ b^{*}_{\kappa}(i \Omega_n) + b_{\kappa}(- i \Omega_n) \right]~, 
\end{align}
where we have used Eq.~\eqref{dimensionless q}.
 
The partition-function path integral is written as
\begin{align}
{\cal Z}_{{\bm t}, {\bm U}, {\bm f}, {\bm I}} \equiv \int  D(\overline{c}, c) D(b^*, b) \,   e^{A^{(c, \overline{c}, b, b^*)}_{{\bm t}, {\bm U}, {\bm f}, {\bm I}}}~.
\end{align}
The fully-interacting one-body electron and phonon Green's functions (GFs) are given, respectively, by 
\begin{align}
  G_{{\bm t}, {\bm U}, {\bm f}, {\bm I}; \alpha, \beta}(i \omega_n)   \equiv \frac{-T^{-1}}{{\cal Z}_{{\bm t}, {\bm U}, {\bm f}, {\bm I}}} \int  D(\overline{c}, c) D(b^*, b) \, c_{\alpha}(i \omega_n) \overline{c}_{\beta}(i \omega_n) e^{A^{(c, \overline{c}, b, b^*)}_{{\bm t}, {\bm U}, {\bm f}, {\bm I}}}
\label{full G}
\end{align}
and 
\begin{align}
 P_{{\bm t}, {\bm U}, {\bm f}, {\bm I}; \kappa, \lambda}(i \Omega_n)  \equiv \frac{-T^{-1}}{{\cal Z}_{{\bm t}, {\bm U}, {\bm f}, {\bm I}}} \int D(\overline{c}, c) D(b^*, b) \, b_{\kappa}(i \Omega_n) b^*_{\lambda}(i \Omega_n) e^{A^{(c, \overline{c}, b, b^*)}_{{\bm t}, {\bm U}, {\bm f}, {\bm I}}}~.
\label{full P}
\end{align}
We also introduce the free one-body electron GF,
\begin{align}
G_{{\bm t}, {\bf 0}, {\bm f}, {\bf 0}; \alpha, \beta}(i \omega_n) = \left( \frac{1}{  i \omega_n   {\mathbb{1}} + {\boldsymbol \mu} - {\bm t}} \right)_{\alpha, \beta}~,
\label{free G}
\end{align}
and the free one-body phonon GF,
\begin{align}
P_{{\bm t}, {\bf 0}, {\bm f}, {\bf 0}; \kappa, \lambda}(i \Omega_n) = \left( \frac{1}{  i \Omega_n   {\mathbb{1}} - {\bm f}} \right)_{\kappa, \lambda}~,
\label{free P}
\end{align}
which are obtained from Eqs.~\eqref{full G} and \eqref{full P}, respectively, by setting ${\bm U} = {\bm 0}$ and ${\bm I} = {\bm 0}$. In Eq.~\eqref{free G}, we introduced the identity matrix $\mathbb{1}$ and allowed the chemical potential $\boldsymbol{\mu}$ to be a diagonal matrix in the spin indices. This allows us to describe spin-polarized systems, and/or do calculations in the Nambu-Gor'kov representation.

All GFs are matrices in the particle Greek indices. To denote the full matrices (as opposed to their elements), we use notations such as ${\bm G}_{{\bm t}, {\bm U}, {\bm f}, {\bm I}}(i \omega_n)$ and ${\bm P}_{{\bm t}, {\bm U}, {\bm f}, {\bm I}}(i \Omega_n)$. When the dependence on the frequency is not made explicit, e.g., as in ${\bm G}_{{\bm t}, {\bm U}, {\bm f}, {\bm I}}$ and ${\bm P}_{{\bm t}, {\bm U}, {\bm f}, {\bm I}}$, we mean that the matrix symbols include the dependence on the Matsubara frequencies as well.

The following important identity holds:
\begin{align}
\Omega_{{\bm t}, {\bm 0}, {\bm f}, {\bm 0}}  
  =     T {\rm Tr} \ln \left( - T {\bm G}_{{\bm t}, {\bm 0}, {\bm f}, {\bm 0}} \right)
       - T {\rm Tr} \ln \left( - T {\bm P}_{{\bm t}, {\bm 0}, {\bm f}, {\bm 0}}  \right)~,
       \label{noninteracting Omega Tr ln}
\end{align}
where the notation ${\rm Tr} {\bm X} \equiv   \sum_n \sum_{\alpha} e^{i \omega_n 0^+} X_{\alpha, \alpha}(i \omega_n)$ was used. The property \eqref{noninteracting Omega Tr ln} is proven in~\ref{app: Proof nonint Omega Tr ln}.

The electronic/phonon self-energy matrices are defined, respectively, as
\begin{align}
{\bm \Sigma}_{{\bm t}, {\bm U}, {\bm f}, {\bm I}}(i \omega_n) \equiv {\bm G}^{-1}_{{\bm t}, {\bf 0}, {\bm f}, {\bf 0}}(i \omega_n) -  {\bm G}^{-1}_{{\bm t}, {\bm U}, {\bm f}, {\bm I}}(i \omega_n)~,
\end{align}
and
\begin{align}
{\bm \Lambda}_{{\bm t}, {\bm U}, {\bm f}, {\bm I}}(i \Omega_n) \equiv {\bm P}^{-1}_{{\bm t}, {\bf 0}, {\bm f}, {\bf 0}}(i \Omega_n) -  {\bm P}^{-1}_{{\bm t}, {\bm U}, {\bm f}, {\bm I}}(i \Omega_n)~.
\label{phonon self-energy}
\end{align}
\subsection{Auxiliary functionals}

We introduce the following functionals:
\begin{align}
\widetilde{\Omega}_{{\bm U}, {\bm I}}\left[{\bm G}^{-1}_{\bm 0} ; {\bm P}^{-1}_{\bm 0}\right] \equiv - T \ln \widetilde{\cal Z}_{{\bm U}, {\bm I}}\left[{\bm G}^{-1}_{\bm 0} ; {\bm P}^{-1}_{\bm 0}\right]~,
\label{Omega functional}
\end{align}
\begin{align}
  \widetilde{\cal Z}_{{\bm U}, {\bm I}}\left[{\bm G}^{-1}_{\bm 0} ; {\bm P}^{-1}_{\bm 0}\right]  \equiv \int D(\overline{c}, c) D(b^*, b) \, e^{\widetilde{A}^{(c, \overline{c}, b, b^*)}_{ {\bm U},  {\bm I}}  \left[{\bm G}^{-1}_{\bm 0} ; {\bm P}^{-1}_{\bm 0}\right]}~,
\end{align}
and
\begin{align}
 \widetilde{A}^{(c, \overline{c}, b, b^*)}_{  {\bm U},  {\bm I}}\left[{\bm G}^{-1}_{\bm 0} ; {\bm P}^{-1}_{\bm 0}\right]   
& \equiv \frac{1}{T}\sum_n \sum_{\alpha, \beta} \overline{c}_{\alpha}(i \omega_n) G^{-1}_{0; \alpha, \beta}(i \omega_n) c_{\beta}(i \omega_n) \nonumber \\
& \quad + \frac{1}{T}\sum_n \sum_{\kappa, \lambda}  b^*_{\kappa}(i \Omega_n) P^{-1}_{0; \kappa, \lambda}(i \Omega_n) b_{\lambda}(i \Omega_n) \nonumber \\
& \quad - \frac{1}{2} \sum_{\alpha, \beta, \gamma, \delta} U_{\alpha, \beta, \delta, \gamma} \int_0^{1/T} d\tau \, \overline{c}_{\alpha}(\tau) \overline{c}_{\beta}(\tau) c_{\gamma}(\tau) c_{\delta}(\tau)  \nonumber \\
& \quad - \sum_{\alpha, \beta, \lambda} I_{\alpha, \beta}^{(\lambda)} \int_0^{1/T} d\tau \, q_{\lambda}(\tau) \overline{c}_{\alpha}(\tau)  c_{\beta}(\tau)~.
\end{align}
The above quantities are functionals of the matrices ${\bm G}^{-1}_{\bm 0}$ and ${\bm P}^{-1}_{\bm 0}$, which ought to be considered as {\it free variables}. They should not be confused with the quantities ${\bm G}^{-1}_{{\bm t}, {\bm 0}, {\bm f}, {\bm 0}}(i \omega_n)$ and ${\bm P}^{-1}_{{\bm t}, {\bm 0}, {\bm f}, {\bm 0}}(i\Omega_n)$, which have a precise physical meaning. If the substitution $\left[ {\bm G}^{-1}_{\bm 0} ; {\bm P}^{-1}_{\bm 0} \right]\rightarrow \left[ {\bm G}^{-1}_{{\bm t}, {\bm 0}, {\bm f}, {\bm 0}}; {\bm P}^{-1}_{{\bm t}, {\bm 0}, {\bm f}, {\bm 0}} \right] $ is made, then  
\begin{align}
\widetilde{\Omega}_{{\bm U}, {\bm M}}\left[ {\bm G}^{-1}_{{\bm t}, {\bm 0}, {\bm f}, {\bm 0}} ; {\bm P}^{-1}_{{\bm t}, {\bm 0}, {\bm f}, {\bm 0}} \right] = \Omega_{{\bm t}, {\bm U}, {\bm f}, {\bm M}}  
\end{align}
follows by construction (this explains the adopted notation).

We now define the following functional,
\begin{align}
\widetilde{\bm G}_{{\bm U}, {\bm I}}\left[{\bm G}^{-1}_{\bm 0}; {\bm P}^{-1}_{\bm 0} \right](i \omega_n) \equiv - \frac{1}{T} \frac{\delta \widetilde{\Omega}_{{\bm U}, {\bm I}}\left[{\bm G}^{-1}_{\bm 0} ; {\bm P}^{-1}_{\bm 0} \right]}{\delta   {\bm G}^{-1 T}_{\bm 0}(i \omega_n)}~,
\end{align}
where ${\bm G}^{-1 T} _{\bm 0}$ denotes the transpose~\cite{note2} of the matrix ${\bm G}^{-1}_{\bm 0}$. The functional $\widetilde{\bm G}_{{\bm U}, {\bm I}}$ has the following property:
\begin{align}
\widetilde{\bm G}_{{\bm U}, {\bm I}}\left[ {\bm G}^{-1}_{{\bm t}, {\bm 0}, {\bm f}, {\bm 0}}; {\bm P}^{-1}_{{\bm t}, {\bm 0}, {\bm f}, {\bm 0}}\right](i \omega_n) = {\bm G}_{{\bm t}, {\bm U}, {\bm f}, {\bm I}}(i \omega_n)~.
\end{align}
Analogously, we introduce the functional 
\begin{align}
\widetilde{\bm P}_{{\bm U}, {\bm I}}\left[{\bm G}^{-1}_{\bm 0} ; {\bm P}^{-1}_{\bm 0} \right](i \Omega_n) \equiv   \frac{1}{T} \frac{\delta \widetilde{\Omega}_{{\bm U}, {\bm I}}\left[{\bm G}^{-1}_{\bm 0}; {\bm P}^{-1}_{\bm 0}\right]}{\delta   {\bm P}^{-1 T}_{\bm 0}(i \Omega_n)}~,
\end{align}
with the property
\begin{align}
\widetilde{\bm P}_{{\bm U}, {\bm I}}\left[ {\bm G}^{-1}_{{\bm t}, {\bm 0}, {\bm f}, {\bm 0}}; {\bm P}^{-1}_{{\bm t}, {\bm 0}, {\bm f}, {\bm 0}}\right](i \Omega_n) = {\bm P}_{{\bm t}, {\bm 0}, {\bm f}, {\bm 0}}(i \Omega_n)~.
\end{align}
Importantly, the functionals $\widetilde{\bm G}_{{\bm U}, {\bm I}}$ and $\widetilde{\bm P}_{{\bm U}, {\bm I}}$ do not depend on the parameters ${\bm t}$ and ${\bm f}$, characterizing the independent-particle system.

\subsection{Variable substitutions}

The functionals introduced above depend on the variables ${\bm G}^{-1}_{\bm 0}$, ${\bm P}^{-1}_{\bm 0}$. We now make two variable substitutions, choosing ${\bm G}$, ${\bm \Sigma}$, ${\bm P}$, and ${\bm \Lambda}$ in such a way that
\begin{align}
\widetilde{\bm G}_{{\bm U}, {\bm I}}\left[{\bm G}^{-1} + {\bm \Sigma} ; {\bm P }^{-1} + {\bm \Lambda} \right](i \omega_n) =  {\bm G}(i \omega_n)
\end{align}
and
\begin{align}
\widetilde{\bm P}_{{\bm U}, {\bm I}}\left[{\bm G}^{-1} + {\bm \Sigma} ; {\bm P }^{-1} + {\bm \Lambda} \right](i \Omega_n) =  {\bm P}(i \Omega_n)~.
\end{align}
These two relations can be used to define the following functionals of the new independent variables ${\bm \Sigma}$ and ${\bm \Lambda}$,
\begin{align}
\widetilde{\bm G}_{{\bm U},  {\bm I}}[{\bm \Sigma} ; {\bm \Lambda}](i \omega_n)   \equiv \widetilde{\bm G}_{{\bm U}, {\bm I}}\left[\widetilde{\bm G}^{-1}_{{\bm U}, {\bm I}}[{\bm \Sigma} ;  {\bm \Lambda}] +{\bm \Sigma} ; \widetilde{\bm P}^{-1}_{{\bm U}, {\bm I}}[{\bm \Sigma} ; {\bm \Lambda}] + {\bm \Lambda} \right](i \omega_n)~,
\end{align}
\begin{align}
 \widetilde{\bm P}_{{\bm U},  {\bm I}}[{\bm \Sigma} ; {\bm \Lambda}](i \Omega_n)   \equiv \widetilde{\bm P}_{{\bm U}, {\bm I}}\left[\widetilde{\bm G}^{-1}_{{\bm U}, {\bm I}}[{\bm \Sigma} ;  {\bm \Lambda}] + {\bm \Sigma} ; \widetilde{\bm P}^{-1}_{{\bm U}, {\bm I}}[{\bm \Sigma} ; {\bm \Lambda}]+{\bm \Lambda} \right](i \Omega_n)~.
\end{align}
When the substitution $\left[ {\bm \Sigma} ; {\bm \Lambda} \right] \rightarrow \left[ {\bm \Sigma}_{{\bm t}, {\bm U}, {\bm f}, {\bm I}} ; {\bm \Lambda}_{{\bm t}, {\bm U}, {\bm f}, {\bm I}} \right]$ is made, the relations
\begin{align}
\widetilde{\bm G}_{{\bm U}, {\bm I}}[{\bm \Sigma}_{{\bm t}, {\bm U}, {\bm f}, {\bm I}} ; {\bm \Lambda}_{{\bm t}, {\bm U}, {\bm f}, {\bm I}} ](i \omega_n) =   {\bm G}_{{\bm t}, {\bm U}, {\bm f}, {\bm I}}(i \omega_n)  
\end{align}
and
\begin{align}
\widetilde{\bm P}_{{\bm U}, {\bm I}}[{\bm \Sigma}_{{\bm t}, {\bm U}, {\bm f}, {\bm I}} ; {\bm \Lambda}_{{\bm t}, {\bm U}, {\bm f}, {\bm I}} ](i \Omega_n) =   {\bm P}_{{\bm t}, {\bm U}, {\bm f}, {\bm I}}(i \Omega_n)  
\end{align}
follow by construction.

\subsection{Functional of the self-energies}
We now introduce the functional
\begin{align}
  \widetilde{F}_{{\bm U},  {\bm I}}[{\bm \Sigma} ; {\bm \Lambda}]  
& \equiv \widetilde{\Omega}_{{\bm U},  {\bm I}}[ \widetilde{\bm G}_{{\bm U},  {\bm I}}[{\bm \Sigma} ; {\bm \Lambda}]^{-1}  + {\bm \Sigma} ;  \widetilde{\bm P}_{{\bm U},  {\bm I}}[{\bm \Sigma} ; {\bm \Lambda}]^{-1} + {\bm \Lambda}] \nonumber \\
& \quad - T {\rm Tr} \ln \left( - T \widetilde{\bm G}_{{\bm U},  {\bm I}}[{\bm \Sigma} ; {\bm \Lambda}] \right)   + T {\rm Tr} \ln \left( - T \widetilde{\bm P}_{{\bm U},  {\bm I}}[{\bm \Sigma} ; {\bm \Lambda}] \right)~,
\label{F functional}
\end{align}
which is written in terms of the other functionals introduced above.

The following important properties hold:
\begin{align}
- \frac{1}{T} \frac{\delta \widetilde{F}_{{\bm U},  {\bm I}}[{\bm \Sigma} ; {\bm \Lambda}]}{ \delta {\bm \Sigma}^T(i \omega_n)} =   \widetilde{\bm G}_{{\bm U},  {\bm I}}[{\bm \Sigma} ; {\bm \Lambda}](i \omega_n)
\label{Sigma derivative}
\end{align}
and
\begin{align}
 \frac{1}{T} \frac{\delta \widetilde{F}_{{\bm U},  {\bm I}}[{\bm \Sigma} ; {\bm \Lambda}]}{ \delta {\bm \Lambda}^T(i \Omega_n)} =   \widetilde{\bm P}_{{\bm U},  {\bm I}}[{\bm \Sigma} ; {\bm \Lambda}](i \Omega_n)~.
 \label{Lambda derivative}
\end{align}
The proof of Eqs.~\eqref{Sigma derivative} and \eqref{Lambda derivative} is given in~\ref{app: Proof 0}.

\subsection{Generalized LWF}

We assume that the pair of functionals $\left( \widetilde{\bm G}_{{\bm U},  {\bm I}}[{\bm \Sigma} ; {\bm \Lambda}]  , \widetilde{\bm P}_{{\bm U},  {\bm I}}[{\bm \Sigma} ; {\bm \Lambda}] \right)$ 
can be inverted, yielding the pair $\left( \widetilde{\bm \Sigma}_{{\bm U},  {\bm I}}[{\bm G} ; {\bm P}]  , \widetilde{\bm \Lambda}_{{\bm U},  {\bm I}}[{\bm G} ; {\bm P}] \right)$. We then define the generalized LWF as the Legendre transform of $\widetilde{F}$, i.e.
\begin{align}
\widetilde{\Phi}_{{\bm U},  {\bm I}}[{\bm G} ; {\bm P}] & \equiv  \widetilde{F}_{{\bm U},  {\bm I}}\left[\widetilde{\bm \Sigma}_{{\bm U},  {\bm I}}[{\bm G} ; {\bm P}]  ; \widetilde{\bm \Lambda}_{{\bm U},  {\bm I}}[{\bm G} ; {\bm P}] \right] \nonumber \\
& \quad  + T {\rm Tr} \left( \widetilde{\bm \Sigma}_{{\bm U},  {\bm I}}[{\bm G} ; {\bm P}] \,  {\bm G}   \right)   - T {\rm Tr} \left( \widetilde{\bm \Lambda}_{{\bm U},  {\bm I}}[{\bm G} ; {\bm P}] \,  {\bm P}   \right)~.
\label{Definition LWF}
\end{align}
Taking Eq.~\eqref{F functional} into account, we write $\widetilde{\Phi}_{{\bm U},  {\bm I}}[{\bm G} ; {\bm P}] $ as 
\begin{align}
 \widetilde{\Phi}_{{\bm U},  {\bm I}}[{\bm G} ; {\bm P}]  
& =  \widetilde{\Omega}_{{\bm U}, {\bm I}}\!\left[  {\bm G}^{-1} \! + \widetilde{\bm \Sigma}_{{\bm U},  {\bm I}}[{\bm G} ; {\bm P}]  ;  {\bm P}^{-1}   + \widetilde{\bm \Lambda}_{{\bm U},  {\bm I}}[{\bm G} ; {\bm P}]  \right] - T {\rm Tr} \ln \left( - T  {\bm G}  \right)  \nonumber \\
& \quad + T {\rm Tr} \left( \widetilde{\bm \Sigma}_{{\bm U},  {\bm I}}[{\bm G} ; {\bm P}] \,  {\bm G}   \right)    + T {\rm Tr} \ln \left( - T  {\bm P}  \right)       - T {\rm Tr} \left( \widetilde{\bm \Lambda}_{{\bm U},  {\bm I}}[{\bm G} ; {\bm P}] \,  {\bm P}   \right)~.
\label{Explicit LWF}
\end{align}
By construction, $\widetilde{\Phi}_{{\bm U},  {\bm I}}[{\bm G} ; {\bm P}]$ is independent of ${\bm t}$ and ${\bm f}$. It has the following properties:
\begin{itemize}
\item[1)] When the replacement $[{\bm G} ; {\bm P}] \rightarrow [{\bm G}_{{\bm t}, {\bm U}, {\bm f}, {\bm I}} ; {\bm P}_{{\bm t}, {\bm U}, {\bm f}, {\bm I}}]$ is made, the grand potential of the electron-phonon system is obtained as
\begin{align}
  {\Omega}_{{\bm t}, {\bm U}, {\bm f}, {\bm I}}
& =   \widetilde{\Phi}_{{\bm U},  {\bm I}}[{\bm G}_{{\bm t}, {\bm U}, {\bm f}, {\bm I}} ; {\bm P}_{{\bm t}, {\bm U}, {\bm f}, {\bm I}}]   + T {\rm Tr} \ln \left( - T  {\bm G}_{{\bm t}, {\bm U}, {\bm f}, {\bm I}} \right) \nonumber \\
& \quad    - T {\rm Tr} \left(  {\bm \Sigma}_{{\bm t}, {\bm U}, {\bm f}, {\bm I}} \,  {\bm G}_{{\bm t}, {\bm U}, {\bm f}, {\bm I}}   \right)   - T {\rm Tr} \ln \left( - T {\bm P}_{{\bm t}, {\bm U}, {\bm f}, {\bm I}} \right) \nonumber \\
& \quad + T {\rm Tr} \left(  {\bm \Lambda}_{{\bm t}, {\bm U}, {\bm f}, {\bm I}}  \,  {\bm P}_{{\bm t}, {\bm U}, {\bm f}, {\bm I}}   \right)~.
\end{align}

\item[2)] The functionals corresponding to the self-energies are obtained from the LWF via functional differentiation, i.e.
\begin{align}
\frac{1}{T} \frac{\delta \widetilde{\Phi}_{{\bm U},  {\bm I}}[{\bm G} ; {\bm P}] }{\delta {\bm G}^T } = \widetilde{\bm \Sigma}_{{\bm U},  {\bm I}}[{\bm G} ; {\bm P}]
\label{G derivative}
\end{align}
and
\begin{align}
- \frac{1}{T} \frac{\delta \widetilde{\Phi}_{{\bm U},  {\bm I}}[{\bm G} ; {\bm P}] }{\delta {\bm P}^T } = \widetilde{\bm \Lambda}_{{\bm U},  {\bm I}}[{\bm G} ; {\bm P}]~.
\label{P derivative}
\end{align}
The proof of Eqs.~\eqref{G derivative} and \eqref{P derivative} is given in~\ref{app: Proof 1}. These properties are often referred to as the {\it $ {\Phi}$-derivability} of the self-energies. Any approximation on $\widetilde{\Phi}_{{\bm U},  {\bm I}}[{\bm G} ; {\bm P}]$ that preserves its symmetry properties (such as invariance under gauge transformations and time translations) yields a conserving approximation for the self-energy~\cite{BaymKadanoff, Baym62, StefanuccivanLeeuwen}.

\item[3)] From the definitions of ${\bm \Sigma}$ and ${\bm \Lambda}$, we have ${\bm \Sigma}_{{\bm t}, {\bm 0}, {\bm f}, {\bm 0}} = {\bm 0}$ and ${\bm \Lambda}_{{\bm t}, {\bm 0}, {\bm f}, {\bm 0}} = {\bm 0}$. Therefore, in the non-interacting case we have
\begin{align}
\Omega_{{\bm t}, {\bm 0}, {\bm f}, {\bm 0}}  
& =   \widetilde{\Phi}_{{\bm 0},  {\bm 0}}[{\bm G}_{{\bm t}, {\bm 0}, {\bm f}, {\bm 0}} ; {\bm P}_{{\bm t}, {\bm 0}, {\bm f}, {\bm 0}}]   +  T {\rm Tr} \ln \left( - T {\bm G}_{{\bm t}, {\bm 0}, {\bm f}, {\bm 0}} \right) \nonumber \\
& \quad - T {\rm Tr} \ln \left( - T {\bm P}_{{\bm t}, {\bm 0}, {\bm f}, {\bm 0}}  \right)~.
\label{Omega physical}
\end{align}
Substituting Eq.~\eqref{noninteracting Omega Tr ln} into \eqref{Omega physical}, we find that the LWF vanishes for a non-interacting system:
\begin{align}
   \widetilde{\Phi}_{{\bm 0},  {\bm 0}}[{\bm G}_{{\bm t}, {\bm 0}, {\bm f}, {\bm 0}} ; {\bm P}_{{\bm t}, {\bm 0}, {\bm f}, {\bm 0}}] = 0~.
   \label{nonintPhi}
\end{align}
\end{itemize}

\section{Expansion of the LWF}
\label{sec: Expansion}
We now study the dependence of the LWF on the parameters ${\bm I}$, by means of a perturbative expansion in the EPI, close to ${\bm I} = {\bm 0}$. We expand the LWF in powers of ${\bm I}$, up to second order:
\begin{align}
\widetilde{\Phi}_{{\bm U},  {\bm I}}[{\bm G} ; {\bm P}]   \approx \widetilde{\Phi}_{{\bm U},  {\bf 0}}[{\bm G} ; {\bm P}] + \widetilde{\Phi}^{(1)}_{{\bm U},  {\bm I}}[{\bm G} ; {\bm P}]   + \widetilde{\Phi}^{(2)}_{{\bm U},  {\bm I}}[{\bm G} ; {\bm P}]~,
\label{Expansion of LWF}
\end{align}
where
\begin{align}
  \widetilde{\Phi}^{(1)}_{{\bm U},  {\bm I}}[{\bm G} ; {\bm P}]    =  \sum_{\lambda, \alpha, \beta} I^{(\lambda)}_{\alpha, \beta} \left. \frac{\partial \widetilde{\Phi}_{{\bm U},  {\bm I}}[{\bm G} ; {\bm P}]}{\partial I^{(\lambda)}_{\alpha, \beta}} \right|_{{\bm I} = {\bm 0}}~,
  \label{first order in M}
\end{align}
and
\begin{align}
  \widetilde{\Phi}^{(2)}_{{\bm U},  {\bm I}}[{\bm G} ; {\bm P}]    = \frac{1}{2} \sum_{\lambda, \alpha, \beta} \sum_{\lambda', \alpha', \beta'} I^{(\lambda)}_{\alpha, \beta} I^{(\lambda')}_{\alpha', \beta'}  \left. \frac{\partial^2 \widetilde{\Phi}_{{\bm U},  {\bm I}}[{\bm G} ; {\bm P}]}{\partial I^{(\lambda)}_{\alpha, \beta} \partial I^{(\lambda')}_{\alpha', \beta'}} \right|_{{\bm I} = {\bm 0}}~.
  \label{second order in M}
\end{align}
It should be noted that the LWF depends on ${\bm I}$ via the parametric dependences of $\widetilde{\bm \Sigma}_{{\bm U},  {\bm I}}$, $\widetilde{\bm \Lambda}_{{\bm U},  {\bm I}}$, and $\widetilde{\Omega}_{{\bm U}, {\bm I}}$---see  Eq.~\eqref{Explicit LWF}.

In the following, we derive the three terms of Eq.~\eqref{Expansion of LWF}. We first note that, in the absence of EPIs (i.e.~for~${\bm I} = {\bm 0}$), the electronic system is totally decoupled from the phonons. It follows that
\begin{align}
& \widetilde{\bm \Lambda}_{{\bm U},  {\bm 0}}[{\bm G} ; {\bm P}] \equiv {\bm 0}~,
& \widetilde{\bm \Sigma}_{{\bm U},  {\bm 0}}[{\bm G} ; {\bm P}] \equiv \widetilde{\bm \Sigma}_{\bm U}[{\bm G}]~,
\label{no EPI case}
\end{align}
where $\widetilde{\bm \Sigma}_{\bm U}[{\bm G}]$ is the {\it universal} self-energy functional for a system of electrons interacting via the Hamiltonian defined in~\eqref{H_EEI}.

\subsection{Zeroth-order term}
The zeroth-order term is
\begin{align}
  \widetilde{\Phi}_{{\bm U},  {\bm 0}}[{\bm G} ; {\bm P}]  
& = - T \ln \widetilde{\cal Z}_{{\bm U}, {\bm 0}}\!\left[  {\bm G}^{-1} \! + \widetilde{\bm \Sigma}_{{\bm U},  {\bm 0}}[{\bm G} ; {\bm P}]  ;  {\bm P}^{-1} + \widetilde{\bm \Lambda}_{{\bm U},  {\bm 0}}[{\bm G} ; {\bm P}]  \right] \nonumber \\  
& \quad  + T {\rm Tr} \left( \widetilde{\bm \Sigma}_{{\bm U},  {\bm 0}}[{\bm G} ; {\bm P}] {\bm G}   \right) - T {\rm Tr} \ln \left( - T  {\bm G}  \right) \nonumber \\
& \quad - T {\rm Tr} \left( \widetilde{\bm \Lambda}_{{\bm U},  {\bm 0}}[{\bm G} ; {\bm P}] {\bm P}   \right) + T {\rm Tr} \ln \left( - T  {\bm P}  \right)~.
\end{align}
At ${\bm I} = {\bm 0}$, we have
\begin{align}
  \widetilde{\cal Z}_{{\bm U}, {\bm 0}}\left[{\bm G}^{-1}_{\bm 0}; {\bm P}^{-1}_{\bm 0}\right]  
  & = \int D(\overline{c}, c) D(b^*, b) e^{\widetilde{A}^{(c, \overline{c}, b, b^*)}_{  {\bm U},  {\bm 0}}  \left[{\bm G}^{-1}_{\bm 0} ; {\bm P}^{-1}_{\bm 0} \right]} \nonumber \\
  & =  \frac{ 1   }{\prod_{ \Omega_n} \det\left[ - \frac{1}{T} {\bm P}^{-1}_{{\bf   0}}(i \Omega_n)   \right]   }   \widetilde{\cal Z}_{\bm U}\left[{\bm G}^{-1}_{\bm 0}\right]~,
  \label{partition function functional at M = 0}
\end{align}
where
\begin{align}
  \widetilde{\cal Z}_{\bm U}\left[{\bm G}^{-1}_{\bm 0}\right] \equiv \int D(\overline{c}, c)e^{\widetilde{A}^{(c, \overline{c}, 0, 0)}_{  {\bm U},  {\bm 0}}  \left[{\bm G}^{-1}_{\bm 0} ; {\bm 0} \right]}
\end{align}
is the partition-function functional for the system of interacting electrons alone. We now apply the identity $\ln \det = {\rm tr} \ln$, as well as Eq.~\eqref{no EPI case}, and we obtain
\begin{align}
  \widetilde{\Phi}_{{\bm U},  {\bm 0}}[{\bm G} ; {\bm P}]  
& = - T \ln \widetilde{\cal Z}_{\bm U} \left[  {\bm G}^{-1}+ \widetilde{\bm \Sigma}_{{\bm U} }[{\bm G}  ]    \right]   + T {\rm Tr} \left( \widetilde{\bm \Sigma}_{{\bm U} }[{\bm G}  ] {\bm G}   \right) - T {\rm Tr} \ln \left( - T  {\bm G}  \right)   \nonumber \\
& \equiv \widetilde{\Phi}^{\rm (E)}_{{\bm U} }[{\bm G}]~,
\end{align}
which is the LWF for the system of interacting electrons, completely decoupled from the phonon system. This is equivalent to the LWF for the electronic system derived in Ref.~\cite{Potthoff06}. In this limit, where the EPI is absent, the properties of the non-interacting phonon system still enter the physical grand potential via the term $- T {\rm Tr} \ln \left( - T {\bm P}_{{\bm t}, {\bm 0}, {\bm f}, {\bm 0}}\right)$, as can be seen from Eq.~\eqref{Omega physical}.

\subsection{First-order term}

We observe that
\begin{align}
& \frac{\partial \widetilde{\Omega}_{{\bm U}, {\bm I}} \left[  {\bm G}^{-1}   + \widetilde{\bm \Sigma}_{{\bm U},  {\bm I}}[{\bm G} ; {\bm P}]  ;  {\bm P}^{-1}   + \widetilde{\bm \Lambda}_{{\bm U},  {\bm I}}[{\bm G} ; {\bm P}]  \right]}{\partial I^{(\lambda)}_{\alpha, \beta}} \nonumber \\
& = - T {\rm Tr} \left( {\bm G} \frac{\partial \widetilde{\bm \Sigma}_{{\bm U}, {\bm I}}[{\bm G}, {\bm P}]}{\partial I^{(\lambda)}_{\alpha, \beta} } - {\bm P} \frac{\partial \widetilde{\bm \Lambda}_{{\bm U}, {\bm I}}[{\bm G}, {\bm P}]}{\partial I^{(\lambda)}_{\alpha, \beta} } \right) \nonumber \\
& \quad + T \frac{1}{\widetilde{\cal Z}_{{\bm U}, {\bm I}}\left[  {\bm G}^{-1}   + \widetilde{\bm \Sigma}_{{\bm U},  {\bm I}}[{\bm G} ; {\bm P}]  ;  {\bm P}^{-1}   + \widetilde{\bm \Lambda}_{{\bm U},  {\bm I}}[{\bm G} ; {\bm P}]  \right]}  \nonumber \\
& \quad \times  \int D(\overline{c}, c) D(b^*, b)\int_0^{1/T} d\tau q_{\lambda}(\tau) \overline{c}_{\alpha}(\tau)  c_{\beta}(\tau)  
\nonumber \\
& \quad \times e^{\widetilde{A}^{(c, \overline{c}, b, b^*)}_{  {\bm U},  {\bm I}}  \left[  {\bm G}^{-1}   + \widetilde{\bm \Sigma}_{{\bm U},  {\bm I}}[{\bm G} ; {\bm P}]  ;  {\bm P}^{-1}   + \widetilde{\bm \Lambda}_{{\bm U},  {\bm I}}[{\bm G} ; {\bm P}]  \right]}~.
\end{align}
Therefore, the first derivative of Eq.~\eqref{Explicit LWF} is
\begin{align}
 \frac{\partial \widetilde{\Phi}_{{\bm U},  {\bm I}}[{\bm G} ; {\bm P}] }{\partial I^{(\lambda)}_{\alpha, \beta}}  
& = T \frac{1}{\widetilde{\cal Z}_{{\bm U}, {\bm I}}\left[  {\bm G}^{-1}   + \widetilde{\bm \Sigma}_{{\bm U},  {\bm I}}[{\bm G} ; {\bm P}]  ;  {\bm P}^{-1}   + \widetilde{\bm \Lambda}_{{\bm U},  {\bm I}}[{\bm G} ; {\bm P}]  \right]}  \nonumber \\
& \quad \times  \int D(\overline{c}, c) D(b^*, b) \int_0^{1/T} d\tau q_{\lambda}(\tau) \overline{c}_{\alpha}(\tau)  c_{\beta}(\tau)  \nonumber \\
& \quad \times e^{\widetilde{A}^{(c, \overline{c}, b, b^*)}_{  {\bm U},  {\bm I}}  \left[  {\bm G}^{-1}   + \widetilde{\bm \Sigma}_{{\bm U},  {\bm I}}[{\bm G} ; {\bm P}]  ;  {\bm P}^{-1}   + \widetilde{\bm \Lambda}_{{\bm U},  {\bm I}}[{\bm G} ; {\bm P}]  \right]}~.
\label{First M derivative of LWF}
\end{align}
At ${\bm I} = {\bm 0}$, this functional vanishes. In fact, it is proportional to the ensemble average of $q_{\lambda}(\tau)$, which is linear in the bosonic fields [see Eq.~\eqref{dimensionless q}], evaluated with an action that is quadratic in the boson fields. Therefore, 
\begin{align}
\widetilde{\Phi}^{(1)}_{{\bm U},  {\bm I}}[{\bm G} ; {\bm P}] = 0~.
\end{align}
\subsection{Second-order term}
\label{sec: second-order selfenergy}
Starting from Eq.~\eqref{First M derivative of LWF}, we directly evaluate the second derivative of the LWF at ${\bm I} = {\bm 0}$, using the fact that the average of an odd-power combination of bosonic operators vanishes if the bosonic action is quadratic. We obtain
\begin{align}
 \left. \frac{\partial^2 \widetilde{\Phi}_{{\bm U},  {\bm I}}[{\bm G} ; {\bm P}] }{\partial I^{(\lambda)}_{\alpha, \beta}   \partial I^{(\lambda')}_{\alpha', \beta'}} \right|_{{\bm I} = {\bm 0}}  
& = - T \frac{1}{\widetilde{\cal Z}_{{\bm U}, {\bm 0}}\left[  {\bm G}^{-1}   + \widetilde{\bm \Sigma}_{ \bf U }[{\bm G}  ]  ;  {\bm P}^{-1}     \right]}  \nonumber \\
& \quad \times  \int D(\overline{c}, c) D(b^*, b)\int_0^{1/T} d\tau \, q_{\lambda}(\tau) \overline{c}_{\alpha}(\tau)  c_{\beta}(\tau)  \nonumber \\
& \quad \times \int_0^{1/T} d\tau' q_{\lambda'}(\tau') \overline{c}_{\alpha'}(\tau')  c_{\beta'}(\tau') \nonumber \\
& \quad \times e^{\widetilde{A}^{(c, \overline{c}, b, b^*)}_{  {\bm U},  {\bm 0}}  \left[  {\bm G}^{-1}   + \widetilde{\bm \Sigma}_{\bm U}[{\bm G}  ]  ;  {\bm P}^{-1}     \right]}~,
\label{Second M derivative of LWF}
\end{align}
where the partition-function functional at ${\bm I} = {\bm 0}$ can be factorized as shown in Eq.~\eqref{partition function functional at M = 0} as
\begin{align}
  \widetilde{\cal Z}_{{\bm U}, {\bm 0}}\left[  {\bm G}^{-1}   + \widetilde{\bm \Sigma}_{ \bm U }[{\bm G}  ]  ;  {\bm P}^{-1}     \right]   = \frac{\widetilde{\cal Z}_{ \bm U  }\left[{\bm G}^{-1}   + \widetilde{\bm \Sigma}_{ \bm U }[{\bm G}  ]  \right]}{\prod_{ \Omega_n} \det\left[ - \frac{1}{T} {\bm P}^{-1}(i \Omega_n)   \right]   }~.
\end{align}
We now perform the functional integration over the bosonic fields in Eq.~\eqref{Second M derivative of LWF}:
\begin{align}
&  \frac{ 1   }{\prod_{ \Omega_n} \! \det\left[ - \frac{1}{T} {\bm P}^{-1}(i \Omega_n)   \right]   } \! \int \!  D(b^*, b)     q_{\lambda}(\tau)      \, q_{\lambda'}(\tau')      e^{\frac{1}{T}\sum_n \sum_{\kappa, \lambda}  b^*_{\kappa}(i \Omega_n) P^{-1}_{  \kappa, \lambda}(i \Omega_n) b_{\lambda}(i \Omega_n) } \nonumber \\
& =  \frac{ 1   }{\prod_{ \Omega_n} \det\left[ - \frac{1}{T} {\bm P}^{-1}(i \Omega_n)   \right]   }     \sum_{i \Omega_n}  \frac{e^{i \Omega_n \left( \tau - \tau' \right) }}{2}    \nonumber \\
& \quad \times \int   D(b^*, b)     \left[ b^{*}_{\lambda}(i \Omega_n) b_{\lambda'}(  i \Omega_n) + b_{\lambda}(- i \Omega_n) b^{*}_{\lambda'}(-i \Omega_n) \right]     \nonumber \\
& \quad \times    e^{\frac{1}{T}\sum_m \sum_{\kappa'', \lambda''}  b^*_{\kappa''}(i \Omega_m) P^{-1}_{  \kappa'', \lambda''}(i \Omega_m) b_{\lambda''}(i \Omega_m) } \nonumber \\ 
& = - T \sum_{i \Omega_n}   e^{i \Omega_n \left( \tau - \tau' \right) }        \frac{1}{2}    \left[   P_{\lambda', \lambda}(i \Omega_n)   + P_{\lambda , \lambda'}(-i \Omega_n) \right]~. 
\end{align}
Replacing this result into Eq.~\eqref{Second M derivative of LWF}, transforming the fermionic fields to the Matsubara frequency representation, and performing the integrals over the time variables $\tau$ and $\tau'$, we arrive at the following result: 
\begin{align}
  \left. \frac{\partial^2 \widetilde{\Phi}_{{\bm U},  {\bm M}}[{\bm G} ; {\bm P}] }{\partial I^{(\lambda)}_{\alpha, \beta}   \partial I^{(\lambda')}_{\alpha', \beta'}} \right|_{{\bm I} = {\bm 0}}  
& =     \frac{1}{\widetilde{\cal Z}_{\bm U }\left[  {\bm G}^{-1}   + \widetilde{\bm \Sigma}_{ \bm U }[{\bm G}  ]     \right]}  \nonumber \\
& \quad \times \sum_{i \Omega_n}  \sum_{ n', m }  \int D(\overline{c}, c) c_{\beta}(i \omega_m) c_{\beta'}(i \omega_{n'} - i \Omega_n)    \nonumber \\
& \quad \times \overline{c}_{\alpha'}(i \omega_{n'}) \overline{c}_{\alpha}(i \omega_m - i \Omega_n)  \, e^{\widetilde{A}^{(c, \overline{c} )}_{ \bm U }  \left[  {\bm G}^{-1}   + \widetilde{\bm \Sigma}_{ \bm U  }[{\bm G}  ]        \right]}  \nonumber \\
& \quad \times       \frac{1}{2}    \left[   P_{\lambda', \lambda}(i \Omega_n)   + P_{\lambda , \lambda'}(-i \Omega_n) \right]~,
\label{Second M derivative of LWF, bosons done}
\end{align}
where we have introduced the electronic action functional in the absence of phonons,
\begin{align}
&  \widetilde{A}^{(c, \overline{c} )}_{  \bm U  }  \left[  {\bm G}^{-1}   + \widetilde{\bm \Sigma}_{ \bm U  }[{\bm G}  ]    \right]   \equiv \widetilde{A}^{(c, \overline{c}, 0, 0)}_{  {\bm U},  {\bm 0}}  \left[  {\bm G}^{-1}   + \widetilde{\bm \Sigma}_{ \bm U  }[{\bm G}  ]  ;  {\bm 0}     \right] \nonumber \\
& =   \frac{1}{T}\sum_n \sum_{\alpha, \beta} \overline{c}_{\alpha}(i \omega_n) \left\{ G^{-1}_{  \alpha, \beta}(i \omega_n) + \widetilde{  \Sigma}_{ \bm U ; \alpha, \beta }[{\bm G}  ](i \omega_n)  \right\}   c_{\beta}(i \omega_n) \nonumber \\
& \quad - \frac{1}{2} \sum_{\alpha, \beta, \gamma, \delta} U_{\alpha, \beta, \delta, \gamma} \int_0^{1/T} d\tau \overline{c}_{\alpha}(\tau) \overline{c}_{\beta}(\tau) c_{\gamma}(\tau) c_{\delta}(\tau)~.   
\end{align}

Eq.~\eqref{Second M derivative of LWF, bosons done} requires the evaluation of a two-particle correlator, in the presence of a non-Gaussian weight. Due to the form of the action, this correlator is equivalent to a two-electron GF. More precisely, its functional dependence on $\bf G$ (here an independent variable of arbitrary value) is the same as the functional dependence of the two-electron GF on the one-electron GF. We can then apply the usual concepts of many-body perturbation theory, provided that the GF is replaced by the independent variable ${\bm G}$. Only at the end of the derivation, ${\bm G}$ will be identified with the physical one-electron GF. Using a compact notation, we write
\begin{align}
& T^{-2} \left<  c_{\beta}(i \omega_m)    \,     c_{\beta'}(i \omega_{n'} - i \Omega_n) \, \overline{c}_{\alpha'}(i \omega_{n'})  \,   \overline{c}_{\alpha}(i \omega_m - i \Omega_n) \right> \nonumber \\
& =   \delta_{n, 0} \, G_{\beta, \alpha}(i \omega_m)  \,  G_{\beta', \alpha'}(i \omega_{n'}  )    - \delta_{m, n'} \, G_{\beta, \alpha'}(i \omega_m)  \, G_{\beta', \alpha}(i \omega_{m} - i \Omega_n)    \nonumber \\
& \quad + \delta_{n,0} \sum_{\nu, \mu , \nu', \mu'} G_{\beta, \nu}(i \omega_m) \, G_{\mu', \alpha'}(i \omega_{n'})  \, G_{\mu, \alpha}(i \omega_m) \, G_{\beta', \nu'}(i \omega_{n'})  \nonumber \\
& \quad \times \widetilde{\Gamma}_{{\bm U}; (\nu, \mu ; \nu', \mu')}[{\bm G}](i \omega_m, i \omega_{n'}) \nonumber \\
& \quad - \delta_{m, n'} \sum_{\nu, \mu , \nu', \mu'} G_{\beta, \nu}(i \omega_m) \,  G_{\mu, \alpha}(i \omega_m - i \Omega_n)  \, G_{\mu', \alpha'}(i \omega_m) \, G_{\beta', \nu'}(i \omega_m - i \Omega_n)  \nonumber \\
& \quad \times \widetilde{\Gamma}_{{\bm U}; (\nu, \mu' ; \nu', \mu)}[{\bm G}](i \omega_m, i \omega_m - i \Omega_n)~,
\end{align}
where $\widetilde{\Gamma}_{{\bm U}; (\nu, \mu ; \nu', \mu')}[{\bm G}](i \omega_m, i \omega_{n'})$ is the {\it reducible} four-leg vertex, a functional of $\bm G$, which depends parametrically on $\bm U$. By inserting this expansion into Eq.~\eqref{Second M derivative of LWF, bosons done}, and inserting the result into Eq.~\eqref{second order in M}, we finally obtain
\begin{align}
  \widetilde{\Phi}^{(2)}_{{\bm U},  {\bm I}}[{\bm G} ; {\bm P}]  & = \frac{T^2}{2} \sum_{ \alpha, \beta} \sum_{  \alpha', \beta'} \sum_{\kappa, \kappa'}  I^{(\kappa)}_{\alpha, \beta} I^{(\kappa')}_{\alpha', \beta'}       \frac{1}{2}    \left[   P_{\kappa', \kappa}(i \Omega_n)   + P_{\kappa , \kappa'}(-i \Omega_n) \right] \nonumber \\
& \quad \times \sum_{i \Omega_n}  \sum_{ i \omega_{n'}, i \omega_m }    \Bigg\{      \delta_{n, 0} \, G_{\beta, \alpha}(i \omega_m)  \,  G_{\beta', \alpha'}(i \omega_{n'}  )  \nonumber \\
& \quad   - \delta_{m, n'} \, G_{\beta, \alpha'}(i \omega_m)  \, G_{\beta', \alpha}(i \omega_{m} - i \Omega_n)    \nonumber \\
& \quad + \delta_{n,0} \sum_{\nu, \mu , \nu', \mu'} G_{\beta, \nu}(i \omega_m) \, G_{\mu', \alpha'}(i \omega_{n'})  \, G_{\mu, \alpha}(i \omega_m) \, G_{\beta', \nu'}(i \omega_{n'})  \nonumber \\
& \quad \times \widetilde{\Gamma}_{{\bm U}; (\nu, \mu ; \nu', \mu')}[{\bm G}](i \omega_m, i \omega_{n'}) \nonumber \\
& \quad - \delta_{m, n'} \sum_{\nu, \mu , \nu', \mu'} G_{\beta, \nu}(i \omega_m) \,  G_{\mu, \alpha}(i \omega_m - i \Omega_n)  \, G_{\mu', \alpha'}(i \omega_m)  \nonumber \\
& \quad \times G_{\beta', \nu'}(i \omega_m - i \Omega_n)  \, \widetilde{\Gamma}_{{\bm U}; (\nu, \mu' ; \nu', \mu)}[{\bm G}](i \omega_m, i \omega_m - i \Omega_n)   \Bigg\}~.
  \label{second order in M, simplification}
\end{align}

It is now convenient to make a change of variables: from now on, instead of $({\bm I}, {\bm P})$, we will use $({\bm M}, {\bm D})$, where 
\begin{align}
D_{\lambda, \lambda'}(i \Omega_n)    \equiv \sum_{ \kappa , \kappa'}   f_{ \lambda, -\kappa }^{-1/2}        
                   f_{ \lambda', -\kappa' }^{-1/2}               \frac{1}{2}    \left[   P_{  \kappa', \kappa}(i \Omega_n)   + P_{  \kappa , \kappa'}(-i \Omega_n) \right]~,
                   \label{definition full D}
\end{align}
and the $\bm M$ parameters are the EPI parameters introduced in Section~\ref{sec: Hamiltonian}. The following property holds,
\begin{align}
 {D}_{\lambda, \lambda'}( i \Omega_n) = D_{\lambda', \lambda} (- i \Omega_n)~. 
\end{align}
The quantity in the first line of Eq.~\eqref{second order in M, simplification} transforms as
\begin{align}
 \sum_{\kappa, \kappa'}  I^{(\kappa)}_{\alpha, \beta} I^{(\kappa')}_{\alpha', \beta'}       \frac{1}{2}    \left[   P_{\kappa', \kappa}(i \Omega_n)   + P_{\kappa , \kappa'}(-i \Omega_n) \right]    =         \sum_{\lambda, \lambda'} M^{(\lambda)}_{\alpha, \beta} M^{(\lambda')}_{\alpha', \beta'}     D_{\lambda, \lambda'}(i \Omega_n)~.
\end{align}
When evaluated at the physical {\it non-interacting} phonon GF [see Eq.~\eqref{free P}], Eq.~\eqref{definition full D} yields
\begin{align}
& \sum_{ \kappa}     f_{ \lambda, -\kappa }^{-1/2}        
              \sum_{ \kappa'}     f_{ \lambda', -\kappa' }^{-1/2}               \frac{1}{2}    \left[   P_{{\bm t}, {\bm 0}, {\bm f}, {\bm 0}; \kappa', \kappa}(i \Omega_n)   + P_{{\bm t}, {\bm 0}, {\bm f}, {\bm 0}; \kappa , \kappa'}(-i \Omega_n) \right] \nonumber \\
 & =  \sum_{ \kappa}     f_{ \lambda, -\kappa }^{-1/2}        
              \sum_{ \kappa'}     f_{ \lambda', -\kappa' }^{-1/2}               \frac{1}{2}    \left[   \left( \frac{1}{  i \Omega_n   \mathbb{1} - {\bm f}} \right)_{\kappa', \kappa}     + \left( \frac{1}{ -  i \Omega_n   \mathbb{1} - {\bm f}} \right)_{\kappa , \kappa'} \right] \nonumber \\
  & =  \delta_{s_{\lambda}, s_{\lambda'}} \sum_{ i }     f_{ i_{\lambda}, -i  }^{-1/2}(s_{\lambda})        
              \sum_{ j }     f_{ i_{\lambda'}, -j }^{-1/2}(s_{\lambda})               \frac{1}{2}  \sum_{\boldsymbol{q} }  \left[  F^*_{\boldsymbol{q}, j} \left( \frac{1}{  i \Omega_n    - \omega_{\boldsymbol{q}, s_{\lambda}} } \right)  F_{ \boldsymbol{q} , i}    \right. \nonumber \\
              & \quad \left. + \, F^*_{\boldsymbol{q}, i} \left( \frac{1}{ -  i \Omega_n     - \omega_{\boldsymbol{q}, s_{\lambda}}} \right)  F_{ \boldsymbol{q} , j} \right]   \nonumber \\
   & =  - \delta_{s_{\lambda}, s_{\lambda'}} \sum_{ i, j }  \sum_{\boldsymbol{q}, \boldsymbol{q}', \boldsymbol{q}''} F^*_{\boldsymbol{q}', i_{\lambda}}  \, \omega^{-1/2}_{\boldsymbol{q}', s_{\lambda}} F^*_{ \boldsymbol{q}' ,  i}        
                       F^*_{\boldsymbol{q}'', i_{\lambda'}}  \, \omega^{-1/2}_{\boldsymbol{q}'', s_{\lambda}} F^*_{ \boldsymbol{q}'' ,  j}                  \,    \frac{\omega_{\boldsymbol{q}, s_{\lambda}}}{   \Omega^2_n    + \omega^2_{\boldsymbol{q}, s_{\lambda}} }            \,  F^*_{\boldsymbol{q}, i} F_{ \boldsymbol{q} , j}  \nonumber \\
    & =  - \delta_{s_{\lambda}, s_{\lambda'}}  \sum_{\boldsymbol{q}  }    F_{\boldsymbol{q}, i_{\lambda}}  \,        
                       F^*_{\boldsymbol{q} , i_{\lambda'}}  \,                     \,    \frac{1}{   \Omega^2_n    + \omega^2_{\boldsymbol{q}, s_{\lambda}} }~,
\end{align}
where we have used the Fourier coefficients defined in Eq.~\eqref{Fourier matrix}, Eq.~\eqref{Fourier diagonalization}, and the property $\omega_{\boldsymbol{q}, s_{\lambda}} = \omega_{-\boldsymbol{q}, s_{\lambda}}$. In the last line, we recognize the bare phonon GF associated with the displacement operator, 
\begin{align}
D^{(0)}_{\lambda, \lambda'}(i \Omega_n)   \equiv -    \delta_{s_{\lambda}, s_{\lambda'}}  \frac{1}{N} \sum_{ \boldsymbol{q}}  e^{i \boldsymbol{q} \cdot (\boldsymbol{R}_{i_{\lambda}} - \boldsymbol{R}_{i_{\lambda'}})}  \frac{ 1 }{ \Omega^2_n + \omega^2_{\boldsymbol{q}, s_{\lambda}} } \equiv D_{{\bm t}, {\bm 0}, {\bm f}, {\bm 0}; \lambda, \lambda'}(i \Omega_n)~,
\label{bare phonon D(0Q)}
\end{align}
which motivates our notation.

The second-order term of the LWF is finally written as
\begin{align}
  \widetilde{\Phi}^{(2)}_{{\bm U},  {\bm M}}[{\bm G} ; {\bm D}]  \equiv    \widetilde{\Phi}^{(\rm F)}_{{\bm U},  {\bm M}}[{\bm G} ; {\bm D}] + \widetilde{\Phi}^{(\rm H)}_{{\bm U},  {\bm M}}[{\bm G} ; {\bm D}]~,
  \label{second order in M, final}
\end{align}
where we have separated a Fock term,
\begin{align}
  \widetilde{\Phi}^{(\rm F)}_{{\bm U},  {\bm M}}[{\bm G} ; {\bm D}]  & = - \frac{T^2}{2} \sum_{\lambda, \alpha, \beta} \sum_{\lambda', \alpha', \beta'} M^{(\lambda)}_{\alpha, \beta} M^{(\lambda')}_{\alpha', \beta'} \sum_{i \Omega_n} D_{\lambda, \lambda'}(i \Omega_n)   \nonumber \\
  & \quad \times \sum_{   i \omega_m }    \Big\{           G_{\beta, \alpha'}(i \omega_m)  \, G_{\beta', \alpha}(i \omega_{m} - i \Omega_n)    \nonumber \\
& \quad +   \sum_{\nu, \mu , \nu', \mu'} G_{\beta, \nu}(i \omega_m) \,  G_{\mu, \alpha}(i \omega_m - i \Omega_n)  \, G_{\mu', \alpha'}(i \omega_m) \nonumber \\
& \quad \times G_{\beta', \nu'}(i \omega_m - i \Omega_n) \, \widetilde{\Gamma}_{{\bm U}; (\nu, \mu' ; \nu', \mu)}[{\bm G}](i \omega_m, i \omega_m - i \Omega_n)  \Big\}~,
  \label{LFW, Fock term}
\end{align}
and a Hartree term,
\begin{align}
  \widetilde{\Phi}^{(\rm H)}_{{\bm U},  {\bm M}}[{\bm G} ; {\bm D}]  & = \frac{T^2}{2} \sum_{\lambda, \alpha, \beta} \sum_{\lambda', \alpha', \beta'} M^{(\lambda)}_{\alpha, \beta} M^{(\lambda')}_{\alpha', \beta'}   D_{\lambda, \lambda'}(0)  \nonumber \\
  & \quad \times \sum_{ i \omega_n , i \omega_m }    \Big\{        G_{\beta, \alpha}(i \omega_m)  \,  G_{\beta', \alpha'}(i \omega_n  )        \nonumber \\
& \quad +   \sum_{\nu, \mu , \nu', \mu'} G_{\beta, \nu}(i \omega_m) \, G_{\mu', \alpha'}(i \omega_n)  \, G_{\mu, \alpha}(i \omega_m) \, G_{\beta', \nu'}(i \omega_n)      \nonumber \\
& \quad \times \widetilde{\Gamma}_{{\bm U}; (\nu, \mu ; \nu', \mu')}[{\bm G}](i \omega_m, i \omega_n)   \Big\}~.
  \label{LWF, Hartree term}
\end{align}
Functionals analogous to \eqref{LFW, Fock term} and \eqref{LWF, Hartree term} were first derived in Ref.~\cite{Anokhin96} for the problem of phonon-mediated superconductivity in a {\it non-interacting} but disordered electron system. The most profound difference between our results \eqref{LFW, Fock term} and \eqref{LWF, Hartree term} and those of Ref.~\cite{Anokhin96} is in the nature of the reducible four-leg vertex $\widetilde{\Gamma}_{{\bm U}; (\nu, \mu ; \nu', \mu')}[{\bm G}](i \omega_m, i \omega_n)$. In our case, this is the universal reducible four-leg vertex for a system of electrons interacting via the Hamiltonian defined in~\eqref{H_EEI}. In the case of Ref.~\cite{Anokhin96}, the analogous four-leg vertex refers to a system of {\it non-interacting} disordered electrons. The four-leg vertex functional introduced in our work is therefore fundamentally different from that introduced in Ref.~\cite{Anokhin96}. Evident implications of this difference will be discussed below when we calculate the electronic self-energies---see below Section~\ref{Hartree and Fock terms: simplifications}. One minor difference between our formulas and the corresponding ones in Ref.~\cite{Anokhin96} lies in the temperature prefactors in Eqs.~\eqref{LFW, Fock term} and~\eqref{LWF, Hartree term}, which stem from the use of a different convention. (In this work we have followed the convention of Ref.~\cite{Potthoff06}.)

In summary, in this Section we have been able to prove that the LWF for the fully interacting electron-phonon system is given, up to the second order in the EPI matrix elements, by
\begin{align}
\widetilde{\Phi}_{{\bm U},  {\bm M}}[{\bm G} ; {\bm D}] & \approx \widetilde{\Phi}^{\rm (E)}_{{\bm U} }[{\bm G}  ] + \widetilde{\Phi}^{(\rm F)}_{{\bm U},  {\bm M}}[{\bm G} ; {\bm D}]   + \widetilde{\Phi}^{(\rm H)}_{{\bm U},  {\bm M}}[{\bm G} ; {\bm D}]~.
\label{terms of the LWF}
\end{align}
\section{Electronic self-energy from the expansion of the LWF}
\label{sect:self-energy}

We now derive the electronic self-energy by applying Eq.~\eqref{G derivative} to the terms of the LWF listed in Eq.~\eqref{terms of the LWF}. We obtain
\begin{align}
\widetilde{ \Sigma}_{{\bm U},  {\bm M}; \phi, \theta}[{\bm G} ; {\bm D}](i \omega_n)  & \approx \widetilde{\Sigma}^{(\rm E)}_{{\bm U};  \phi, \theta }[{\bm G}  ](i \omega_n) + \widetilde{\Sigma}^{(\rm F)}_{{\bm U},  {\bm M}; \phi, \theta}[{\bm G} ; {\bm D}](i \omega_n)  \nonumber \\
& \quad + \widetilde{\Sigma}^{(\rm H)}_{{\bm U},  {\bm M}; \phi, \theta}[{\bm G} ; {\bm D}](i \omega_n)~,
\label{Self-energy, functional notation}
\end{align}
where
\begin{align}
\widetilde{ \Sigma}^{(\rm E)}_{{\bm U} ; \phi, \theta}[{\bm G} ](i \omega_n) \equiv \frac{1}{T} \frac{\delta \widetilde{\Phi}^{\rm (E)}_{{\bm U} }[{\bm G}  ] }{\delta {  G}_{\theta, \phi}(i \omega_n) }~, 
\label{purely electronic Sigma}
\end{align}
\begin{align}
& \widetilde{ \Sigma}^{(\rm F)}_{{\bm U},  {\bm M}; \phi, \theta}[{\bm G} ; {\bm D}](i \omega_n)   \equiv \frac{1}{T} \frac{\delta \widetilde{\Phi}^{(\rm F)}_{{\bm U},  {\bm M}}[{\bm G} ; {\bm D}] }{\delta {  G}_{\theta, \phi}(i \omega_n) }  \nonumber \\
& = - T   \sum_{\lambda, \lambda'}  \sum_{i \Omega } D_{\lambda, \lambda'} (i \Omega )       \Bigg\{      \sum_{ \alpha ,   \beta }  M^{(\lambda)}_{\phi, \alpha }  M^{(\lambda')}_{\beta, \theta}   \, G_{\alpha , \beta}(i \omega_n - i \Omega ) \nonumber \\
& \quad +     \sum_{ \alpha, \beta ,  \alpha' }  M^{(\lambda)}_{\alpha, \beta } M^{(\lambda')}_{\alpha', \theta} \sum_{ \mu , \nu , \mu'} G_{\mu , \alpha}(i \omega_n)  \, G_{\nu, \alpha'}(i \omega_n - i \Omega ) \, G_{\beta , \mu' }(i \omega_n - i \Omega )  \nonumber \\
& \quad \times     \widetilde{\Gamma}_{{\bm U}; (\phi , \mu  ; \mu' , \nu)}[{\bm G}](i \omega_n, i \omega_n - i \Omega )   \nonumber \\  
& \quad +    \sum_{  \beta , \alpha' , \beta'} M^{(\lambda)}_{\phi, \beta} M^{(\lambda')}_{\alpha', \beta'} \sum_{\nu,   \nu', \mu'} G_{\beta, \mu'}(i \omega_n - i \Omega) \,      G_{\nu, \alpha'}(i \omega_n - i \Omega) \, G_{\beta', \nu'}(i \omega_n  )         \nonumber \\
& \quad \times \widetilde{\Gamma}_{{\bm U}; ( \mu' , \nu ; \nu', \theta)}[{\bm G}](i \omega_n - i \Omega, i \omega_n   ) \nonumber \\
& \quad + \frac{1}{2} \sum_{   i \omega_m }  \sum_{ \alpha, \beta} \sum_{  \alpha', \beta'} M^{(\lambda)}_{\alpha, \beta} M^{(\lambda')}_{\alpha', \beta'} \sum_{\nu, \mu , \nu', \mu'} G_{\beta, \mu }(i \omega_m) \,  G_{\mu', \alpha}(i \omega_m - i \Omega ) \,  G_{\nu, \alpha'}(i \omega_m)  \nonumber \\
& \quad \times    G_{\beta', \nu'}(i \omega_m - i \Omega )       \,  \frac{  \delta \widetilde{\Gamma}_{{\bm U}; (\mu  , \nu ; \nu', \mu')}[{\bm G}](i \omega_m, i \omega_m - i \Omega )  }{\delta {  G}_{\theta, \phi}(i \omega_n) }   \Bigg\}~, 
\label{Sigma Fock before simplification}
\end{align}
and
\begin{align}
&  \widetilde{ \Sigma}^{(\rm H)}_{{\bm U},  {\bm M}; \phi, \theta}[{\bm G} ; {\bm D}](i \omega_n) \equiv \frac{1}{T} \frac{\delta \widetilde{\Phi}^{(\rm H)}_{{\bm U},  {\bm M}}[{\bm G} ; {\bm D}] }{\delta {G}_{\theta, \phi}(i \omega_n) }  \nonumber \\
& =  T \sum_{\lambda, \lambda'} D_{\lambda, \lambda'} (0)      \sum_{ i \omega_m   }    \Bigg\{        \sum_{   \alpha , \beta }   M^{(\lambda)}_{\phi, \theta} M^{(\lambda')}_{\alpha , \beta }     \,  G_{\beta , \alpha }(i \omega_m  )         \nonumber \\
& \quad +     \sum_{ \alpha,  \alpha', \beta} M^{(\lambda)}_{\alpha, \beta} M^{(\lambda')}_{\alpha', \theta}    \sum_{ \mu , \nu , \mu'}  G_{\beta, \nu}(i \omega_m) \, G_{\mu', \alpha'}(i \omega_n)         \, G_{\mu, \alpha}(i \omega_m)   \nonumber \\
& \quad \times \widetilde{\Gamma}_{{\bm U}; (\phi, \mu' ; \nu, \mu)}[{\bm G}](i \omega_n, i \omega_m)     \nonumber \\
& \quad +     \sum_{\beta, \alpha', \beta'} M^{(\lambda)}_{\phi, \beta} M^{(\lambda')}_{\alpha', \beta'}    \sum_{\nu, \mu , \nu' } G_{\beta', \nu'}(i \omega_m)  \, G_{\mu, \alpha'}(i \omega_m) \, G_{\beta, \nu}(i \omega_n) \nonumber \\
& \quad \times   \widetilde{\Gamma}_{{\bm U}; (\nu', \mu ; \nu, \theta)}[{\bm G}](i \omega_m, i \omega_n)      \nonumber \\
& \quad   +  \frac{1}{2} \sum_{ i \omega_s   } \sum_{ \alpha, \beta , \alpha', \beta'} M^{(\lambda)}_{\alpha, \beta} M^{(\lambda')}_{\alpha', \beta'}   \sum_{\nu, \mu , \nu', \mu'} G_{\beta, \mu}(i \omega_m) \, G_{\mu', \alpha'}(i \omega_s)      \, G_{\nu, \alpha}(i \omega_m)       \nonumber \\
& \quad \times   G_{\beta', \nu'}(i \omega_s)  \, \frac{\delta  \widetilde{\Gamma}_{{\bm U}; (\mu, \nu ; \nu', \mu')}[{\bm G}](i \omega_m, i \omega_s) }{\delta {  G}_{\theta, \phi}(i \omega_n) }     \Bigg\}~. 
\label{Sigma Hartree before simplification}
\end{align}
\subsection{Hartree and Fock terms: simplifications}
\label{Hartree and Fock terms: simplifications}

\begin{figure}
\begin{tikzpicture}
\begin{feynman} 
\vertex (i) at(0,0);
\vertex (I) at(0,-0.3){\( \scriptstyle \phi \)};
\vertex (j) at(1,0);
\vertex (J) at(1,-0.3){\( \scriptstyle \alpha \)};
\vertex (k) at(0.5,0.866);
\vertex (K) at(0.3,1.166){\( \scriptstyle \lambda \)};
\vertex (name1) at(0.5, 0.3){\( \scriptstyle M \)};
\vertex (l) at(3.5,0);
\vertex (L) at(3.5,-0.3){\( \scriptstyle \beta \)};
\vertex (m) at(4.5,0);
\vertex (M) at(4.5,-0.3){\( \scriptstyle \theta \)};
\vertex (n) at(4,0.866);
\vertex (N) at(4.2,1.166){\( \scriptstyle \lambda' \)};
\vertex (name2) at(4, 0.3){\( \scriptstyle M \)};
\vertex(inv1) at(0,-3);
\vertex(inv2) at(0,0);
\diagram* { 
(i) -- (j) -- [fermion, edge label'=\(\scriptstyle i\omega_n - i \Omega\)] (k) -- [fermion, edge label'=\(\scriptstyle i\omega_n\)] (i),
(l) -- (m) -- [fermion, edge label'=\(\scriptstyle i\omega_n\)] (n) -- [fermion, edge label'=\(\scriptstyle i\omega_n - i \Omega\)] (l),
(l) -- [fermion, edge label'=\(\scriptstyle i\omega_n - i \Omega\)] (j),
(k) -- [boson, half left, looseness = 0.75, edge label=\( \scriptstyle  i \Omega\)] (n),
(inv1) -- [draw = none] (inv2),
}; 
\end{feynman}
\end{tikzpicture}
\begin{tikzpicture}
\begin{feynman}
\vertex (plus) at(-2.8,0.9){\( + \,\, \dfrac{1}{2} \)};
\vertex (a) at(0,-2.066);  
\vertex (A) at(0,-2.266){\( \scriptstyle \lambda \)};
\vertex (b) at(0.5,-1.2); 
\vertex (B) at(0.7,-1.2){\( \scriptstyle \alpha \)}; 
\vertex (c) at(-0.5,-1.2); 
\vertex (C) at(-0.7,-1.2){\( \scriptstyle \beta \)};
\vertex (name1) at(0,-1.5){\( \scriptstyle M \)};
\vertex (g) at(0.5,0); 
\vertex (G) at(0.7,0.05){\( \scriptstyle \mu' \)}; 
\vertex (h) at(-0.5,0); 
\vertex (H) at(-0.7,0){\( \scriptstyle \mu \)};  
\vertex (i) at(-1,0.866);  
\vertex (I) at(-1.2,0.866){\( \scriptstyle \nu \)}; 
\vertex (j) at(-0.5,1.732);
\vertex (J) at(-0.5,1.9){\( \scriptstyle \phi \)};
\vertex (k) at(0.5,1.732);
\vertex (K) at(0.5,1.9){\( \scriptstyle \theta \)};
\vertex (l) at(1,0.866); 
\vertex (L) at(1.25,0.916){\( \scriptstyle \nu' \)}; 
\vertex (name2) at(0,0.866){\( \scriptstyle U^{[6]} \)};
\vertex (q) at(0.5,2.932);  
\vertex (Q) at(0.7,2.932){\( \scriptstyle \beta' \)}; 
\vertex (r) at(0,3.798); 
\vertex (R) at(0,4){\( \scriptstyle \lambda' \)};  
\vertex (t) at(-0.5,2.932); 
\vertex (T) at(-0.7,2.982){\( \scriptstyle \alpha' \)};  
\vertex (name3) at(0,3.2){\( \scriptstyle M \)};
\vertex (om1a) at(-0.65,3.3){\( \scriptstyle i \omega_m \)};
\vertex (om1b) at(0.95,3.3){\( \scriptstyle i \omega_m - i \Omega \)};
\vertex (om2a) at(-0.65,-1.6){\( \scriptstyle i \omega_m \)};
\vertex (om2b) at(0.95,-1.6){\( \scriptstyle i \omega_m - i \Omega \)};
\diagram* { 
(a) -- [fermion ] (b)  -- (c)  -- [fermion ] (a)   ,
(g) -- (h)  -- [anti fermion, edge label=\(\scriptstyle i\omega_m\)] (i)  -- (j),
(j) -- [anti fermion, edge label=\(\scriptstyle i\omega_n\)] (k)  -- (l)  -- [anti fermion, edge label=\(\scriptstyle i\omega_m - i \Omega\)] (g),
(q) -- [fermion] (r)  -- [fermion ] (t)  -- (q),
(l) -- [fermion, edge label'=\(\scriptstyle i\omega_m - i \Omega\)] (q),
(t) -- [fermion, edge label'=\(\scriptstyle i\omega_m  \)] (i),
(b) -- [fermion, edge label'=\(\scriptstyle i\omega_m - i \Omega\)] (g),
(h) -- [fermion, edge label'=\(\scriptstyle i\omega_m  \)] (c),
(a) -- [boson, half left, looseness = 1.1, edge label'=\( \scriptstyle i \Omega\) ] (r),
};
\end{feynman}
\end{tikzpicture}

\caption{Diagrammatic representation of the Fock self-energy, $\Sigma^{\rm (F)}_{\phi, \theta}(i \omega_n)$---see Eq.~\eqref{Sigma Fock simplified}. The wavy line appearing in both diagrams corresponds to a fully-dressed (interacting) phonon propagator, namely, $D_{\lambda, \lambda'}(i \Omega)$ [compare with Eq.~\eqref{Sigma Fock simplified}]. In order to recover the algebraic expression for the Fock self-energy, Eq.~\eqref{Sigma Fock simplified}, the diagram in this figure must be intended to be multiplied by a factor $-T$.
\label{Fig: diagram Fock self-energy}}
\end{figure}
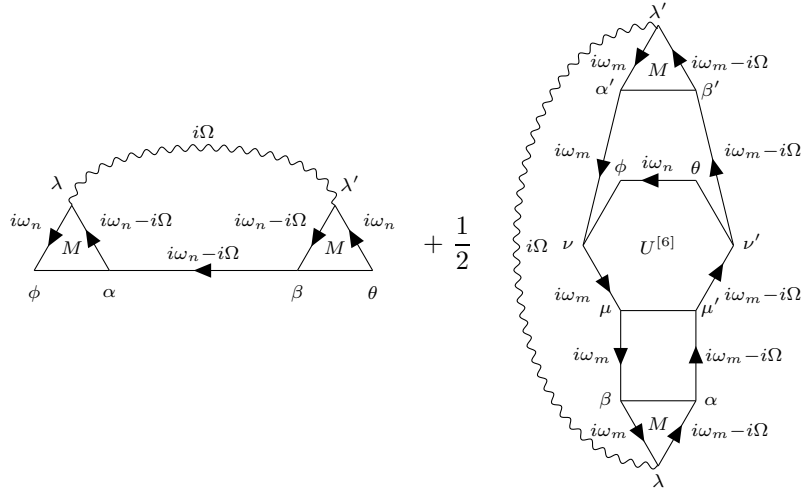

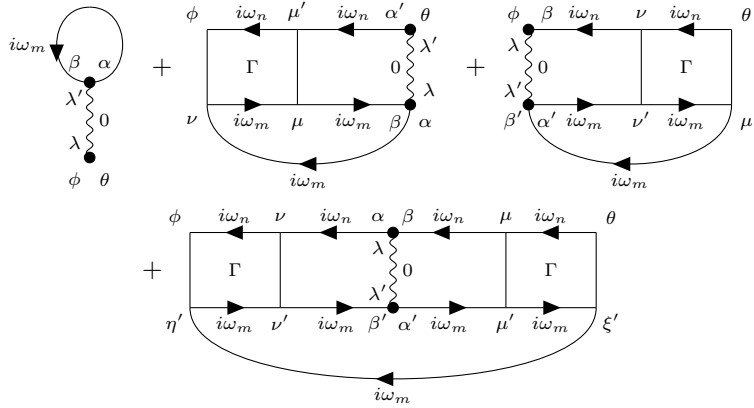
\begin{figure}
\begin{center}
\begin{tikzpicture}
\begin{feynman} 
\vertex (i) at(0,0);
\filldraw (0,0) circle(2pt); 
\vertex (il) at(-0.2,-0.3){\( \scriptstyle \phi \)};
\vertex (ir) at(0.2,-0.3){\( \scriptstyle \theta \)};
\vertex (iu) at(-0.2, 0.2){\( \scriptstyle \lambda \)};
\vertex (j) at(0,1);
\vertex (jb) at(-0.2, 0.8){\( \scriptstyle \lambda' \)};
\vertex (jr) at(0.2, 1.25){\( \scriptstyle \alpha \)};
\vertex (jl) at(-0.2, 1.25){\( \scriptstyle \beta \)};
\filldraw (0,1) circle(2pt);
\vertex (g) at(0,2); 
\diagram* { 
(j) -- [anti fermion, half left, edge label=\(\scriptstyle i \omega_m \)] (g) -- [plain, half left] (j),
(i) -- [boson, edge label'=\(\scriptstyle 0 \)] (j),
}; 
\end{feynman}
\end{tikzpicture}
\begin{tikzpicture}
\begin{feynman} 
\vertex (plus) at(-0.6,0.5){\( + \)}; 
\vertex (a) at(0,0);
\vertex (A) at(-0.2,-0.2){\( \scriptstyle \nu \)};
\vertex (b) at(1.2,0);
\vertex (B) at(1.2,-0.25){\( \scriptstyle \mu \)};
\vertex (c) at(1.2,1);
\vertex (C) at(1.2,1.25){\( \scriptstyle \mu' \)};
\vertex (d) at(0,1);
\vertex (D) at(-0.2,1.2){\( \scriptstyle \phi \)};
\vertex (name) at(0.6,0.5) {\( \scriptstyle \Gamma \)};
\vertex (m) at(2.7,1);
\filldraw (2.7,1) circle(2pt); 
\vertex (mr) at(2.9,1.2){\( \scriptstyle \theta \)}; 
\vertex (ml) at(2.5,1.25){\( \scriptstyle \alpha' \)};
\vertex (md) at(2.95, 0.8){\( \scriptstyle \lambda' \)};
\vertex (l) at(2.7,0);
\filldraw (2.7,0) circle(2pt);
\vertex (lr) at(2.9,-0.25){\( \scriptstyle \alpha \)}; 
\vertex (ll) at(2.5,-0.25){\( \scriptstyle \beta \)};
\vertex (lu) at(2.95, 0.2){\( \scriptstyle \lambda \)};
\diagram* { 
(a) -- [fermion, edge label'=\(\scriptstyle i\omega_m \)] (b) -- (c) -- [fermion, edge label'=\(\scriptstyle i\omega_n \)] (d) -- (a),
(c) -- [anti fermion, edge label=\(\scriptstyle i\omega_n \)] (m),
(l) -- [fermion, half left, looseness = 1, edge label=\(\scriptstyle i\omega_m \)] (a),
(m) -- [boson, edge label'=\(\scriptstyle 0 \)] (l),
(l) -- [anti fermion, edge label=\(\scriptstyle i\omega_m \)] (b), 
}; 
\end{feynman}
\end{tikzpicture}
\begin{tikzpicture}
\begin{feynman} 
\vertex (plus) at(-0.7,0.5){\( + \)}; 
\vertex (a) at(1.5,0);
\vertex (A) at(1.5,-0.2){\( \scriptstyle \nu' \)};
\vertex (b) at(2.7,0);
\vertex (B) at(2.9,-0.25){\( \scriptstyle \mu \)};
\vertex (c) at(2.7,1);
\vertex (C) at(2.9,1.2){\( \scriptstyle \theta \)};
\vertex (d) at(1.5,1);
\vertex (D) at(1.5,1.2){\( \scriptstyle \nu \)};
\vertex (name) at(2.1,0.5) {\( \scriptstyle \Gamma \)};
\vertex (m) at(0,1);
\filldraw (0,1) circle(2pt);
\vertex (mr) at(0.25, 1.2){\( \scriptstyle \beta \)};
\vertex (ml) at(-0.2, 1.2){\( \scriptstyle \phi \)};
\vertex (mb) at(-0.2, 0.8){\( \scriptstyle \lambda \)}; 
\vertex (l) at(0,0);
\filldraw (0,0) circle(2pt);
\vertex (ll) at(-0.2,-0.2){\( \scriptstyle \beta' \)};
\vertex (lr) at(0.25,-0.2){\( \scriptstyle \alpha' \)};
\vertex (lu) at(-0.2, 0.2){\( \scriptstyle \lambda' \)}; 
\diagram* { 
(a) -- [fermion, edge label'=\(\scriptstyle i\omega_m \)] (b) -- (c) -- [fermion, edge label'=\(\scriptstyle i\omega_n \)] (d) -- (a),
(d) -- [fermion, edge label'=\(\scriptstyle i\omega_n \)] (m), 
(l) -- [anti fermion, half right, looseness = 1, edge label'=\(\scriptstyle i\omega_m \)] (b),
(m) -- [boson, edge label=\(\scriptstyle 0 \)] (l),
(l) -- [fermion, edge label'=\(\scriptstyle i\omega_m \)] (a), 
}; 
\end{feynman}
\end{tikzpicture}

\begin{tikzpicture}
\begin{feynman} 
\vertex (plus) at(-0.5,0.5){\( + \)}; 
\vertex (a) at(0,0);
\vertex (A) at(-0.2,-0.2){\( \scriptstyle \eta' \)};
\vertex (b) at(1.2,0);
\vertex (B) at(1.2,-0.2){\( \scriptstyle \nu' \)};
\vertex (c) at(1.2,1);
\vertex (C) at(1.2,1.2){\( \scriptstyle \nu \)};
\vertex (d) at(0,1);
\vertex (D) at(-0.2,1.2){\( \scriptstyle \phi \)};
\vertex (name1) at(0.6,0.5) {\( \scriptstyle \Gamma \)};
\vertex (e) at(4.2,0);
\vertex (E) at(4.2,-0.2){\( \scriptstyle \mu' \)};
\vertex (f) at(5.4,0);
\vertex (F) at(5.6,-0.2){\( \scriptstyle \xi' \)};
\vertex (g) at(5.4,1);
\vertex (G) at(5.6,1.2){\( \scriptstyle \theta \)};
\vertex (h) at(4.2,1);
\vertex (H) at(4.2,1.2){\( \scriptstyle \mu \)};
\vertex (name2) at(4.8,0.5) {\( \scriptstyle \Gamma \)};
\vertex (m1) at(2.7,0);
\vertex (M1l) at(2.5,-0.2){\( \scriptstyle \beta' \)};
\vertex (M1r) at(2.9,-0.2){\( \scriptstyle \alpha' \)};
\vertex (M1u) at(2.5,0.2){\( \scriptstyle \lambda' \)};
\vertex (m2) at(2.7,1);
\vertex (M2l) at(2.5,1.2){\( \scriptstyle \alpha \)};
\vertex (M2r) at(2.9,1.2){\( \scriptstyle \beta \)};
\vertex (M2b) at(2.5,0.8){\( \scriptstyle \lambda \)};
\filldraw (2.7,0) circle(2pt); 
\filldraw (2.7,1) circle(2pt); 
\diagram* { 
(a) -- [fermion, edge label'=\(\scriptstyle i\omega_m \)] (b) -- (c) -- [fermion, edge label'=\(\scriptstyle i\omega_n \)] (d) -- (a),
(e) -- [fermion, edge label'=\(\scriptstyle i\omega_m \)] (f) -- (g) -- [fermion, edge label'=\(\scriptstyle i\omega_n \)] (h) -- (e),
(m1) -- [boson, edge label'=\(\scriptstyle 0 \)] (m2),
(e) -- [anti fermion, edge label=\(\scriptstyle i\omega_m \)] (m1) -- [anti fermion, edge label=\(\scriptstyle i\omega_m \)] (b), 
(c) -- [anti fermion, edge label=\(\scriptstyle i\omega_n \)] (m2) -- [anti fermion, edge label=\(\scriptstyle i\omega_n \)] (h),
(a) -- [anti fermion, half right, looseness = 0.6, edge label'=\(\scriptstyle i\omega_m \)] (f),
}; 
\end{feynman}
\end{tikzpicture}
\end{center}
\caption{Diagrammatic representation of the first four terms of the Hartree self-energy, $\Sigma^{\rm (H)}_{\phi, \theta}(i \omega_n)$---see Eq.~\eqref{Sigma Hartree simplified}. 
In order to recover the algebraic expression for the Hartree self-energy, Eq.~\eqref{Sigma Hartree simplified}, the diagram in this figure must be intended to be multiplied by a factor $T$. The different sign with respect to the Fock case is due to the presence of an additional fermionic loop. This remark applies to the diagrams of Figure \ref{Fig: diagram Hartree self-energy - 2} as well. 
\label{Fig: diagram Hartree self-energy - 1}}
\end{figure}

\begin{figure}
\begin{center}
\begin{tikzpicture}
\begin{feynman} 
\vertex (plus) at(-2.5,-0.734){\( \dfrac{1}{2} \,\, \)}; 
\vertex (a) at(-0.5,0);
\vertex (A) at(-0.7,-0.2){\( \scriptstyle \mu \)};
\vertex (b) at(-1,0.866);
\vertex (B) at(-1.1,1){\( \scriptstyle \xi \)};
\vertex (c) at(-0.5,1.732);
\vertex (C) at(-0.7,1.9){\( \scriptstyle \phi \)};
\vertex (d) at(0.5,1.732);
\vertex (D) at(0.6,1.9){\( \scriptstyle \theta \)};
\vertex (e) at(1,0.866);
\vertex (E) at(1.2,1){\( \scriptstyle \xi' \)};
\vertex (f) at(0.5,0);
\vertex (F) at(0.7,-0.15){\( \scriptstyle \mu' \)};
\vertex (name1) at(0,0.866) {\( \scriptstyle U^{[6]} \)};
\vertex (w) at(-0.5,-2);
\vertex (W) at(-0.7,-2){\( \scriptstyle \nu \)};
\vertex (x) at(-0.5,-3.2);
\vertex (X) at(-0.5,-3.45){\( \scriptstyle \eta \)};
\vertex (y) at(0.5,-3.2);
\vertex (Y) at(0.5,-3.4){\( \scriptstyle \eta' \)};
\vertex (z) at(0.5,-2);
\vertex (Z) at(0.7,-1.95){\( \scriptstyle \nu' \)};
\vertex (name2) at(0,-2.6) {\( \scriptstyle \Gamma \)};
\vertex (m1) at(-0.5,-1);
\vertex (M1u) at(-0.7,-0.85){\( \scriptstyle \beta \)};
\vertex (M1d) at(-0.7,-1.15){\( \scriptstyle \alpha \)};
\vertex (M1r) at(-0.3,-0.8){\( \scriptstyle \lambda \)};
\vertex (m2) at(0.5,-1);
\vertex (M2u) at(0.7,-0.8){\( \scriptstyle \alpha' \)};
\vertex (M2d) at(0.7,-1.2){\( \scriptstyle \beta' \)};
\vertex (M2l) at(0.3,-0.8){\( \scriptstyle \lambda' \)};
\filldraw (-0.5,-1) circle(2pt); 
\filldraw (0.5,-1) circle(2pt); 
\vertex (inv1) at(0,0);
\vertex (inv2) at(0,-3.9);
\diagram* { 
(a) -- [anti fermion, edge label=\(\scriptstyle i\omega_m \)] (b) -- (c) -- [anti fermion, edge label=\(\scriptstyle i\omega_n \)] (d) -- (e) -- [anti fermion, edge label=\(\scriptstyle i\omega_s \)] (f) -- (a),
(a) -- [fermion, edge label'=\(\scriptstyle i\omega_m \)] (m1),
(m2) -- [fermion, edge label'=\(\scriptstyle i\omega_s \)] (f),
(m1) -- [boson, edge label'=\(\scriptstyle 0 \)] (m2),
(w) -- [fermion, edge label'=\(\scriptstyle i\omega_m \)] (x) -- (y) -- [fermion, edge label'=\(\scriptstyle i\omega_s \)] (z) -- (w),
(m1) -- [fermion, edge label'=\(\scriptstyle i\omega_m \)] (w),
(z) -- [fermion, edge label'=\(\scriptstyle i\omega_s \)] (m2),
(x) -- [fermion, half left, looseness = 0.65, edge label=\(\scriptstyle i\omega_m \)] (b), 
(e) -- [fermion, half left, looseness = 0.65, edge label=\(\scriptstyle i\omega_s \)] (y),
(inv1) --[draw = none] (inv2),
}; 
\end{feynman}
\end{tikzpicture}
\begin{tikzpicture}
\begin{feynman} 
\vertex (plus) at(-3.2,-0.734){\( + \,\, \dfrac{1}{2}\)}; 
\vertex (a) at(-0.5,0);
\vertex (A) at(-0.7,-0.2){\( \scriptstyle \eta \)};
\vertex (b) at(-1,0.866);
\vertex (B) at(-1.1,1){\( \scriptstyle \nu \)};
\vertex (c) at(-0.5,1.732);
\vertex (C) at(-0.7,1.9){\( \scriptstyle \phi \)};
\vertex (d) at(0.5,1.732);
\vertex (D) at(0.6,1.9){\( \scriptstyle \theta \)};
\vertex (e) at(1,0.866);
\vertex (E) at(1.2,1.05){\( \scriptstyle \nu' \)};
\vertex (f) at(0.5,0);
\vertex (F) at(0.7,-0.15){\( \scriptstyle \eta' \)};
\vertex (name1) at(0,0.866) {\( \scriptstyle U^{[6]} \)};
\vertex (w) at(-0.5,-1.2);
\vertex (W) at(-0.7,-1.2){\( \scriptstyle \xi \)};
\vertex (x) at(-0.5,-2.4);
\vertex (X) at(-0.5,-2.65){\( \scriptstyle \mu \)};
\vertex (y) at(0.5,-2.4);
\vertex (Y) at(0.5,-2.6){\( \scriptstyle \mu' \)};
\vertex (z) at(0.5,-1.2);
\vertex (Z) at(0.7,-1.15){\( \scriptstyle \xi' \)};
\vertex (name2) at(0,-1.8) {\( \scriptstyle \Gamma \)};
\vertex (m1) at(-2.2,-1);
\vertex (M1u) at(-1.8,-0.85){\( \scriptstyle \alpha \)};
\vertex (M1d) at(-1.8,-1.15){\( \scriptstyle \beta \)};
\vertex (M1r) at(-2.5,-0.9){\( \scriptstyle \lambda \)};
\vertex (m2) at(2.2,-1);
\vertex (M2u) at(1.8,-0.85){\( \scriptstyle \beta' \)};
\vertex (M2d) at(1.8,-1.15){\( \scriptstyle \alpha' \)};
\vertex (M2l) at(2.5,-0.9){\( \scriptstyle \lambda' \)};
\filldraw (-2.2,-1) circle(2pt); 
\filldraw (2.2,-1) circle(2pt); 
\vertex (omm1) at(-1.4,-2){\( \scriptstyle i \omega_m \)};
\vertex (omm2) at(-1,0.25){\( \scriptstyle i \omega_m \)};
\vertex (oms1) at(1.4,-2){\( \scriptstyle i \omega_s \)};
\vertex (oms2) at(1,0.25){\( \scriptstyle i \omega_s \)};
\diagram* { 
(a) -- [anti fermion] (b) -- (c) -- [anti fermion, edge label=\(\scriptstyle i\omega_n \)] (d) -- (e) -- [anti fermion ] (f) -- (a),
(a) -- [fermion, edge label'=\(\scriptstyle i\omega_m \)] (w),
(z) -- [fermion, edge label'=\(\scriptstyle i\omega_s \)] (f),
(m1) -- [boson, half right, looseness = 1.67, edge label'=\(\scriptstyle 0 \)] (m2), 
(w) -- [fermion, edge label'=\(\scriptstyle i\omega_m \)] (x) -- (y) -- [fermion, edge label'=\(\scriptstyle i\omega_s \)] (z) -- (w),
(m1) -- [fermion, edge label=\(\scriptstyle i\omega_m \)] (b),
(e) -- [fermion, edge label=\(\scriptstyle i\omega_s \)] (m2),
(x) -- [fermion ] (m1), 
(m2) -- [fermion ] (y), 
}; 
\end{feynman}
\end{tikzpicture}

\begin{tikzpicture}
\begin{feynman} 
\vertex (plus) at(-3,-1.834){\( + \,\,\dfrac{1}{2}\)}; 
\vertex (a) at(-0.5,0);
\vertex (A) at(-0.7,-0.2){\( \scriptstyle \eta \)};
\vertex (b) at(-1,0.866);
\vertex (B) at(-1.1,1){\( \scriptstyle \zeta \)};
\vertex (c) at(-0.5,1.732);
\vertex (C) at(-0.7,1.9){\( \scriptstyle \phi \)};
\vertex (d) at(0.5,1.732);
\vertex (D) at(0.6,1.9){\( \scriptstyle \theta \)};
\vertex (e) at(1,0.866);
\vertex (E) at(1.2,1){\( \scriptstyle \zeta' \)};
\vertex (f) at(0.5,0);
\vertex (F) at(0.7,-0.15){\( \scriptstyle \eta' \)};
\vertex (name1) at(0,0.866) {\( \scriptstyle U^{[6]} \)};
\vertex (w) at(-0.5,-4.2);
\vertex (W) at(-0.7,-4.2){\( \scriptstyle \nu \)};
\vertex (x) at(-0.5,-5.4);
\vertex (X) at(-0.5,-5.65){\( \scriptstyle \chi \)};
\vertex (y) at(0.5,-5.4);
\vertex (Y) at(0.5,-5.6){\( \scriptstyle \chi' \)};
\vertex (z) at(0.5,-4.2);
\vertex (Z) at(0.7,-4.15){\( \scriptstyle \nu' \)};
\vertex (name2) at(0,-4.8) {\( \scriptstyle \Gamma \)};
\vertex (i) at(-0.5,-1);
\vertex (I) at(-0.7,-1){\( \scriptstyle \xi \)};
\vertex (j) at(-0.5,-2.2);
\vertex (J) at(-0.7,-2.2){\( \scriptstyle \mu \)};
\vertex (k) at(0.5,-2.2);
\vertex (K) at(0.7,-2.15){\( \scriptstyle \mu' \)};
\vertex (l) at(0.5,-1);
\vertex (L) at(0.7,-1){\( \scriptstyle \xi' \)};
\vertex (name3) at(0,-1.6) {\( \scriptstyle \Gamma \)};
\vertex (m1) at(-0.5,-3.2);
\vertex (M1u) at(-0.7,-3.05){\( \scriptstyle \beta \)};
\vertex (M1d) at(-0.7,-3.35){\( \scriptstyle \alpha \)};
\vertex (M1r) at(-0.3,-3){\( \scriptstyle \lambda \)};
\vertex (m2) at(0.5,-3.2);
\vertex (M2u) at(0.7,-3){\( \scriptstyle \alpha' \)};
\vertex (M2d) at(0.7,-3.35){\( \scriptstyle \beta' \)};
\vertex (M2l) at(0.3,-3){\( \scriptstyle \lambda' \)};
\filldraw (-0.5,-3.2) circle(2pt); 
\filldraw (0.5,-3.2) circle(2pt); 
\diagram* { 
(a) -- [anti fermion, edge label=\(\scriptstyle i\omega_m \)] (b) -- (c) -- [anti fermion, edge label=\(\scriptstyle i\omega_n \)] (d) -- (e) -- [anti fermion, edge label=\(\scriptstyle i\omega_s \)] (f) -- (a),
(a) -- [fermion, edge label'=\(\scriptstyle i\omega_m \)] (i),
(j) -- [fermion, edge label'=\(\scriptstyle i\omega_m \)] (m1),
(m2) -- [fermion, edge label'=\(\scriptstyle i\omega_s \)] (k),
(l) -- [fermion, edge label'=\(\scriptstyle i\omega_s \)] (f),
(m1) -- [boson, edge label'=\(\scriptstyle 0 \)] (m2),
(w) -- [fermion, edge label'=\(\scriptstyle i\omega_m \)] (x) -- (y) -- [fermion, edge label'=\(\scriptstyle i\omega_s \)] (z) -- (w),
(i) -- [fermion, edge label'=\(\scriptstyle i\omega_m \)] (j) -- (k) -- [fermion, edge label'=\(\scriptstyle i\omega_s \)] (l) -- (i),
(m1) -- [fermion, edge label'=\(\scriptstyle i\omega_m \)] (w),
(z) -- [fermion, edge label'=\(\scriptstyle i\omega_s \)] (m2),
(x) -- [fermion, half left, looseness = 0.55, edge label=\(\scriptstyle i\omega_m \)] (b), 
(e) -- [fermion, half left, looseness = 0.55, edge label=\(\scriptstyle i\omega_s \)] (y),
}; 
\end{feynman}
\end{tikzpicture}
\begin{tikzpicture}
\begin{feynman} 
\vertex (plus) at(-2.5,0.366){\( + \,\, \dfrac{1}{2}\)}; 
\vertex (a) at(-0.5,0);
\vertex (A) at(-0.7,-0.2){\( \scriptstyle \mu \)};
\vertex (b) at(-1,0.866);
\vertex (B) at(-1.1,1){\( \scriptstyle \nu \)};
\vertex (c) at(-0.5,1.732);
\vertex (C) at(-0.7,1.9){\( \scriptstyle \phi \)};
\vertex (d) at(0.5,1.732);
\vertex (D) at(0.6,1.9){\( \scriptstyle \theta \)};
\vertex (e) at(1,0.866);
\vertex (E) at(1.2,1.05){\( \scriptstyle \nu' \)};
\vertex (f) at(0.5,0);
\vertex (F) at(0.7,-0.15){\( \scriptstyle \mu' \)};
\vertex (name1) at(0,0.866) {\( \scriptstyle U^{[6]} \)};
\vertex (m1) at(-0.5,-1);
\vertex (M1u) at(-0.7,-0.85){\( \scriptstyle \beta \)};
\vertex (M1d) at(-0.7,-1.15){\( \scriptstyle \alpha \)};
\vertex (M1r) at(-0.3,-0.8){\( \scriptstyle \lambda \)};
\vertex (m2) at(0.5,-1);
\vertex (M2u) at(0.7,-0.8){\( \scriptstyle \alpha' \)};
\vertex (M2d) at(0.7,-1.2){\( \scriptstyle \beta' \)};
\vertex (M2l) at(0.3,-0.8){\( \scriptstyle \lambda' \)};
\filldraw (-0.5,-1) circle(2pt); 
\filldraw (0.5,-1) circle(2pt); 
\vertex (inv1) at(0,0);
\vertex (inv2) at(0,-3.6);
\diagram* { 
(a) -- [anti fermion, edge label=\(\scriptstyle i\omega_m \)] (b) -- (c) -- [anti fermion, edge label=\(\scriptstyle i\omega_n \)] (d) -- (e) -- [anti fermion, edge label=\(\scriptstyle i\omega_s \)] (f) -- (a),
(a) -- [fermion, edge label'=\(\scriptstyle i\omega_m \)] (m1),
(m2) -- [fermion, edge label'=\(\scriptstyle i\omega_s \)] (f),
(m1) -- [boson, edge label'=\(\scriptstyle 0 \)] (m2),
(m1) -- [fermion, half left, looseness = 1.2, edge label=\(\scriptstyle i\omega_m \)] (b), 
(e) -- [fermion, half left, looseness = 1.2, edge label=\(\scriptstyle i\omega_s \)] (m2),
(inv1) --[draw = none] (inv2),
}; 
\end{feynman}
\end{tikzpicture}

\end{center} 
\caption{Diagrammatic representation of the last four terms of the Hartree self-energy, $\Sigma^{\rm (H)}_{\phi, \theta}(i \omega_n)$---see Eq.~\eqref{Sigma Hartree simplified}. }
\label{Fig: diagram Hartree self-energy - 2}
\end{figure}
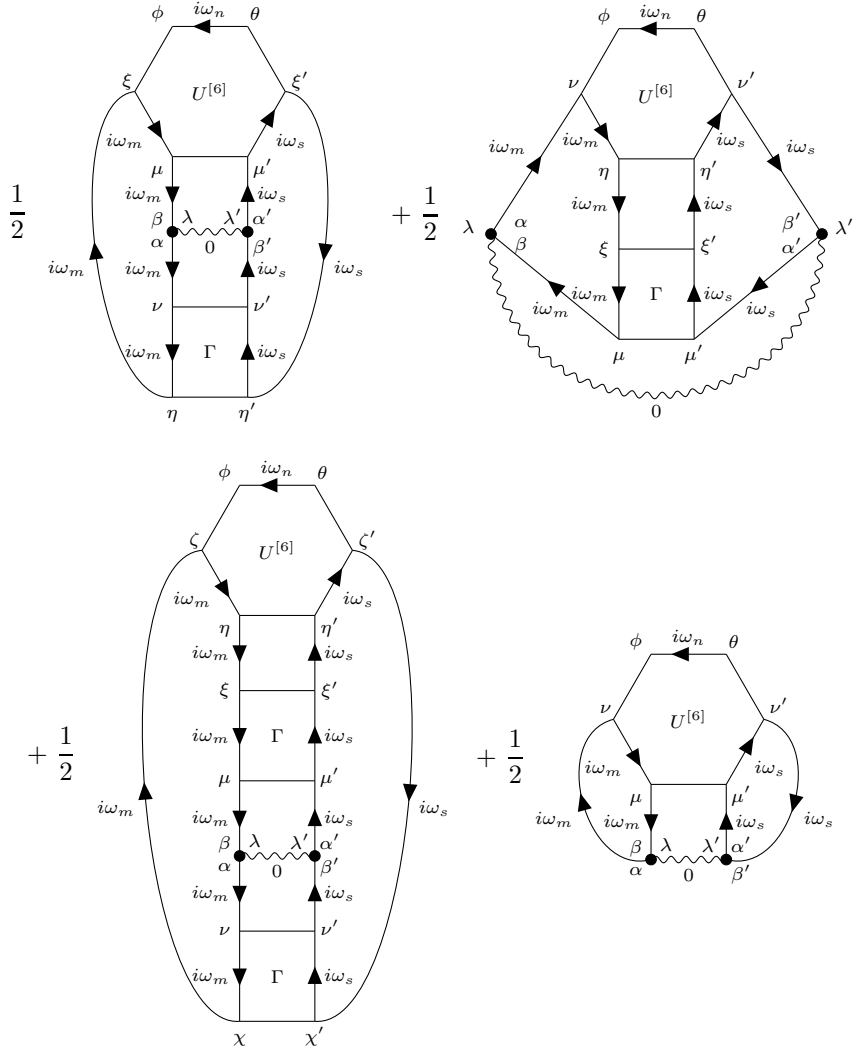

Eqs. \eqref{Sigma Fock before simplification} and \eqref{Sigma Hartree before simplification} require the evaluation of the functional derivative of the reducible four-leg vertex, i.e.
\begin{align}
\widetilde{\Gamma}^{[6]}_{{\bm U}; (\mu  , \nu ; \phi, \theta; \nu', \mu')}[{\bm G}](i \omega_m, i \omega_n, i \omega_s ) \equiv \frac{  \delta \widetilde{\Gamma}_{{\bm U}; (\mu  , \nu ;  \nu', \mu')}[{\bm G}](i \omega_m, i \omega_s )  }{\delta {  G}_{\theta, \phi}(i \omega_n) }~.
\label{reducible hexagon}
\end{align}
We hasten to emphasize that the quantity $\widetilde{\Gamma}^{[6]}_{{\bm U}; (\mu  , \nu ; \phi, \theta; \nu', \mu')}[{\bm G}](i \omega_m, i \omega_n, i \omega_s )$ introduced in Eq.~\eqref{reducible hexagon}, which is the {\it reducible} six-leg vertex, contains three distinct frequency arguments, in contrast to the analogous quantity for a system of {\it non-interacting} disordered electrons~\cite{Anokhin96}. Its expression can be simplified by means of the Bethe-Salpeter equation, which connects the four-leg reducible ($\Gamma$) and four-leg {\it irreducible} ($U^{[4]}$) kernels of the two-particle GFs~\cite{BetheSalpeter, StefanuccivanLeeuwen, GiulianiVignale}:  
\begin{align}
 \widetilde{\Gamma}_{{\bm U}; (\mu  , \nu ; \nu', \mu')}[{\bm G}](i \omega_m, i \omega_s )  
 &  =  \widetilde{U}^{[4]}_{{\bm U}; (\mu  , \nu ; \nu', \mu')}[{\bm G}](i \omega_m, i \omega_s ) \nonumber \\
 & \quad + \sum_{\xi, \xi', \eta, \eta'} \widetilde{U}^{[4]}_{{\bm U}; (\mu  , \xi ; \xi', \mu')}[{\bm G}](i \omega_m, i \omega_s ) \, G_{\xi, \eta}(i \omega_m) \nonumber \\
 & \quad \times G_{\eta', \xi'}(i \omega_s) \,  \widetilde{\Gamma}_{{\bm U}; (\eta  , \nu ; \nu', \eta')}[{\bm G}](i \omega_m, i \omega_s )~.
 \label{Bethe-Salpeter}
\end{align}
By applying the functional derivative in Eq.~\eqref{reducible hexagon} to Eq.~\eqref{Bethe-Salpeter}, and carrying out a few algebraic steps detailed in~\ref{app: simplification}, we can express Eq.~\eqref{reducible hexagon} in terms of $\Gamma$ and the irreducible quantities $U^{[4]}$ and $U^{[6]}$ only, where
\begin{align}
\widetilde{U}^{[6]}_{{\bm U}; (\mu  , \nu ; \phi, \theta; \nu', \mu')}[{\bm G}](i \omega_m, i \omega_n, i \omega_s ) \equiv \frac{  \delta \widetilde{U}^{[4]}_{{\bm U}; (\mu  , \nu ;  \nu', \mu')}[{\bm G}](i \omega_m, i \omega_s )  }{\delta {  G}_{\theta, \phi}(i \omega_n) }~.
\label{irreducible hexagon}
\end{align}
The result is given in~\ref{app: simplification}. Applying it to Eq.~\eqref{Sigma Fock before simplification}, and using the property 
\begin{align}
\widetilde{\Gamma}_{{\bm U}; (\mu  , \nu ;  \nu', \mu')}[{\bm G}](i \omega_m, i \omega_s ) = \widetilde{\Gamma}_{{\bm U}; (\nu', \mu'; \mu  , \nu )}[{\bm G}](i \omega_s, i \omega_m )   
\end{align}
of the two-particle reducible vertex, we obtain the Fock self-energy
\begin{align}
& \widetilde{ \Sigma}^{(\rm F)}_{{\bm U},  {\bm M}; \phi, \theta}[{\bm G} ; {\bm D}](i \omega_n)     \nonumber \\
& = - T   \sum_{\lambda, \lambda'}  \sum_{i \Omega } D_{\lambda, \lambda'} (i \Omega )       \Bigg\{      \sum_{ \alpha ,   \beta }  \widetilde{M}^{(\lambda) }_{{\bm U}; (\phi, \alpha) }[{\bm G}](i \omega_n, i \omega_n - i \Omega)  \, G_{\alpha , \beta}(i \omega_n - i \Omega ) \nonumber \\
& \quad \times  \widetilde{M}^{(\lambda')}_{{\bm U}; (\beta, \theta)}[{\bm G}](i \omega_n - i \Omega, i \omega_n)        \nonumber \\
& \quad + \frac{1}{2} \sum_{   i \omega_m }  \sum_{ \alpha, \beta} \sum_{  \alpha', \beta'} \widetilde{M}^{(\lambda)}_{{\bm U}; (\alpha, \beta)}[{\bm G}](i \omega_m - i \Omega, i \omega_m) \, \widetilde{M}^{(\lambda')}_{{\bm U}; (\alpha', \beta')}[{\bm G}](i \omega_m, i \omega_m - i \Omega) \nonumber \\
& \quad \times  \sum_{\nu, \mu , \nu', \mu'} G_{\beta, \mu }(i \omega_m) \,  G_{\mu', \alpha}(i \omega_m - i \Omega ) \, G_{\nu, \alpha'}(i \omega_m) \, G_{\beta', \nu'}(i \omega_m - i \Omega )   \nonumber \\
& \quad \times       \widetilde{U}^{[6]}_{{\bm U}; (\mu  , \nu ; \phi, \theta; \nu', \mu')}[{\bm G}](i \omega_m, i \omega_n, i \omega_m - i \Omega )     \Bigg\}~, 
\label{Sigma Fock simplified}
\end{align}
where we have introduced the renormalized EPI vertex,
\begin{align}
&  \widetilde{M}^{(\lambda)}_{{\bm U}; (\alpha, \beta)}[{\bm G}](i \omega_m, i \omega_s) \nonumber \\
&  \equiv M^{(\lambda)}_{\alpha, \beta} +  \sum_{\nu, \nu', \mu, \mu'} \widetilde{\Gamma}_{{\bm U}; (\alpha  , \nu ; \nu', \beta)}[{\bm G}](i \omega_m, i \omega_s ) \, G_{\nu, \mu}(i \omega_m) \, G_{\mu', \nu'}(i \omega_s) \, M^{(\lambda)}_{\mu, \mu'}~.
\label{M dressed}
\end{align}
The same treatment, applied to Eq.~\eqref{Sigma Hartree before simplification}, yields the Hartree self-energy 
\begin{align}
&  \widetilde{ \Sigma}^{(\rm H)}_{{\bm U},  {\bm M}; \phi, \theta}[{\bm G} ; {\bm D}](i \omega_n)   \nonumber \\
& =  T \sum_{\lambda, \lambda'} D_{\lambda, \lambda'} (0)      \sum_{ i \omega_m   }    \Bigg\{        \sum_{   \alpha , \beta }   M^{(\lambda)}_{\phi, \theta} M^{(\lambda')}_{\alpha , \beta }     \,  G_{\beta , \alpha }(i \omega_m  )         \nonumber \\
& \quad +     \sum_{ \alpha,  \alpha', \beta} M^{(\lambda)}_{\alpha, \beta} M^{(\lambda')}_{\alpha', \theta}    \sum_{ \mu , \nu , \mu'}  G_{\beta, \mu}(i \omega_m) \, G_{\mu', \alpha'}(i \omega_n)         \, G_{\nu, \alpha}(i \omega_m)   \nonumber \\
& \quad \times \widetilde{\Gamma}_{{\bm U}; (\phi, \mu' ; \mu, \nu)}[{\bm G}](i \omega_n, i \omega_m)     \nonumber \\
& \quad +     \sum_{\beta, \alpha', \beta'} M^{(\lambda)}_{\phi, \beta} M^{(\lambda')}_{\alpha', \beta'}    \sum_{\nu, \mu , \nu' } G_{\beta', \mu}(i \omega_m)  \, G_{\nu', \alpha'}(i \omega_m) \, G_{\beta, \nu}(i \omega_n) \nonumber \\
& \quad \times   \widetilde{\Gamma}_{{\bm U}; (\nu, \theta; \mu, \nu'  )}[{\bm G}](i \omega_n, i \omega_m)      \nonumber \\
& \quad   +     \sum_{ \alpha, \beta , \alpha', \beta'} M^{(\lambda)}_{\alpha, \beta} M^{(\lambda')}_{\alpha', \beta'}   \sum_{\nu, \mu , \nu', \mu'} G_{\beta, \mu}(i \omega_n) \, G_{\mu', \alpha'}(i \omega_m)      \, G_{\nu, \alpha}(i \omega_n) \, G_{\beta', \nu'}(i \omega_m)        \nonumber \\
& \quad \times \sum_{ \xi',   \eta'}   \widetilde{\Gamma}_{{\bm U}; (\mu  , \theta ; \xi', \mu')}[{\bm G}](i \omega_n, i \omega_m )  G_{\eta', \xi'}(i \omega_m) \,  \widetilde{\Gamma}_{{\bm U}; (\phi  , \nu ; \nu', \eta')}[{\bm G}](i \omega_n, i \omega_m )   \nonumber \\ 
& \quad   +  \frac{1}{2} \sum_{ i \omega_s   } \sum_{ \alpha, \beta , \alpha', \beta'} M^{(\lambda)}_{\alpha, \beta} M^{(\lambda')}_{\alpha', \beta'}   \sum_{\nu, \mu , \nu', \mu'} G_{\beta, \mu}(i \omega_m) \, G_{\mu', \alpha'}(i \omega_s)      \, G_{\nu, \alpha}(i \omega_m) \,     \nonumber \\
& \quad \times G_{\beta', \nu'}(i \omega_s)  \sum_{\xi, \xi', \eta, \eta'}   \widetilde{U}^{[6]}_{{\bm U}; (\mu  , \xi ; \phi, \theta ; \xi', \mu' )}[{\bm G}](i \omega_m, i \omega_n, i \omega_s )  \, G_{\xi, \eta}(i \omega_m) \, G_{\eta', \xi'}(i \omega_s) \nonumber \\
& \quad \times  \widetilde{\Gamma}_{{\bm U}; (\eta  , \nu ; \nu', \eta')}[{\bm G}](i \omega_m, i \omega_s )   \nonumber \\ 
& \quad   +  \frac{1}{2} \sum_{ i \omega_s   } \sum_{ \alpha, \beta , \alpha', \beta'} M^{(\lambda)}_{\alpha, \beta} M^{(\lambda')}_{\alpha', \beta'}   \sum_{\nu, \mu , \nu', \mu'} G_{\beta, \mu}(i \omega_m) \, G_{\mu', \alpha'}(i \omega_s)      \, G_{\nu, \alpha}(i \omega_m)   \nonumber \\
& \quad \times G_{\beta', \nu'}(i \omega_s) \sum_{\xi, \xi', \eta, \eta'} \widetilde{\Gamma}_{{\bm U}; (\mu  , \xi ; \xi', \mu')}[{\bm G}](i \omega_m, i \omega_s ) \, G_{\xi, \eta}(i \omega_m) \, G_{\eta', \xi'}(i \omega_s) \nonumber \\
& \quad \times \widetilde{U}^{[6]}_{{\bm U}; (\eta  , \nu ; \phi, \theta; \nu', \eta')}[{\bm G}](i \omega_m, i \omega_n, i \omega_s )    \nonumber \\ 
& \quad   +  \frac{1}{2} \sum_{ i \omega_s   } \sum_{ \alpha, \beta , \alpha', \beta'} M^{(\lambda)}_{\alpha, \beta} M^{(\lambda')}_{\alpha', \beta'}   \sum_{\nu, \mu , \nu', \mu'} G_{\beta, \mu}(i \omega_m) \, G_{\mu', \alpha'}(i \omega_s)      \, G_{\nu, \alpha}(i \omega_m) \,     \nonumber \\
& \quad \times G_{\beta', \nu'}(i \omega_s)  \sum_{\xi, \xi', \eta, \eta', \zeta, \zeta', \chi, \chi'} \widetilde{\Gamma}_{{\bm U}; (\mu  , \xi ; \xi', \mu')}[{\bm G}](i \omega_m, i \omega_s ) \, G_{\xi, \eta}(i \omega_m) \, G_{\eta', \xi'}(i \omega_s) \,        \nonumber \\
& \quad   \times  \widetilde{U}^{[6]}_{{\bm U}; (\eta  , \zeta; \phi, \theta ; \zeta', \eta' )}[{\bm G}](i \omega_m , i \omega_n , i \omega_s )  G_{\zeta, \chi}(i \omega_m) \, G_{\chi', \zeta' }(i \omega_s) \nonumber \\
& \quad \times \widetilde{\Gamma}_{{\bm U}; (\chi  , \nu ; \nu', \chi')}[{\bm G}](i \omega_m, i \omega_s )    \nonumber \\ 
& \quad    +  \frac{1}{2} \sum_{ i \omega_s   } \! \sum_{ \alpha, \beta , \alpha', \beta'} \! M^{(\lambda)}_{\alpha, \beta} M^{(\lambda')}_{\alpha', \beta'} \! \!  \sum_{\nu, \mu , \nu', \mu'}  G_{\beta, \mu}(i \omega_m) \, G_{\mu', \alpha'}(i \omega_s)      \, G_{\nu, \alpha}(i \omega_m)    \nonumber \\ 
& \quad \times G_{\beta', \nu'}(i \omega_s) \, \widetilde{U}^{[6]}_{{\bm U}; (\mu  , \nu ; \phi, \theta; \nu', \mu')}[{\bm G}](i \omega_m, i \omega_n, i \omega_s )   \Bigg\}   ~  . 
\label{Sigma Hartree simplified}
\end{align}
It is useful to look at the diagrammatic representation of the quantities that we have just derived. In Fig.~\ref{Fig: diagram Renormalized EPI vertex}, we represent the renormalized EPI vertex as a triangle with two oriented sides (corresponding to two distinct fermionic frequencies), and we give its diagrammatic equation. In Fig.~\ref{Fig: diagram Fock self-energy}, we represent the two terms contributing to the Fock self-energy in Eq.~\eqref{Sigma Fock simplified}. We want to stress the representation of the $U^{[6]}$ term as a hexagon with three oriented sides and {\it three distinct fermionic frequencies}. This is different from the analogous term derived in Ref.~\cite{Anokhin96} for the case of a disordered electron system in the absence of EEIs. In the latter, two sides of the hexagon have the same fermionic frequency~\cite{Anokhin96}. In the case of EEIs, which we are treating here, we cannot make such an assumption, and the three frequencies can, in principle, be all different [recall the definition in Eq.~\eqref{irreducible hexagon}]. For this reason, we need a different diagrammatic representation of the six-leg vertex, which produces Feynman diagrams with distinct topological features with respect to those of Ref.~\cite{Anokhin96}. This is evident also in the representation of the Hartree self-energy terms, Figs.~\ref{Fig: diagram Hartree self-energy - 1} and \ref{Fig: diagram Hartree self-energy - 2}. In Fig.~\ref{Fig: diagram Hartree self-energy - 1}, we represent the first four terms of Eq.~\eqref{Sigma Hartree simplified}: these are topologically equivalent to the analogous ones derived in Ref.~\cite{Anokhin96}, since they do not involve $U^{[6]}$. The last four terms of Eq.~\eqref{Sigma Hartree simplified}, represented in Fig.~\ref{Fig: diagram Hartree self-energy - 2}, are instead topologically different from their counterparts in Ref.~\cite{Anokhin96}. Their structure is a peculiar effect of EEIs.

\begin{figure}
\begin{center}
\begin{tikzpicture}
\begin{feynman}
\vertex (plus) at(4.5,0.5){\( +  \)};
\vertex (equal) at(2,0.5){\( =  \)};
\vertex (i) at(0,0);
\vertex (I) at(0,-0.3){\(  \scriptstyle  \alpha \)};
\vertex (j) at(1,0);
\vertex (J) at(1,-0.3){\( \scriptstyle \beta \)};
\vertex (k) at(0.5,0.866);
\vertex (K) at(0.5,1.166){\( \scriptstyle \lambda \)};
\vertex (name1) at(0.5, 0.3){\( \scriptstyle M \)};
\vertex (a) at(6,0); 
\vertex (A) at(6,-0.3){\(  \scriptstyle \alpha \)};
\vertex (b) at(7,0); 
\vertex (B) at(7,-0.3){\( \scriptstyle \beta \)};
\vertex (c) at(7,1.2); 
\vertex (C) at(7.3,1.2){\( \scriptstyle \nu' \)};
\vertex (d) at(6,1.2); 
\vertex (D) at(5.7,1.2){\( \scriptstyle \nu \)};
\vertex (name2) at(6.5,0.6){\( \scriptstyle \Gamma \)};
\vertex (m) at(6.5,2.3);
\vertex (Mup) at(6.5,2.6){\( \scriptstyle \lambda \)};
\vertex (Ml) at(6.2,2.2){\( \scriptstyle \mu \)};
\vertex (Mr) at(6.8,2.25){\( {\scriptstyle \mu'} \)};
\filldraw (6.5,2.3) circle(2pt);
\filldraw (3.5,0.5) circle(2pt); 
\vertex (up) at(3.5,0.8){\( \scriptstyle \lambda \)}; 
\vertex (l) at(3.2,0.2){\( \scriptstyle \alpha \)}; 
\vertex (r) at(3.8,0.2){\( \scriptstyle \beta \)}; 
\diagram* { 
(i) -- (j) -- [fermion, edge label'=\( \scriptstyle i\omega_s  \)] (k) -- [fermion, edge label'=\( \scriptstyle i\omega_m \)] (i),
(a) -- (b) -- [fermion, edge label'=\( \scriptstyle i\omega_s  \)] (c)  --  (d)  -- [fermion, edge label'=\( \scriptstyle i\omega_m\)] (a),
(c) -- [fermion, edge label'=\(\scriptstyle i\omega_s\)] (m) [dot] -- [fermion, edge label'=\(\scriptstyle i\omega_m\)] (d),
}; 
\end{feynman}
\end{tikzpicture}
\end{center}
\caption{Diagrammatic representation of the renormalized EPI vertex, Eq.~\eqref{M dressed}. The dot with three Greek indices in the right-hand side represents the bare EPI vertex $M^{(\lambda)}_{\alpha, \beta}$. Oriented lines connecting two Greek indices and carrying a frequency label represent fully-dressed (interacting) fermionic propagators; for example, in the second term on the right-hand side, the line connecting $\mu$ to $\nu$ with frequency label $i \omega_m$ represents the GF $G_{\nu, \mu}(i \omega_m)$ [compare with Eq.~\eqref{M dressed}]. These oriented lines are also features of vertex functionals with fermionic character (i.e.~that should be connected to fermionic propagators), such as $\widetilde{M}^{(\lambda)}_{{\bm U}; (\alpha, \beta)}[{\bm G}](i \omega_m, i \omega_s)$ and $\widetilde{\Gamma}_{{\bm U}; (\alpha  , \nu ; \nu', \beta)}[{\bm G}](i \omega_m, i \omega_s )$. For simplicity, in all diagrams we remove the $\widetilde{...}$ symbols over the functionals, so we write, e.g., $M$ for $\widetilde{M}$ and $\Gamma$ for $\widetilde{\Gamma}$. 
\label{Fig: diagram Renormalized EPI vertex}}
\end{figure}
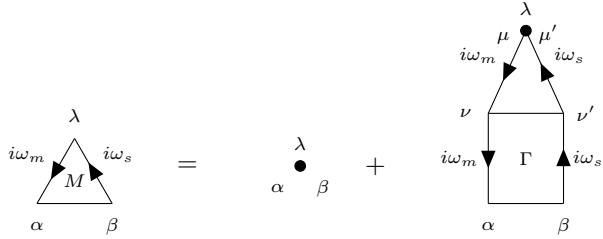

It should be emphasized that the expansion of the LWF and self-energy functionals up to the second order in the EPI matrix elements effectively produces infinite-order approximations for the corresponding {\it physical quantities}. In fact, the physical self-energy depends on the physical GF, which should be inserted in the place of the independent variable ${\bm G}$ into Eqs.~\eqref{purely electronic Sigma}, \eqref{Sigma Fock simplified}, and \eqref{Sigma Hartree simplified}. As the physical GF depends on the EPI matrix elements up to all orders, so does the physical self-energy. This is a common feature of many-body perturbation theory schemes based on the LWF, which when used with the physical GFs yields self-consistent equations that lead formally to all-order results. An example in the context of electron systems interacting via long-range Coulomb interactions is the so-called GW self-consistent scheme.

Before concluding this Section, we note that our treatment of the electronic self-energy is different from the one presented in Ref.~\cite{Giustino17}. It is worthwhile to comment on the differences between the two approaches. The author of Ref.~\cite{Giustino17} introduces the electronic self-energy evaluated {\it at clamped nuclei} (${\bm \Sigma}^{\rm cn}$), which leads to the definition of the electronic GF at clamped nuclei (${\bm G}^{\rm cn}$). The two are connected via the corresponding Dyson equation. These quantities are evaluated within the context of the most general (Hedin-Baym) formalism, by setting to zero all the GFs associated with the phonon displacement ${\bm D}$. This procedure corresponds to switching off the EPI completely. As a second step, a higher-level Dyson equation is written for the full problem (in the presence of EPIs), by considering ${\bm G}^{\rm cn}$ as the ``non-interacting" GF, and introducing a self-energy that connects it to the fully interacting GF. 

Our approach is different, as we write a single electronic GF (and the corresponding Dyson equation) for the fully interacting problem, and our partition of the self-energy results from an expansion of the LWF in powers of the EPI matrix elements. Therefore, our electronic self-energy term ${\bm \Sigma}^{\rm (E)}$, given by Eq.~\eqref{purely electronic Sigma}, does not coincide with the quantity ${\bm \Sigma}^{\rm cn}$ introduced in Ref.~\cite{Giustino17}, since the former---as well as the other terms that we have derived, ${\bm \Sigma}^{\rm (H)}$ and ${\bm \Sigma}^{\rm (F)}$---is a functional of the fully interacting GF, while the latter is a functional of ${\bm G}^{\rm cn}$. The complicated structure of our Hartree and Fock self-energy terms does not appear in Ref.~\cite{Giustino17}. It would probably appear in a treatment of the explicit form of the quantity ${\bm \Sigma}^{\rm (dGW)}$ introduced in Ref.~\cite{Giustino17} (which is not discussed there). Establishing the exact correspondence between the partition introduced in Ref.~\cite{Giustino17} and ours is well beyond the scope of this work.

\section{Phononic self-energy from the expansion of the LWF}
\label{sect:phononic-self-energy}

We now derive the phononic self-energy, by applying Eq.~\eqref{P derivative} to the terms of the LWF listed in Eq.~\eqref{terms of the LWF}. We first restore the $({\bm I}, {\bm P})$ representation, by rewriting Eq.~\eqref{terms of the LWF} as
\begin{align}
\widetilde{\Phi}_{{\bm U},  {\bm I}}[{\bm G} ; {\bm P}] & \approx \widetilde{\Phi}^{\rm (E)}_{{\bm U} }[{\bm G}  ] + \widetilde{\Phi}^{(\rm F)}_{{\bm U},  {\bm I}}[{\bm G} ; {\bm P}]   + \widetilde{\Phi}^{(\rm H)}_{{\bm U},  {\bm I}}[{\bm G} ; {\bm P}]~.
\end{align}
The corresponding phononic self-energy functionals are given by
\begin{align}
\widetilde{ \Lambda}_{{\bm U},  {\bm I}; \phi, \theta}[{\bm G} ; {\bm P}](i \Omega_n)  \approx   \widetilde{\Lambda}^{(\rm F)}_{{\bm U},  {\bm I}; \phi, \theta}[{\bm G}  ](i \Omega_n)   + \widetilde{\Lambda}^{(\rm H)}_{{\bm U},  {\bm I}; \phi, \theta}[{\bm G}  ](i \Omega_n)~,
\end{align}
where
\begin{align}
&  \widetilde{ \Lambda}^{(\rm F)}_{{\bm U},  {\bm I}; \phi, \theta}[{\bm G}  ](i \Omega_n)    \equiv - \frac{1}{T} \frac{\delta \widetilde{\Phi}^{(\rm F)}_{{\bm U},  {\bm I}}[{\bm G} ; {\bm P}] }{\delta {  P}_{\theta, \phi}(i \Omega_n) }  \nonumber \\
&   =      \frac{T }{2} \sum_{  \alpha, \beta} \sum_{  \alpha', \beta'}  \sum_{i \Omega_{n'}}         \frac{1}{2}    \left(  
 \delta_{n',n}       I^{(\phi)}_{\alpha, \beta} I^{(\theta)}_{\alpha', \beta'}              + 
 \delta_{n', -n}      I^{(\theta)}_{\alpha, \beta} I^{(\phi)}_{\alpha', \beta'}       \right)  \nonumber \\
 & \quad \times \sum_{   i \omega_m }    \Big\{           G_{\beta, \alpha'}(i \omega_m)  \, G_{\beta', \alpha}(i \omega_{m} - i \Omega_{n'})    \nonumber \\
& \quad +   \sum_{\nu, \mu , \nu', \mu'} G_{\beta, \nu}(i \omega_m) \,  G_{\mu, \alpha}(i \omega_m - i \Omega_{n'}) \, \widetilde{\Gamma}_{{\bm U}; (\nu, \mu' ; \nu', \mu)}[{\bm G}](i \omega_m, i \omega_m - i \Omega_{n'}) \nonumber \\
& \quad \times G_{\mu', \alpha'}(i \omega_m)  G_{\beta', \nu'}(i \omega_m - i \Omega_{n'})   \Big\}~,
\label{Lambda Fock}
\end{align}
and
\begin{align}
& \widetilde{ \Lambda}^{(\rm H)}_{{\bm U},  {\bm I}; \phi, \theta}[{\bm G}  ](i \Omega_n)   \equiv - \frac{1}{T} \frac{\delta \widetilde{\Phi}^{(\rm H)}_{{\bm U},  {\bm I}}[{\bm G} ; {\bm P}] }{\delta {  P}_{\theta, \phi}(i \Omega_n) }  \nonumber \\
&   =   - \delta_{n,0}  \frac{T }{2} \sum_{ \alpha, \beta} \sum_{  \alpha', \beta'}     
                  \frac{1}{2}    \left(          I^{(\phi)}_{\alpha, \beta} I^{(\theta)}_{\alpha', \beta'}      +       I^{(\theta)}_{\alpha, \beta} I^{(\phi)}_{\alpha', \beta'}    \right)   \sum_{ i \omega_n , i \omega_m }    \Big\{        G_{\beta, \alpha}(i \omega_m)  \,  G_{\beta', \alpha'}(i \omega_n  )        \nonumber \\
& \quad +   \sum_{\nu, \mu , \nu', \mu'} G_{\beta, \nu}(i \omega_m) \, G_{\mu', \alpha'}(i \omega_n) \, \widetilde{\Gamma}_{{\bm U}; (\nu, \mu ; \nu', \mu')}[{\bm G}](i \omega_m, i \omega_n) \, G_{\mu, \alpha}(i \omega_m) \nonumber \\
& \quad \times G_{\beta', \nu'}(i \omega_n)         \Big\}~.
\label{Lambda Hartree}
\end{align}
The following properties hold:
\begin{align}
\widetilde{ \Lambda}^{(\rm F / H)}_{{\bm U},  {\bm I}; \phi, \theta}[{\bm G}  ](i \Omega_n) = \widetilde{ \Lambda}^{(\rm F / H )}_{{\bm U},  {\bm I};  \theta , \phi}[{\bm G}  ](- i \Omega_n)~.
\end{align}
\section{Extended Eliashberg equations}
\label{sec: Eliashberg}

We now derive the extended Eliashberg equations for the anomalous components of the electronic GF and self-energy. We consider the particular case of a crystal with a single orbital per unit cell. In this case the spin-orbital index $\sigma$ introduced for fermions reduces to the spin index, $\sigma = \pm 1$. We neglect magnetic impurities and relativistic effects (such as spin-orbit and magnetic anisotropies). We then choose a basis of real single-electron wave functions for the definition of the Hamiltonian parameters.

For the sake of simplicity, in this Section we drop the functional notation that we have adopted in the previous Sections. It is intended that self-energies and vertex functions are functionals of the GFs.

For the derivation of extended Eliashberg equations, we use the Gor'kov-Nambu (G-N) representation for the fermionic sector of the Hamiltonian. As discussed in detail in~\ref{app:Nambu}, the G-N Hamiltonian can be directly obtained from our general formulation provided that the following choices are made:
\begin{align}
U_{\alpha, \beta, \delta, \gamma} \rightarrow  U_{i_\alpha, i_\beta, i_\delta, i_\gamma} \delta_{\sigma_{\alpha}, \sigma_{\delta}}   \delta_{\sigma_{\beta}, \sigma_{\gamma}} \sigma_{\alpha} \sigma_{\beta}~,
\label{U Nambu}
\end{align}
\begin{align}
M^{(\lambda)}_{\alpha, \beta}   \rightarrow  M^{(i_{\lambda})}_{ i_{\alpha}, i_{\beta}}(s_{\lambda}) \, \delta_{\sigma_{\alpha}, \sigma_{\beta}} \sigma_{\alpha}~,
\label{M Nambu}
\end{align} 
and
\begin{align}
t_{\alpha, \beta} & \rightarrow \left[ t_{i_{\alpha}, i_{\beta}}  + \sum_k U_{i_{\alpha}, k, i_{\beta}, k} -  \sum_{k, s} I^{(k)}_{i_{\alpha}, i_{\beta}}(s) 
\sum_m f^{-1}_{k, m}(s) \sum_l I^{(m)}_{l,l}(s) \right] \delta_{\sigma_{\alpha}, \sigma_{\beta}} \sigma_{\alpha} \nonumber \\
& \equiv h_{i_{\alpha}, i_{\beta}} \delta_{\sigma_{\alpha}, \sigma_{\beta}} \sigma_{\alpha}~.
\label{t Nambu}
\end{align}
Therefore, our derivation of the LWF and related self-energies holds in the G-N representation as well. GFs and self-energies acquire a $2 \times 2$ matrix structure in spin space; the non-zero solutions of the equations for the non-diagonal matrix elements signal the onset of the superconducting phase. Although we will keep using, for the sake of brevity, the word {\it electrons} (as well as terms such as EEI and EPI), it should be kept in mind that the fermions appearing in this derivation are actually {\it Nambu fermions}. 

In the G-N formalism, it is natural to introduce a $2 \times 2$ matrix representation for the spin sector. We denote such matrices by means of double-underlined symbols, i.e.,
\begin{align}
& \underline{\underline{G}}_{\, i,j}(i \omega_n) \equiv \left( \begin{matrix}
G_{(i, \uparrow), (j, \uparrow)}(i \omega_n) & G_{(i, \uparrow), (j, \downarrow)}(i \omega_n) \\
G_{(i, \downarrow), (j, \uparrow)}(i \omega_n) & G_{(i, \downarrow), (j, \downarrow)}(i \omega_n) 
 \end{matrix}\right)  ~, \nonumber \\
 & \underline{\underline{\Sigma}}_{\, i,j}(i \omega_n) \equiv \left( \begin{matrix}
\Sigma_{(i, \uparrow), (j, \uparrow)}(i \omega_n) & \Sigma_{(i, \uparrow), (j, \downarrow)}(i \omega_n) \\
\Sigma_{(i, \downarrow), (j, \uparrow)}(i \omega_n) & \Sigma_{(i, \downarrow), (j, \downarrow)}(i \omega_n) 
 \end{matrix}\right)  ~,
\end{align}
where the matrix elements $G_{(i, \sigma), (j, \sigma')}(i \omega_n)$ and $\Sigma_{(i, \sigma), (j, \sigma')}(i \omega_n)$ are shorthands for the quantities $G_{{\bm t}, {\bm U}, {\bm f}, {\bm I}; \alpha, \beta}(i \omega_n)$ and $\Sigma_{{\bm t}, {\bm U}, {\bm f}, {\bm I}; \alpha, \beta}(i \omega_n)$ introduced earlier, respectively, where $\alpha = (i , \sigma)$ and $\beta = (j, \sigma')$.

We switch to the reciprocal lattice representation, by applying various Fourier transforms (see Section~\ref{sec: Hamiltonian}). Namely, the transformations from direct to reciprocal lattice for the GF and the self-energy read, respectively, as
\begin{align}\label{eq:FT G-N matrices}
& \underline{\underline{G}}_{\, i,j}(i \omega_n) \equiv \sum_{\boldsymbol{k}} F_{\boldsymbol{k}, i} F^*_{\boldsymbol{k}, j} \, \underline{\underline{G}}_{\, \boldsymbol{k}}(i \omega_n) ~, \nonumber \\
& \underline{\underline{\Sigma}}_{\, i,j}(i \omega_n) \equiv \sum_{\boldsymbol{k}} F_{\boldsymbol{k}, i} F^*_{\boldsymbol{k}, j} \, \underline{\underline{\Sigma}}_{\, \boldsymbol{k}}(i \omega_n) ~.
\end{align} 
We expand the matrices in Eq.~(\ref{eq:FT G-N matrices}) over the set of standard spin-$1/2$ Pauli matrices:
\begin{align}
\underline{\underline{\tau_{\,0}}} = \left( \begin{matrix} 1 & 0 \\ 0 & 1 \end{matrix}\right)~, \quad 
\underline{\underline{\tau_{\,1}}} = \left( \begin{matrix} 0 & 1 \\ 1 & 0 \end{matrix}\right)~, \quad 
\underline{\underline{\tau_{\,2}}} = \left( \begin{matrix} 0 & -i \\ i & 0 \end{matrix}\right)~, \quad 
\underline{\underline{\tau_{\,3}}} = \left( \begin{matrix} 1 & 0 \\ 0 & -1 \end{matrix}\right)~.
\end{align}

The electronic Dyson equation reads as
\begin{align}
\underline{\underline{G}}^{-1}_{\,\boldsymbol{k}}(i \omega_n) = i \omega_n \underline{\underline{\tau_{\,0}}} - \xi_{\boldsymbol{k}} \underline{\underline{\tau_{\,3}}} - \underline{\underline{\Sigma}}_{\,\boldsymbol{k}}(i \omega_n)~,
\label{Dyson Eliashberg}
\end{align}
where $\xi_{\boldsymbol{k}} = h_{\boldsymbol{k}} - \mu$, and $h_{\boldsymbol{k}}$ is an eigenvalue of the matrix $h_{i,j}$ introduced in Eq.~\eqref{t Nambu}. For the electronic self-energy, we take Eq.~\eqref{Self-energy, functional notation}, which we rewrite in the Nambu representation as
\begin{align}
\underline{\underline{\Sigma}}_{\,\boldsymbol{k}}(i \omega_n) = \underline{\underline{\Sigma}}^{\rm (E)}_{\,\boldsymbol{k}}(i \omega_n)
+ \underline{\underline{\Sigma}}^{\rm (H)}_{\,\boldsymbol{k}}(i \omega_n) + \underline{\underline{\Sigma}}^{\rm (F)}_{\,\boldsymbol{k}}(i \omega_n) \equiv \underline{\underline{\Sigma}}^{\rm (E)}_{\,\boldsymbol{k}}(i \omega_n)
+ \underline{\underline{\Sigma}}^{\rm (EP)}_{\,\boldsymbol{k}}(i \omega_n)~,
\label{separation self-energy}
\end{align}
where ``E" refers to the purely electronic contribution in Eq.~\eqref{purely electronic Sigma}, while ``H" and ``F" label the Hartree and Fock terms due to the EPI, respectively. Finally, ``EP" labels the total term due to the EPI. 

The self-energy terms are decomposed similarly to what is done in Ref.~\cite{Anokhin96}, except that we are here considering the case of EEIs rather than disorder. With respect to their derivation, we also include the terms $\propto \underline{\underline{\tau_{2}}}$. We write
\begin{align}
\underline{\underline{\Sigma}}^{\rm (E)}_{\,\boldsymbol{k}}(i \omega_n) &\equiv  
i \omega_n \left[ 1 - \gamma_{\boldsymbol{k}}(i \omega_n)\right] \underline{\underline{\tau_{\,0}}} + \phi^{\rm (E)}_{\boldsymbol{k}}(i \omega_n)\, \underline{\underline{\tau_{\,1}}} + \overline{\phi}^{\rm (E)}_{\boldsymbol{k}}(i \omega_n)\, \underline{\underline{\tau_{\,2}}} \nonumber\\
& \quad +  \chi^{\rm (E)}_{\boldsymbol{k}}(i \omega_n)\,  \underline{\underline{\tau_{\,3}}}
\label{Sigma E Eliashberg}
\end{align}
and
\begin{align}
\underline{\underline{\Sigma}}^{\rm (EP)}_{\,\boldsymbol{k}}(i \omega_n) & \equiv  
i \omega_n \gamma_{\boldsymbol{k}}(i \omega_n) \left[ 1 - Z_{\boldsymbol{k}}(i \omega_n)\right] \underline{\underline{\tau_{\,0}}} + \phi^{\rm (EP)}_{\boldsymbol{k}}(i \omega_n)\, \underline{\underline{\tau_{\,1}}} + \overline{\phi}^{\rm (EP)}_{\boldsymbol{k}}(i \omega_n)\, \underline{\underline{\tau_{\,2}}} 
\nonumber \\
& \quad +  \chi^{\rm (EP)}_{\boldsymbol{k}}(i \omega_n)\,  \underline{\underline{\tau_{\,3}}}~.
\label{Sigma EP Eliashberg}
\end{align}
The total self-energy therefore reads as following
\begin{align}
\underline{\underline{\Sigma}}_{\,\boldsymbol{k}}(i \omega_n) & \equiv 
i \omega_n \left[ 1 - \gamma_{\boldsymbol{k}}(i \omega_n) Z_{\boldsymbol{k}}(i \omega_n)\right] \underline{\underline{\tau_{\,0}}} + \phi_{\boldsymbol{k}}(i \omega_n)\, \underline{\underline{\tau_{\,1}}} + \overline{\phi}_{\boldsymbol{k}}(i \omega_n)\, \underline{\underline{\tau_{\,2}}} 
\nonumber \\
& \quad + \chi_{\boldsymbol{k}}(i \omega_n)\,  \underline{\underline{\tau_{\,3}}}~,
\label{Sigma Eliashberg}
\end{align}
where 
\begin{align}
& \phi_{\boldsymbol{k}}(i \omega_n) \equiv \phi^{\rm (E)}_{\boldsymbol{k}}(i \omega_n) + \phi^{\rm (EP)}_{\boldsymbol{k}}(i \omega_n)~, \nonumber \\
& \overline{\phi}_{\boldsymbol{k}}(i \omega_n) \equiv \overline{\phi}^{\rm (E)}_{\boldsymbol{k}}(i \omega_n) + \overline{\phi}^{\rm (EP)}_{\boldsymbol{k}}(i \omega_n)~, \nonumber \\
& \chi_{\boldsymbol{k}}(i \omega_n) \equiv \chi^{\rm (E)}_{\boldsymbol{k}}(i \omega_n) + \chi^{\rm (EP)}_{\boldsymbol{k}}(i \omega_n)~.
\end{align}
The quantities introduced above can be expressed in terms of the self-energy components as follows:
\begin{align}
&  \gamma_{\boldsymbol{k}}(i \omega_n) = 1 - \frac{\Sigma^{\rm (E)}_{\boldsymbol{k}; \uparrow, \uparrow}(i \omega_n) + \Sigma^{\rm (E)}_{\boldsymbol{k}; \downarrow, \downarrow}(i \omega_n)}{2 i \omega_n}~, \nonumber \\
&  Z_{\boldsymbol{k}}(i \omega_n) = 1 - \frac{\Sigma^{\rm (EP)}_{\boldsymbol{k}; \uparrow, \uparrow}(i \omega_n) + \Sigma^{\rm (EP)}_{\boldsymbol{k}; \downarrow, \downarrow}(i \omega_n)}{2 i \omega_n \gamma_{\boldsymbol{k}}(i \omega_n) }~, \nonumber \\
&  \chi^{\rm (E / EP)}_{\boldsymbol{k}}(i \omega_n) = \frac{\Sigma^{\rm (E / EP)}_{\boldsymbol{k}; \uparrow, \uparrow}(i \omega_n) - \Sigma^{\rm (E / EP)}_{\boldsymbol{k}; \downarrow, \downarrow}(i \omega_n)}{2}~, \nonumber \\
&  \phi^{\rm (E / EP)}_{\boldsymbol{k}}(i \omega_n) = \frac{\Sigma^{\rm (E / EP)}_{\boldsymbol{k}; \uparrow, \downarrow}(i \omega_n) + \Sigma^{\rm (E / EP)}_{\boldsymbol{k}; \downarrow, \uparrow}(i \omega_n)}{2}~, \nonumber \\
&  \overline{\phi}^{\rm (E / EP)}_{\boldsymbol{k}}(i \omega_n) = \frac{\Sigma^{\rm (E / EP)}_{\boldsymbol{k}; \uparrow, \downarrow}(i \omega_n) - \Sigma^{\rm (E / EP)}_{\boldsymbol{k}; \downarrow, \uparrow}(i \omega_n)}{-2i}~. 
\label{Eliashberg list}
\end{align}

We now insert Eq.~\eqref{Sigma Eliashberg} into Eq.~\eqref{Dyson Eliashberg}, and solve for the GF via matrix inversion. This yields the expression for the GF as a function of the self-energy components,
\begin{align}
\underline{\underline{G}}_{\,\boldsymbol{k}}(i \omega_n) & = \frac{1}{\Theta_{\boldsymbol{k}}(i \omega_n)} \left\{ i \omega_n \gamma_{\boldsymbol{k}}(i \omega_n) Z_{\boldsymbol{k}}(i \omega_n) \, \underline{\underline{\tau_{\,0}}} + \phi_{\boldsymbol{k}}(i \omega_n)\, \underline{\underline{\tau_{\,1}}} + \overline{\phi}_{\boldsymbol{k}}(i \omega_n)\, \underline{\underline{\tau_{\,2}}} \right. \nonumber \\
& \quad \left. \, + \left[ \xi_{\boldsymbol{k}} + \chi_{\boldsymbol{k}}(i \omega_n) \right] \underline{\underline{\tau_{\,3}}} \, \right\}~,
\label{G Eliashberg}
\end{align}
where
\begin{align}
\Theta_{\boldsymbol{k}}(i \omega_n) \equiv \left[ i \omega_n \gamma_{\boldsymbol{k}}(i \omega_n) Z_{\boldsymbol{k}}(i \omega_n) \right]^2 - \left[ \xi_{\boldsymbol{k}} + \chi_{\boldsymbol{k}}(i \omega_n) \right]^2 - \phi^2_{\boldsymbol{k}}(i \omega_n) -  \overline{\phi}^2_{\boldsymbol{k}}(i \omega_n)~.
\end{align}

The quantity in Eq.~\eqref{Sigma Eliashberg} must be identified with the total electronic self-energy, which we have derived as a functional of the GF, see Eq.~\eqref{Self-energy, functional notation}. 

The Eliashberg equations are obtained by inserting the expansion \eqref{G Eliashberg} into Eq.~\eqref{Self-energy, functional notation}. In principle, this procedure yields a self-consistent set of eight equations that determine the eight unknown quantities listed in Eq.~\eqref{Eliashberg list}.

\subsection{Explicit form of the EPI self-energy functional}
To make further progress, we need to adopt an explicit expression for the self-energy $\underline{\underline{\Sigma}}_{\,\boldsymbol{k}}(i \omega_n)$ as a functional of the (fully interacting) electronic GF. Historically, the work of Scalapino, Schrieffer and Wilkins~\cite{Scalapino66} has been a fundamental milestone with respect to the inclusion of EEIs in the Eliashberg equations. However, the self-energy that we have derived here from the LWF has a much more complicated form than the simple one that they postulated in their work. As we will see momentarily, this leads to several difficulties in our case. In brief, if the self-energy functional is too complicated, it is impossible to derive a set of self-consistent equations that can be used in practice, unless further approximations are adopted. 

We start by separating the self-energy according to Eq.~\eqref{separation self-energy}. We deal with the EPI self-energy first: in our derivation, this is given by the sum of Fock and Hartree contributions. As we have seen in Section~\ref{Hartree and Fock terms: simplifications}, our derivation based on the LWF produces 1) a Hartree self-energy that includes various terms depending on the reducible four-leg and irreducible six-leg vertices related to EEIs (see Figs.~\ref{Fig: diagram Hartree self-energy - 1} and \ref{Fig: diagram Hartree self-energy - 2}), and 2) a Fock self-energy that contains both EEI-renormalized EPI vertices and a new term that depends on the six-leg irreducible vertex related to EEIs (see Fig.~\ref{Fig: diagram Fock self-energy}). We note that the expression that is usually assumed for the EPI self-energy functional~\cite{Scalapino66,ScalapinoParks} is of the Fock type but does not include the second term, i.e.~the one that depends on the six-leg irreducible vertex. Furthermore, usual theories~\cite{Scalapino66,ScalapinoParks} neglect all Hartree-type contributions and rely on a static approximation for the renormalized vertex (analogous to the approximation made on the screened potential)~\cite{ScalapinoParks}, without taking into account explicitly the full functional dependence of the vertex on the electronic GFs. On the other hand, all the new features we have discovered in the electronic self-energy cannot be taken into account in a derivation of extended Eliashberg equations, as there is no closed-form expression for the four-leg reducible and six-leg irreducible EEI vertices. Further progress could be made, for example, by adopting an explicit expression---based e.g.~on Dynamical Mean-Field Theory (DMFT)~\cite{DMFT}---for the EEI self-energy functional $\underline{\underline{\Sigma}}^{({\rm E})}_{\, \boldsymbol{k}}(i \omega_n)$ and, consequently, for the EEI vertices. This is well beyond the scope of this work. 

In this work, in order to derive Eliashberg-type equations, we make the simplest approximation: we neglect all the EEI vertices terms appearing in the Hartree and Fock self-energy terms. (This is analogous to the approximation made in Ref.~\cite{Anokhin96} for the case of non-interacting electrons in the presence of disorder.) The Hartree and Fock self-energies reduce, respectively, to
\begin{align}
\underline{\underline{\Sigma}}^{\rm (H)}_{\, \boldsymbol{k}}(i \omega_n) & \approx \underline{\underline{\tau_3}} \, T   \sum_{ s} M^{(\boldsymbol{0}, s)}_{\boldsymbol{k}} D_{\boldsymbol{0}, s}(0) \sum_{\boldsymbol{k}', i \omega_m} M^{(\boldsymbol{0}, s)}_{\boldsymbol{k}'} \Big[ G_{  \boldsymbol{k}', \uparrow \uparrow}(i \omega_m)  - G_{  \boldsymbol{k}', \downarrow \downarrow}(i \omega_m) \Big] \nonumber \\
& \equiv \underline{\underline{\tau_3}} \Sigma^{\rm (H)}_{\boldsymbol{k}}
\label{Hartree Eliashberg}
\end{align}
and
\begin{align}
\underline{\underline{\Sigma}}^{\rm (F)}_{\, \boldsymbol{k}}(i \omega_n)  \approx - T \sum_{i \Omega} \sum_{\boldsymbol{q}, s} M^{(-\boldsymbol{q} , s)}_{\boldsymbol{k}} M^{(\boldsymbol{q} , s)}_{\boldsymbol{k} - \boldsymbol{q}}  D_{\boldsymbol{q}, s}(i \Omega) \, \underline{\underline{\tau_3}} \cdot \underline{\underline{G}}_{\, \boldsymbol{k} - \boldsymbol{q}}(i \omega_n - i \Omega) \cdot \underline{\underline{\tau_3}}~,
\label{Fock Eliashberg}
\end{align}
where we have used the Fourier representation of the EPI matrix elements, as discussed in detail in~\ref{app: Derivation Hamiltonian} [see, in particular, Eq.~\eqref{exact expression for M bare - Fourier}]. In the approximation we made,  the Hartree self-energy~\eqref{Hartree Eliashberg} retains its non-local nature in space, but loses its non-locality in time.

Eqs.~\eqref{Hartree Eliashberg} and \eqref{Fock Eliashberg} are sufficient to carry out the derivation of {\it four} out of the eight Eliashberg equations, which connect the unknown quantities listed in Eq.~\eqref{Eliashberg list}. We proceed with this first part of the derivation in Section~\ref{sec: Eliashberg first part}, and defer the discussion of the EEI self-energy to the second part of the derivation, reported in Section~\ref{sec: Eliashberg second part}.

\subsection{Derivation of extended Eliashberg equations: first part}
\label{sec: Eliashberg first part}

We now substitute Eq.~\eqref{G Eliashberg} into Eqs.~\eqref{Hartree Eliashberg} and \eqref{Fock Eliashberg}, sum the resulting expressions, and set the result equal to the right-hand side of Eq.~\eqref{Sigma EP Eliashberg}. This gives a $2 \times 2$ matrix equation, equivalent to the following set of four equations: 
\begin{align}
i \omega_n \gamma_{\boldsymbol{k}}(i \omega_n) \left[ 1 - Z_{\boldsymbol{k}}(i \omega_n)\right]  &  =   - T \sum_{i \omega_m} \sum_{\boldsymbol{q}, s} M^{( \boldsymbol{q} - \boldsymbol{k} , s)}_{\boldsymbol{k}}
M^{( \boldsymbol{k} - \boldsymbol{q} , s)}_{\boldsymbol{q}} D_{\boldsymbol{k} - \boldsymbol{q}, s}(i \omega_n - i \omega_m) \nonumber \\
& \quad \times \frac{  i \omega_m   \gamma_{  \boldsymbol{q}}(i \omega_m) Z_{ \boldsymbol{q}}(i \omega_m)}{  \Theta_{   \boldsymbol{q}  }(i \omega_m)}~,
\label{Eliashberg I}
\end{align}
\begin{align}
  \chi^{\rm (EP)}_{\boldsymbol{k}}(i \omega_n) & =   - T \sum_{i \omega_m} \sum_{\boldsymbol{q}, s} M^{( \boldsymbol{q} - \boldsymbol{k} , s)}_{\boldsymbol{k}}
M^{( \boldsymbol{k} - \boldsymbol{q} , s)}_{\boldsymbol{q}} D_{\boldsymbol{k} - \boldsymbol{q}, s}(i \omega_n - i \omega_m) \, \frac{\xi_{  \boldsymbol{q}} + \chi_{  \boldsymbol{q}  }(i \omega_m)  }{ \Theta_{  \boldsymbol{q}  }(i \omega_m)} \nonumber \\
& \quad + \Sigma^{\rm (H)}_{\boldsymbol{k}}~,
\label{Eliashberg II}
\end{align}
\begin{align}
\phi^{\rm (EP)}_{\boldsymbol{k}}(i \omega_n)     =   T \sum_{i \omega_m} \sum_{\boldsymbol{q}, s} M^{( \boldsymbol{q} - \boldsymbol{k} , s)}_{\boldsymbol{k}}
M^{( \boldsymbol{k} - \boldsymbol{q} , s)}_{\boldsymbol{q}} D_{\boldsymbol{k} - \boldsymbol{q}, s}(i \omega_n - i \omega_m) \, \frac{\phi_{  \boldsymbol{q}  }(i \omega_m)   }{\Theta_{  \boldsymbol{q}  }(i \omega_m)}~,
\label{Eliashberg III}
\end{align}
and
\begin{align}
  \overline{\phi}^{\rm (EP)}_{\boldsymbol{k}}(i \omega_n)  =   T \sum_{i \omega_m} \sum_{\boldsymbol{q}, s} M^{( \boldsymbol{q} - \boldsymbol{k} , s)}_{\boldsymbol{k}}
M^{( \boldsymbol{k} - \boldsymbol{q} , s)}_{\boldsymbol{q}}  D_{\boldsymbol{k} - \boldsymbol{q}, s}(i \omega_n - i \omega_m) \frac{ \overline{\phi}_{   \boldsymbol{q}  }(i \omega_m) }{\Theta_{  \boldsymbol{q}  }(i \omega_m)}~,
\label{Eliashberg IV}
\end{align}
where the term
\begin{align}
\Sigma^{\rm (H)}_{\boldsymbol{k}} \equiv   2 T   \sum_{ s} M^{(\boldsymbol{0}, s)}_{\boldsymbol{k}} M^{(\boldsymbol{0}, s)}_{\boldsymbol{q}} D_{\boldsymbol{0}, s}(0) \sum_{\boldsymbol{q}, i \omega_m} \frac{   \xi_{  \boldsymbol{q} } + \chi_{  \boldsymbol{q} }(i \omega_m)    }{\Theta_{\boldsymbol{q}}(i \omega_m)}  
\end{align}
appears only in Eq.~\eqref{Eliashberg II}. We note that Eqs.~\eqref{Eliashberg II}, \eqref{Eliashberg III} and \eqref{Eliashberg IV} relate EP self-energy terms (on the left-hand sides) to the respective total self-energy terms (on the right-hand sides). Therefore, these are not yet self-consistent equations. To complete the derivation, we need to take into account the EEI self-energy. This is done in the next Section.

\subsection{Derivation of extended Eliashberg equations: second part}
\label{sec: Eliashberg second part}

For temperatures larger than the critical temperature, $T > T_{\rm c}$, the anomalous terms of the self-energy and GF vanish. For $T \rightarrow T^{-}_{\rm c}$ , the anomalous terms are very small, and the Eliashberg equations can be linearized in the quantities $\phi_{\boldsymbol{k}}(i \omega_n)$ and $\overline{\phi}_{\boldsymbol{k}}(i \omega_n)$. Namely, we put
\begin{align}
\underline{\underline{\Sigma}}_{\,\boldsymbol{k}}(i \omega_n) \equiv\underline{\underline{\Sigma}}^{\rm (N)}_{\,\boldsymbol{k}}(i \omega_n) + \delta \underline{\underline{\Sigma}}_{\,\boldsymbol{k}}(i \omega_n)~,
\end{align}
where
\begin{align}
\underline{\underline{\Sigma}}^{\rm (N)}_{\,\boldsymbol{k}}(i \omega_n)   \equiv 
i \omega_n \left[ 1 - \gamma^{\rm (N)}_{\boldsymbol{k}}(i \omega_n) Z^{\rm (N)}_{\boldsymbol{k}}(i \omega_n)\right] \underline{\underline{\tau_{\,0}}}     + \chi^{\rm (N)}_{\boldsymbol{k}}(i \omega_n)\,  \underline{\underline{\tau_{\,3}}}~.
\label{Sigma Eliashberg normal}
\end{align}
The superscript ``N" denotes quantities that are evaluated in the normal state. The anomalous correction $\delta \underline{\underline{\Sigma}}_{\,\boldsymbol{k}}(i \omega_n)$ vanishes for $T > T_{\rm c}$; for $T < T_{\rm c}$ it includes, in general, both finite off-diagonal {\it and} diagonal terms. For $T \rightarrow T^{-}_{\rm c}$, the change in the self-energy in going from the superconducting to the normal phase can be approximated as
\begin{align}
\delta \underline{\underline{\Sigma}}_{\,\boldsymbol{k}}(i \omega_n) & \approx \sum_{\boldsymbol{q}, i \omega_m} \sum_{\sigma, \sigma'} \left. \frac{\delta \underline{\underline{\Sigma}}_{\,\boldsymbol{k}}(i \omega_n)}{\delta G_{\boldsymbol{q}; \sigma, \sigma'}(i \omega_m) }  \right|_{G = G^{\rm (N)}} \delta G_{\boldsymbol{q}; \sigma, \sigma'}(i \omega_m)~,
\label{Eliashberg linearization}
\end{align}
where $\delta G_{\boldsymbol{q}; \sigma, \sigma'}(i \omega_m)$ denotes the term of the GF which is linear in the anomalous components of the self-energy. We now express this quantity in terms of $\delta \Sigma$, by using the following relations, which are consequences of the Dyson equation:
\begin{align}
& G^{-1} = G_0^{-1} - \Sigma \Rightarrow \delta \Sigma = - \delta G^{-1}~, \nonumber \\
& G^{-1} \cdot G = 1 \Rightarrow \delta G^{-1} \cdot G + G^{-1} \cdot \delta G = 0 \Rightarrow   \delta G^{-1} = - G^{-1} \cdot \delta G \cdot G^{-1} \nonumber \\
& \Rightarrow \delta \Sigma = G^{-1} \cdot \delta G \cdot G^{-1} \Rightarrow \delta G = G \cdot \delta \Sigma \cdot G~,
\end{align}
where we have used a compact notation. We obtain
\begin{align}
& \delta\underline{\underline{\Sigma}}_{\,\boldsymbol{k}}(i \omega_n) \nonumber \\
 & = \sum_{\boldsymbol{q}, i \omega_m} \sum_{\sigma, \sigma'} \left. \frac{\delta \underline{\underline{\Sigma}}_{\,\boldsymbol{k}}(i \omega_n)}{\delta G_{\boldsymbol{q}; \sigma, \sigma'}(i \omega_m) }  \right|_{G = G^{\rm (N)}} \Big[ \underline{\underline{G}}^{\rm (N)}_{\, \boldsymbol{q} }(i \omega_m) \cdot \delta \underline{\underline{\Sigma}}_{\, \boldsymbol{q} }(i \omega_m) \cdot \underline{\underline{G}}^{\rm (N)}_{\, \boldsymbol{q} }(i \omega_m) \Big]_{\sigma, \sigma'} \nonumber \\
 & = \sum_{\boldsymbol{q}, i \omega_m} G^{\rm (N)}_{\, \boldsymbol{q} ; \uparrow, \uparrow }(i \omega_m) \, G^{\rm (N)}_{\, \boldsymbol{q}; \downarrow, \downarrow }(i \omega_m) \sum_{\sigma } \left. \frac{\delta \underline{\underline{\Sigma}}_{\,\boldsymbol{k}}(i \omega_n)}{\delta G_{\boldsymbol{q}; \sigma, - \sigma}(i \omega_m) }  \right|_{G = G^{\rm (N)}}  \,  \delta  \Sigma_{\, \boldsymbol{q}; \sigma, - \sigma }(i \omega_m)   \nonumber \\
& \quad + \sum_{\boldsymbol{q}, i \omega_m}    \sum_{\sigma } \left[ G^{\rm (N)}_{\, \boldsymbol{q}; \sigma, \sigma }(i \omega_m) \right]^2 \left. \frac{\delta \underline{\underline{\Sigma}}_{\,\boldsymbol{k}}(i \omega_n)}{\delta G_{\boldsymbol{q}; \sigma,  \sigma}(i \omega_m) }  \right|_{G = G^{\rm (N)}}  \,  \delta  \Sigma_{\, \boldsymbol{q}; \sigma,  \sigma }(i \omega_m)~,
\label{linearizing Eliashberg anomalous}
\end{align}
where we have used the fact that $\underline{\underline{G}}^{\rm (N)}_{\, \boldsymbol{q} }(i \omega_m)$ is diagonal. 

We should now derive an expression for $\delta \underline{\underline{\Sigma}}_{\,\boldsymbol{k}}(i \omega_n) /\delta G_{\boldsymbol{q}; \sigma,  \sigma}(i \omega_m)$. Referring to the partition of the self-energy given in Eq.~\eqref{separation self-energy}, and using the approximations~\eqref{Hartree Eliashberg} and~\eqref{Fock Eliashberg}, we obtain
\begin{align}
\left. \frac{\delta  \Sigma^{{\rm (H)}}_{\boldsymbol{k}; \mu, \mu'}(i \omega_n)}{\delta G_{\boldsymbol{q}; \sigma,  \sigma'}(i \omega_m) }  \right|_{G = G^{\rm (N)}} \approx \delta_{\mu, \mu'} \mu \, \delta_{\sigma, \sigma'} \sigma \,  T   \sum_{ s} M^{(\boldsymbol{0}, s)}_{\boldsymbol{k}} M^{(\boldsymbol{0}, s)}_{\boldsymbol{q}}  D^{\rm (N)}_{\boldsymbol{0}, s}(0)
\end{align}
and
\begin{align}
\left. \frac{\delta  \Sigma^{{\rm (F)}}_{ \boldsymbol{k} ; \mu, \mu' }(i \omega_n)}{\delta G_{\boldsymbol{q}; \sigma,  \sigma'}(i \omega_m) }  \right|_{G = G^{\rm (N)}} & \approx   - \delta_{\sigma, \mu} \delta_{\sigma', \mu'} \sigma \sigma' \, T   \sum_{ s}  
M^{( \boldsymbol{q} - \boldsymbol{k} , s)}_{\boldsymbol{k}}
M^{( \boldsymbol{k} - \boldsymbol{q} , s)}_{\boldsymbol{q}} \nonumber \\
& \quad \times D^{\rm (N)}_{\boldsymbol{k} - \boldsymbol{q}, s}(i \omega_n - i \omega_m)~.
\end{align}
Next, we consider the EEI self-energy functional, for which we write the Ward identity
\begin{align}
\left. \frac{\delta  \Sigma^{{\rm (E)}}_{\boldsymbol{k}; \mu, \mu'}(i \omega_n)}{\delta G_{\boldsymbol{q}; \sigma,  \sigma'}(i \omega_m) }  \right|_{G = G^{\rm (N)}} \equiv U^{[4] ({\rm N})}_{\boldsymbol{k}, \boldsymbol{q} ; \mu, \sigma; \sigma', \mu'}(i \omega_n, i \omega_m)~,
\end{align}
where $U^{[4] ({\rm N})}_{\boldsymbol{k}, \boldsymbol{q} ; \mu, \sigma; \sigma', \mu'}(i \omega_n, i \omega_m)$ is the irreducible four-leg vertex related to EEIs [see Eq.~\eqref{Bethe-Salpeter}]. In this case, it must be evaluated in the normal state, at $G=G^{\rm (N)}$. By comparison, in Ref.~\cite{Scalapino66} the EEI self-energy was taken in the GW approximation~\cite{GiulianiVignale}, with a statically screened interaction potential. Such potential is itself, in principle, a functional of the electronic GF through the dielectric function; however, in order to obtain tractable equations, in Ref.~\cite{Scalapino66} this dependence is neglected, and the effect of the screened potential is finally embedded in a ``pseudopotential" that does not depend on energy and momentum. By contrast, we keep the more general four-leg irreducible vertex in our derivation. Our procedure leads to a tractable system of equations if the EEI self-energy functional (and the related EEI vertices) can be determined {\it in the normal state} via independent means appropriate e.g.~to strongly correlated electron systems, such as DMFT~\cite{DMFT}.

Substituting all these expressions back into \eqref{linearizing Eliashberg anomalous}, and using the identity 
\begin{align}
G^{\rm (N)}_{\, \boldsymbol{q} ; \uparrow, \uparrow }(i \omega_m) \, G^{\rm (N)}_{\, \boldsymbol{q}; \downarrow, \downarrow }(i \omega_m) = 1 / \Theta^{\rm (N)}_{\boldsymbol{q}}( i \omega_m)~,
\end{align} 
we obtain
\begin{align}
& \delta\Sigma_{ \boldsymbol{k}; \mu, \mu'}(i \omega_n) \nonumber \\
 & = \sum_{\boldsymbol{q}, i \omega_m} \frac{1}{\Theta^{\rm (N)}_{\boldsymbol{q}}( i \omega_m)} \sum_{\sigma } U^{[4] {\rm N}}_{\boldsymbol{k}, \boldsymbol{q} ; \mu, \sigma; -\sigma, \mu'}(i \omega_n, i \omega_m)  \,  \delta  \Sigma_{\, \boldsymbol{q}; \sigma, - \sigma }(i \omega_m)   \nonumber \\
& \quad + \sum_{\boldsymbol{q}, i \omega_m}    \sum_{\sigma } \left[ G^{\rm (N)}_{\, \boldsymbol{q}; \sigma, \sigma }(i \omega_m) \right]^2 U^{[4] {\rm N}}_{\boldsymbol{k}, \boldsymbol{q} ; \mu, \sigma; \sigma, \mu'}(i \omega_n, i \omega_m)  \,  \delta  \Sigma_{\, \boldsymbol{q}; \sigma,  \sigma }(i \omega_m) \nonumber \\
& \quad + \delta_{\mu, \mu'} \mu \sum_{\boldsymbol{q}, i \omega_m}    \sum_{\sigma } \left[ G^{\rm (N)}_{\, \boldsymbol{q}; \sigma, \sigma }(i \omega_m) \right]^2  \,   \sigma \,  T   \sum_{ s} M^{(\boldsymbol{0}, s)}_{\boldsymbol{k}} M^{(\boldsymbol{0}, s)}_{\boldsymbol{q}}  D^{\rm (N)}_{\boldsymbol{0}, s}(0)  \,  \delta  \Sigma_{\, \boldsymbol{q}; \sigma,  \sigma }(i \omega_m) \nonumber \\
 & \quad + \delta_{\mu, -\mu'}\sum_{\boldsymbol{q}, i \omega_m} \frac{1}{\Theta^{\rm (N)}_{\boldsymbol{q}}( i \omega_m)}      T   \sum_{ s} M^{( \boldsymbol{q} - \boldsymbol{k} , s)}_{\boldsymbol{k}}
M^{( \boldsymbol{k} - \boldsymbol{q} , s)}_{\boldsymbol{q}}  D^{\rm (N)}_{\boldsymbol{k} - \boldsymbol{q}, s}(i \omega_n - i \omega_m)
  \nonumber \\
  & \quad \times  \delta  \Sigma_{\, \boldsymbol{q}; \mu, - \mu }(i \omega_m)   \nonumber \\
& \quad - \delta_{\mu, \mu'} \sum_{\boldsymbol{q}, i \omega_m}      \left[ G^{\rm (N)}_{\, \boldsymbol{q}; \mu, \mu }(i \omega_m) \right]^2   T   \sum_{ s} M^{( \boldsymbol{q} - \boldsymbol{k} , s)}_{\boldsymbol{k}}
M^{( \boldsymbol{k} - \boldsymbol{q} , s)}_{\boldsymbol{q}} D^{\rm (N)}_{\boldsymbol{k} - \boldsymbol{q}, s}(i \omega_n - i \omega_m)
  \nonumber \\
  & \quad \times  \delta  \Sigma_{\, \boldsymbol{q}; \mu,  \mu }(i \omega_m)~.
\label{linearizing Eliashberg anomalous - 2}
\end{align}
Regarding the four-leg EEI vertex, it holds that
\begin{align}
U^{[4]({\rm N})}_{\boldsymbol{k}, \boldsymbol{q} ; \mu, \sigma; \sigma', \mu'}(i \omega_n, i \omega_m)   \equiv \delta_{\sigma, \mu} \delta_{\sigma', \mu'} U^{[4] ({\rm N})}_{\boldsymbol{k}, \boldsymbol{q} ; \mu,  \mu'}(i \omega_n, i \omega_m)~.
\end{align}
The previous result stems from the fact that in the normal state all interaction vertices conserve spin and, therefore, the spin indices along the same propagator line must be equal. The self-energy corrections $\delta  \Sigma_{\, \boldsymbol{q}; \mu,  \mu }(i \omega_m)$ and $\delta  \Sigma_{\, \boldsymbol{q}; \mu, - \mu }(i \omega_m)$ are then decoupled: the diagonal terms satisfy
\begin{align}
  \delta\Sigma_{ \boldsymbol{k}; \mu, \mu}(i \omega_n)  
 & =   \sum_{\boldsymbol{q}, i \omega_m}      \left[ G^{\rm (N)}_{\, \boldsymbol{q}; \mu, \mu }(i \omega_m) \right]^2 \Big\{ U^{[4] ({\rm N})}_{\boldsymbol{k}, \boldsymbol{q} ; \mu,   \mu }(i \omega_n, i \omega_m)    
 \nonumber \\
& \quad -      T   \sum_{ s} M^{( \boldsymbol{q} - \boldsymbol{k} , s)}_{\boldsymbol{k}}
M^{( \boldsymbol{k} - \boldsymbol{q} , s)}_{\boldsymbol{q}} D^{\rm (N)}_{\boldsymbol{k} - \boldsymbol{q}, s}(i \omega_n - i \omega_m)
  \Big\}  \delta  \Sigma_{\, \boldsymbol{q}; \mu,  \mu }(i \omega_m) \nonumber \\
& \quad +   \mu \sum_{\boldsymbol{q}, i \omega_m}    \sum_{\sigma } \left[ G^{\rm (N)}_{\, \boldsymbol{q}; \sigma, \sigma }(i \omega_m) \right]^2  \,   \sigma \,  T   \sum_{ s} M^{(\boldsymbol{0}, s)}_{\boldsymbol{k}} M^{(\boldsymbol{0}, s)}_{\boldsymbol{q}} D^{\rm (N)}_{\boldsymbol{0}, s}(0)  \nonumber \\
& \quad \times  \delta  \Sigma_{\, \boldsymbol{q}; \sigma,  \sigma }(i \omega_m)  ~,
\label{linearizing Eliashberg anomalous - 3 diagonal}
\end{align}
while the nondiagonal terms satisfy
\begin{align}
  \delta\Sigma_{ \boldsymbol{k}; \mu, -\mu}(i \omega_n)  
 & = \sum_{\boldsymbol{q}, i \omega_m} \frac{1}{\Theta^{\rm (N)}_{\boldsymbol{q}}( i \omega_m)}  \Big\{ U^{[4] ({\rm N})}_{\boldsymbol{k}, \boldsymbol{q} ; \mu,   - \mu}(i \omega_n, i \omega_m)     \nonumber \\
 & \quad +        T   \sum_{ s} M^{( \boldsymbol{q} - \boldsymbol{k} , s)}_{\boldsymbol{k}}
M^{( \boldsymbol{k} - \boldsymbol{q} , s)}_{\boldsymbol{q}} D^{\rm (N)}_{\boldsymbol{k} - \boldsymbol{q}, s}(i \omega_n - i \omega_m)
  \Big\}  \nonumber \\
  & \quad \times \delta  \Sigma_{\, \boldsymbol{q}; \mu, - \mu }(i \omega_m)~.
\label{linearizing Eliashberg anomalous - 3 nondiagonal}
\end{align}
Note that the normal solution, corresponding to $\delta\Sigma_{ \boldsymbol{k}; \mu, \mu'}(i \omega_n) = 0$ for every choice of $\mu, \mu'$, is always possible. Moreover, we can choose $\delta\Sigma_{ \boldsymbol{k}; \mu, \mu}(i \omega_n) = 0$, while the off-diagonal components must be non-zero. These choices correspond to setting
\begin{align}
\delta \underline{\underline{\Sigma}}_{\, \boldsymbol{k}}(i \omega_n) \equiv \phi_{\boldsymbol{k}}(i \omega_n) \underline{\underline{\tau_1}}
+ \overline{\phi}_{\boldsymbol{k}}(i \omega_n) \underline{\underline{\tau_2}}~,
\label{anomalous is nondiagonal}
\end{align}
which conveniently allows to identify the full correction $\delta \underline{\underline{\Sigma}}_{\, \boldsymbol{k}}(i \omega_n)$ with the anomalous (off-diagonal) components of the self-energy. We adopt Eq.~\eqref{anomalous is nondiagonal} in the rest of the derivation.

Applying Eq.~\eqref{anomalous is nondiagonal} to Eq.~\eqref{linearizing Eliashberg anomalous - 3 nondiagonal}, we obtain the following Eliashberg-type equations for the total anomalous components of the self-energy:
\begin{align}
\phi_{\boldsymbol{k}}(i \omega_n)  
 & = \sum_{\boldsymbol{q}, i \omega_m} \frac{\phi_{\boldsymbol{q}}(i \omega_m)}{\Theta^{\rm (N)}_{\boldsymbol{q}}( i \omega_m)}  \Bigg[ \frac{1}{2}U^{[4] ({\rm N})}_{\boldsymbol{k}, \boldsymbol{q} ; \uparrow,   \downarrow}(i \omega_n, i \omega_m)        + \frac{1}{2} U^{[4] ({\rm N})}_{\boldsymbol{k}, \boldsymbol{q} ; \downarrow,   \uparrow}(i \omega_n, i \omega_m)         \nonumber \\
 & \quad +        T   \sum_{ s} M^{( \boldsymbol{q} - \boldsymbol{k} , s)}_{\boldsymbol{k}}
M^{( \boldsymbol{k} - \boldsymbol{q} , s)}_{\boldsymbol{q}} D^{\rm (N)}_{\boldsymbol{k} - \boldsymbol{q}, s}(i \omega_n - i \omega_m)    
  \Bigg]   \nonumber \\
  & \quad - \frac{ i }{2}  \sum_{\boldsymbol{q}, i \omega_m} \frac{\overline{\phi}_{\boldsymbol{q}}(i \omega_m) }{\Theta^{\rm (N)}_{\boldsymbol{q}}( i \omega_m)}  \left[  U^{[4] ({\rm N})}_{\boldsymbol{k}, \boldsymbol{q} ; \uparrow,   \downarrow}(i \omega_n, i \omega_m)      -   U^{[4] {\rm (N)}}_{\boldsymbol{k}, \boldsymbol{q} ; \downarrow,   \uparrow}(i \omega_n, i \omega_m)    \right]
\label{Eliashberg equation - phi}
\end{align}
and
\begin{align}
\overline{\phi}_{\boldsymbol{k}}(i \omega_n)    
 & = \sum_{\boldsymbol{q}, i \omega_m} \frac{   \overline{\phi}_{\boldsymbol{q}}(i \omega_m)}{\Theta^{\rm (N)}_{\boldsymbol{q}}( i \omega_m)}  \Bigg[ \frac{1}{2} U^{[4] ({\rm N})}_{\boldsymbol{k}, \boldsymbol{q} ; \uparrow,   \downarrow}(i \omega_n, i \omega_m)           + \frac{1}{2} U^{[4] ({\rm N})}_{\boldsymbol{k}, \boldsymbol{q} ; \downarrow,   \uparrow}(i \omega_n, i \omega_m)       \nonumber \\
 & \quad +        T   \sum_{ s} M^{( \boldsymbol{q} - \boldsymbol{k} , s)}_{\boldsymbol{k}}
M^{( \boldsymbol{k} - \boldsymbol{q} , s)}_{\boldsymbol{q}} D^{\rm (N)}_{\boldsymbol{k} - \boldsymbol{q}, s}(i \omega_n - i \omega_m)    
  \Bigg]  \nonumber \\
 & \quad + \frac{i}{2} \sum_{\boldsymbol{q}, i \omega_m} \frac{\phi_{\boldsymbol{q}}(i \omega_m)}{\Theta^{\rm (N)}_{\boldsymbol{q}}( i \omega_m)}  \left[ U^{[4] {\rm (N)}}_{\boldsymbol{k}, \boldsymbol{q} ; \uparrow,   \downarrow}(i \omega_n, i \omega_m) - U^{[4] {\rm (N)}}_{\boldsymbol{k}, \boldsymbol{q} ; \downarrow,   \uparrow}(i \omega_n, i \omega_m)          
  \right]~.
\label{Eliashberg equation - bar phi}
\end{align}
Eqs.~\eqref{Eliashberg equation - phi} and \eqref{Eliashberg equation - bar phi}, together with \eqref{Eliashberg III} and \eqref{Eliashberg IV}, completely determine the anomalous components of the self-energy, for temperatures $T \approx T_{\rm c}$. If we assume that $U^{[4] {\rm (N)}}_{\boldsymbol{k}, \boldsymbol{q} ; \uparrow,   \downarrow}(i \omega_n, i \omega_m) = U^{[4] {\rm (N)}}_{\boldsymbol{k}, \boldsymbol{q} ; \downarrow,   \uparrow}(i \omega_n, i \omega_m)  $, Eqs.~\eqref{Eliashberg equation - phi} and \eqref{Eliashberg equation - bar phi} achieve a simplification that effectively decouples the equation for $\phi_{\boldsymbol{k}}(i \omega_n)$ from that for $\overline{\phi}_{\boldsymbol{k}}(i \omega_n)$. Combining these simplified equations with \eqref{Eliashberg III} and \eqref{Eliashberg IV} evaluated at $\Theta_{\boldsymbol{q}}(i \omega_m) \approx \Theta^{\rm (N)}_{\boldsymbol{q}}(i \omega_m) $ and $D_{\boldsymbol{k} - \boldsymbol{q}, s}(i \omega_n - i \omega_m) \approx D^{\rm (N)}_{\boldsymbol{k} - \boldsymbol{q}, s}(i \omega_n - i \omega_m)$, we obtain
\begin{align}
\phi^{\rm (E)}_{\boldsymbol{k}}(i \omega_n)  
 & = \sum_{\boldsymbol{q}, i \omega_m}     U^{[4] ({\rm N})}_{\boldsymbol{k}, \boldsymbol{q} ; \uparrow,   \downarrow}(i \omega_n, i \omega_m)  \frac{\phi_{\boldsymbol{q}}(i \omega_m)}{\Theta^{\rm (N)}_{\boldsymbol{q}}( i \omega_m)}                  
\label{Eliashberg equation - phi simplified}
\end{align}
and
\begin{align}
\phi^{\rm (EP)}_{\boldsymbol{k}}(i \omega_n)     =   T \sum_{i \omega_m} \sum_{\boldsymbol{q}, s} M^{( \boldsymbol{q} - \boldsymbol{k} , s)}_{\boldsymbol{k}}
M^{( \boldsymbol{k} - \boldsymbol{q} , s)}_{\boldsymbol{q}} D^{\rm (N)}_{\boldsymbol{k} - \boldsymbol{q}, s}(i \omega_n - i \omega_m) \, \frac{\phi_{  \boldsymbol{q}  }(i \omega_m)   }{\Theta^{\rm (N)}_{  \boldsymbol{q}  }(i \omega_m)}~,
\label{Eliashberg III simplified}
\end{align}
which connect $\phi^{\rm (E)}_{\boldsymbol{k}}(i \omega_n)$ and $\phi^{\rm (EP)}_{\boldsymbol{k}}(i \omega_n)$, respectively, to their sum $\phi_{\boldsymbol{k}}(i \omega_n) = \phi^{\rm (E)}_{\boldsymbol{k}}(i \omega_n) + \phi^{\rm (EP)}_{\boldsymbol{k}}(i \omega_n)$, thus forming a system of two coupled matrix equations in two unknown functions. An identical system is obtained for the $\underline{\underline{\tau_2}}$ components of the self-energy, with the replacements $\phi^{\rm (E)} \rightarrow \overline{\phi}^{\rm (E)}$, $\phi^{\rm (EP)} \rightarrow \overline{\phi}^{\rm (EP)}$, and $\phi \rightarrow \overline{\phi}$. Therefore, the $\underline{\underline{\tau_1}}$ components of the self-energy are equal to the $\underline{\underline{\tau_2}}$ components up to a constant, which can also be zero. It is possible to set $\overline{\phi}^{\rm (E)} = \overline{\phi}^{\rm (EP)}= 0 = \overline{\phi}$, while retaining a nonzero solution for $\phi$~\cite{Anokhin96}. 

In the present framework, the remaining ingredients for the total determination of the self-energy are Eqs.~\eqref{Eliashberg I} and \eqref{Eliashberg II} evaluated in the normal state, plus the two equations for $\gamma_{\boldsymbol{k}}(i \omega_n)$ and $\chi^{\rm (E)}_{\boldsymbol{k}}(i \omega_n)$ taken from \eqref{Eliashberg list}, also evaluated in the normal state. In fact, we recall that our working hypothesis is that the EEI self-energy is known in the normal state; therefore, the latter two equations are to be considered as solved.

To summarize, we put together all the above results, and present a set of {\it six} equations---not eight, because of the gauge choice $\overline{\phi}^{\rm (E)} = \overline{\phi}^{\rm (EP)}= 0$---that allow to determine the anomalous corrections to the self-energy in the linearized regime that applies for $T \approx T_{\rm c}$:
\begin{align}\label{Eliashberg I final}
\phi^{\rm (E)}_{\boldsymbol{k}}(i \omega_n)  
    = \sum_{\boldsymbol{q}, i \omega_m}     U^{[4] ({\rm N})}_{\boldsymbol{k}, \boldsymbol{q} ; \uparrow,   \downarrow}(i \omega_n, i \omega_m)  \frac{\phi^{\rm (E)}_{\boldsymbol{q}}(i \omega_m) + \phi^{\rm (EP)}_{\boldsymbol{q}}(i \omega_m)}{\Theta^{\rm (N)}_{\boldsymbol{q}}( i \omega_m)}~,          
\end{align}
\begin{align}\label{Eliashberg II final}
 \phi^{\rm (EP)}_{\boldsymbol{k}}(i \omega_n)  &  =   T \sum_{i \omega_m} \sum_{\boldsymbol{q}, s} M^{( \boldsymbol{q} - \boldsymbol{k} , s)}_{\boldsymbol{k}}
M^{( \boldsymbol{k} - \boldsymbol{q} , s)}_{\boldsymbol{q}} D^{\rm (N)}_{\boldsymbol{k} - \boldsymbol{q}, s}(i \omega_n - i \omega_m) \nonumber \\
& \quad  \times \frac{\phi^{\rm (E)}_{\boldsymbol{q}}(i \omega_m) + \phi^{\rm (EP)}_{\boldsymbol{q}}(i \omega_m)  }{\Theta^{\rm (N)}_{  \boldsymbol{q}  }(i \omega_m)}~,   
\end{align}
\begin{align}\label{Eliashberg III final}
\gamma_{\boldsymbol{k}}(i \omega_n) = 1 - \frac{\Sigma^{\rm (E, N)}_{\boldsymbol{k}; \uparrow, \uparrow}(i \omega_n) + \Sigma^{\rm (E, N)}_{\boldsymbol{k}; \downarrow, \downarrow}(i \omega_n)}{2 i \omega_n}~, 
\end{align}
\begin{align}\label{Eliashberg IV final}
\chi^{\rm (E )}_{\boldsymbol{k}}(i \omega_n) = \frac{\Sigma^{\rm (E, N)}_{\boldsymbol{k}; \uparrow, \uparrow}(i \omega_n) - \Sigma^{\rm (E, N)}_{\boldsymbol{k}; \downarrow, \downarrow}(i \omega_n)}{2}~,  
\end{align}
\begin{align}\label{Eliashberg V final}
  i \omega_n \gamma_{\boldsymbol{k}}(i \omega_n) \left[ 1 - Z_{\boldsymbol{k}}(i \omega_n)\right]   & =   - T \sum_{i \omega_m} \sum_{\boldsymbol{q}, s} M^{( \boldsymbol{q} - \boldsymbol{k} , s)}_{\boldsymbol{k}}
M^{( \boldsymbol{k} - \boldsymbol{q} , s)}_{\boldsymbol{q}} D^{\rm (N)}_{\boldsymbol{k} - \boldsymbol{q}, s}(i \omega_n - i \omega_m) \nonumber \\
& \quad  \times \frac{  i \omega_m   \gamma_{  \boldsymbol{q}}(i \omega_m) Z_{ \boldsymbol{q}}(i \omega_m)}{  \Theta^{\rm (N)}_{   \boldsymbol{q}  }(i \omega_m)}~,
\end{align}
\begin{align}\label{Eliashberg VI final}
  \chi^{\rm (EP)}_{\boldsymbol{k}}(i \omega_n) & =   - T \sum_{i \omega_m} \sum_{\boldsymbol{q}, s} \left[ M^{( \boldsymbol{q} - \boldsymbol{k} , s)}_{\boldsymbol{k}}
M^{( \boldsymbol{k} - \boldsymbol{q} , s)}_{\boldsymbol{q}} D^{\rm (N)}_{\boldsymbol{k} - \boldsymbol{q}, s}(i \omega_n - i \omega_m) \right. \nonumber \\
& \quad \left. - 2      M^{(\boldsymbol{0}, s)}_{\boldsymbol{k}} M^{(\boldsymbol{0}, s)}_{\boldsymbol{q}} D_{\boldsymbol{0}, s}(0)  \right] \frac{\xi_{  \boldsymbol{q}} + \chi_{  \boldsymbol{q}  }(i \omega_m)  }{ \Theta^{\rm (N)}_{  \boldsymbol{q}  }(i \omega_m)}~.
\end{align}
In all of the above,
\begin{align}
\Theta^{\rm (N)}_{\boldsymbol{k}}(i \omega_n) \equiv \left[ i \omega_n \gamma_{\boldsymbol{k}}(i \omega_n) Z_{\boldsymbol{k}}(i \omega_n) \right]^2 - \left[ \xi_{\boldsymbol{k}} + \chi_{\boldsymbol{k}}(i \omega_n) \right]^2~.
\end{align}

The equation for the total anomalous component of the electronic self-energy is obtained by summing Eqs.~\eqref{Eliashberg I final} and \eqref{Eliashberg II final}:
\begin{align}\label{Eliashberg anomalous Sigma}
\phi_{\boldsymbol{k}}(i \omega_n)  
    & =    \sum_{\boldsymbol{q}, i \omega_m} \Bigg[   \sum_s T M^{( \boldsymbol{q} - \boldsymbol{k} , s)}_{\boldsymbol{k}}
M^{( \boldsymbol{k} - \boldsymbol{q} , s)}_{\boldsymbol{q}} D^{\rm (N)}_{\boldsymbol{k} - \boldsymbol{q}, s}(i \omega_n - i \omega_m) \nonumber \\
& \quad + U^{[4] ({\rm N})}_{\boldsymbol{k}, \boldsymbol{q} ; \uparrow,   \downarrow}(i \omega_n, i \omega_m)  \Bigg]   \frac{\phi_{\boldsymbol{q}}(i \omega_m)    }{\Theta^{\rm (N)}_{  \boldsymbol{q}  }(i \omega_m)}  ~.          
\end{align}
The inclusion of the four-leg irreducible vertex $U^{[4]}$ formally allows for a part of the EEI effects that are not captured in widely used approximation schemes, such as the Tolmachev-Morel-Anderson method~\cite{Tolmachev, MorelAnderson}, where the Coulomb potential entering the self-energy equations is approximated as a momentum- and frequency-independent pseudopotential. While the Tolmachev-Morel-Anderson method has allowed for great progress in the history of superconductivity (in particular, it laid the foundation for the derivation of the McMillan formula for $T_{\rm c}$~\cite{McMillan}), it is based on an uncontrolled approximation that neglects Coulomb vertex corrections and the associated retardation effects~\cite{Bauer12}. Within the stated approximations (i.e.~linearization of the self-energy functional for $T$ close to $T_{\rm c}$, Eq.~\eqref{Eliashberg linearization}), our formulas include the formally exact contributions from the EEI self-energy functional; we remind the reader that the rest of our approximations affects the EPI self-energy functional only, see Eqs.~\eqref{Hartree Eliashberg} and~\eqref{Fock Eliashberg}.

\section{Summary and conclusions}
\label{sect:conclusions}

In summary, in this work we have derived the Luttinger-Ward functional for a system of electrons in the presence of both electron-electron and electron-phonon interactions---see Eqs.~\eqref{LFW, Fock term}-\eqref{terms of the LWF}. Without relying on a weak-coupling skeleton-diagram expansion, we have demonstrated that this functional generates two types of contributions to the electronic self-energy: i) a Fock-type contribution, which is reported in Eq.~\eqref{Sigma Fock simplified} and, diagrammatically, in Fig.~\ref{Fig: diagram Fock self-energy}; and ii) a Hartree-type contribution, which can be found in Eq.~\eqref{Sigma Hartree simplified} and Figs.~\ref{Fig: diagram Hartree self-energy - 1}--\ref{Fig: diagram Hartree self-energy - 2}. In both classes of contributions, which are due to electron-phonon interactions, entirely new terms appear, containing the irreducible six-leg vertex related to electron-electron interactions. To the best of our knowledge, these contributions have never been discussed in the literature on electron-phonon interactions and related phonon-mediated superconductivity. Under certain approximations, we have used these results to derive extended Eliashberg equations---see Eqs.~\eqref{Eliashberg I final}--\eqref{Eliashberg VI final}. These include vertex corrections due to electron-electron interactions, which are not accounted for by the Tolmachev-Morel-Anderson pseudopotential~\cite{Tolmachev, MorelAnderson}.

As emphasized throughout this work, much more work is needed to shed light on the physical implications of our theory, especially in the realm of phonon-mediated superconductivity in strongly correlated electron systems (SCESs). In this case, approximations such as DMFT~\cite{DMFT} and beyond-DMFT approaches~\cite{beyondDMFT}, such as dual fermion/bosons~\cite{dual}, which have been proven to be very useful in dealing with SCESs, may also be used in this context to examine $U^{[6]}$, as well as the functionals $\Sigma^{({\rm E})}$ and $U^{[4]}$ which are needed as inputs in our extended Eliashberg equations. Another possibility is to rely on a large-$N$ approximation, using $1/N$ as small parameter to select leading diagrams contributing to $U^{[6]}$, $\Sigma^{({\rm E})}$, and $U^{[4]}$, where $N$ is the number of fermion flavors.

We believe that the theory presented in this work could provide tools for the study of EPIs and phonon-mediated superconductivity in materials where the applicability of Migdal theorem is disputed because of features such as, e.g., large phonon frequencies, van Hove singularities, and strong EEIs. Relevant modern examples of these materials include hydrates~\cite{drozdov_nature_2015,drozdov_nature_2019,somayazulu_prl_2019,errea_prl_2015,
errea_nature_2016,errea_arxiv_2019,quan_prb_2016,sano_prb_2016,liu_prb_2019,ghosh_prb_2019} and magic-angle twisted bilayer graphene~\cite{cao_naturea_2018,cao_natureb_2018,yankowitz_science_2019,
lu_nature_2019,kerelsky_nature_2019,xie_nature_2019,
jiang_nature_2019,cao_arxiv_2019,hesp_arxiv_2019,
polshyn_naturephys_2019,wu_prl_2018,lian_prl_2019,
yudhistira_prb_2019,angeli_prx_2019,bistritzer_pnas_2011,
sanjose_prl_2012,tarnopolsky_prl_2019,koshino_prx_2018,
carr_prr_2019}. We mention, in passing, that our theory can be generalized to systems of electrons interacting with cavity photons~\cite{Schlawin19}, paying particular attention to the ${\bm A}^2$ term and gauge invariance~\cite{andolina_prb_2019}.

\section*{Acknowledgements}
It has been a great pleasure and honor to contribute to the special issue of
Annals of Physics dedicated to G. M. Eliashberg, one of the founding fathers
of the microscopic theory of superconductivity, on the occasion of his ninetieth
birthday.

This work was supported by the European Union's Horizon 2020 research and innovation programme under grant agreement No.~785219 - GrapheneCore2. M. I. K. also acknowledges support from FLAG-ERA JTC2017 Project GRANSPORT.

\appendix

\section{Derivation of the Hamiltonian}
\label{app: Derivation Hamiltonian}

Since the conventions used for the definition of the EPI parameters are not universal in the literature, we present here a full derivation of the electron-phonon Hamiltonian that we have used in this work. This allows to identify the physical definition of the parameters $\lbrace M^{(\lambda)}_{\alpha, \beta} \rbrace$, and to put them in correspondence with the parameters used by other authors. 
 
We use some concepts borrowed from the derivations given in Refs.~\cite{Grosso, Mahan, Cappelluti06}. We write the Hamiltonian for a general system of electrons and nuclei in first quantization, in the position representation, as
\begin{align}
H    & =    T_{\rm e}  + V_{\rm ee}  + V_{\rm eN}  + T_{\rm N}  + V_{\rm NN}  \nonumber \\
& =    \sum_{\boldsymbol{r}} \frac{- \hbar^2 \nabla^2_{\boldsymbol{r}}}{2 m}  + \frac{1}{2} \sum_{\boldsymbol{r}, \boldsymbol{r}' \neq \boldsymbol{r}} V_{\rm ee}(  \boldsymbol{r}  - \boldsymbol{r}'  )   +   \sum_{\boldsymbol{r}, \boldsymbol{R}  } V_{\rm eN}(  \boldsymbol{r}  - \boldsymbol{R}   ) + \sum_{\boldsymbol{R}} \frac{- \hbar^2 \nabla^2_{\boldsymbol{R}}}{2 m_{\boldsymbol{R}}}  \nonumber \\
& \quad + \frac{1}{2} \sum_{\boldsymbol{R}, \boldsymbol{R}' \neq \boldsymbol{R}} V_{\rm NN}(  \boldsymbol{R}  - \boldsymbol{R}'  )~,
\end{align}
where $\boldsymbol{r}$ is an electron coordinate, and $\boldsymbol{R}$ is a nucleus coordinate; $m$ (without subscript) is the electron mass, while $m_{\boldsymbol{R}}$ is the mass of the nucleus at position $\boldsymbol{R}$.

The standard Born-Oppenheimer derivation goes as follows. We introduce the set of nuclear equilibrium position vectors $\boldsymbol{R}_{i,n} \equiv \boldsymbol{R}_{i} + \boldsymbol{B}_{n}$, where $\boldsymbol{R}_{i}$ is a lattice vector and $\boldsymbol{B}_{n}$ is a basis vector that distinguishes the atoms inside a given unit cell. The equilibrium positions satisfy the following set of equations,
\begin{align}
\sum_{ \boldsymbol{R}' \neq \boldsymbol{R}}   \nabla_{\boldsymbol{R}} V_{\rm NN}(  \boldsymbol{R}  - \boldsymbol{R}'  )   = \boldsymbol{0} \, , \quad \forall \boldsymbol{R}~.
\end{align}
Displacements of the nuclei with respect to the equilibrium positions are denoted as $\boldsymbol{Q}_{i,n}  \equiv \boldsymbol{R}  - \boldsymbol{R}_{i, n}$. If they are assumed to be small, then the Hamiltonian can be expanded in powers of $\boldsymbol{Q}_{i,n}$ up to the second order. Denoting with $\mu, \nu$ the Cartesian coordinates, we obtain
\begin{align}
H      & \approx       \sum_{\boldsymbol{r}} \frac{- \hbar^2 \nabla^2_{\boldsymbol{r}}}{2 m} +   \sum_{\boldsymbol{r}, (i, n)  } V_{\rm eN}(  \boldsymbol{r}  - \boldsymbol{R}_{i, n}   ) + \frac{1}{2} \sum_{\boldsymbol{r}, \boldsymbol{r}' \neq \boldsymbol{r}} V_{\rm ee}(  \boldsymbol{r}  - \boldsymbol{r}'  )          \nonumber \\
& \quad   
- \sum_{\boldsymbol{r}, (i,n)  } \sum_{\mu}  Q_{(i,n), \mu}  \left.  \left[        \frac{ \partial V_{\rm eN}( \boldsymbol{R}   )}{\partial R_{\mu}}   - \frac{1}{2}    \sum_{ \nu}   Q_{(i,n), \nu}   \frac{ \partial^2 V_{\rm eN}(     \boldsymbol{R}  ) }{\partial R_{\mu} \partial R_{\nu}}  \right]  \right|_{\boldsymbol{R} = \boldsymbol{r}  - \boldsymbol{R}_{i,n}}   \nonumber \\
& \quad  + \sum_{(i,n)} \Bigg[ \frac{- \hbar^2 \nabla^2_{\boldsymbol{Q}_{i,n}}}{2 m_{n}}   + \frac{1}{4} \sum_{ (i',n') \neq (i,n)} \sum_{\mu, \nu} \left( Q_{(i,n), \mu} - Q_{(i',n'), \mu} \right)     \nonumber \\
& \quad \times   \left( Q_{(i,n), \nu} - Q_{(i',n'), \nu} \right)    \left. \frac{ \partial^2 V_{\rm NN}( \boldsymbol{R}   ) }{\partial R_{\mu} \partial R_{\nu}} \right|_{\boldsymbol{R} = \boldsymbol{R}_{i,n} - \boldsymbol{R}_{i',n'} }  \Bigg] \nonumber \\
& \quad + \frac{1}{2} \sum_{(i,n), (i',n') \neq (i,n)} V_{\rm NN}(  \boldsymbol{R}_{i,n} - \boldsymbol{R}_{i',n'}  )~,
\label{Hamiltonian first quantization}
\end{align}
where $m_{n}$ is the mass of the nucleus at position $\boldsymbol{B}_n$ within a unit cell (due to the lattice periodicity, it does not depend on the unit cell $i$). In the right-hand side of Eq.~\eqref{Hamiltonian first quantization}: 1) the first line is the Hamiltonian for the electronic subsystem alone, including the kinetic energy, a spatially-periodic external potential given by the interaction with the lattice, and the EEI; 2) the second line is the EPI, including a linear and a quadratic term in the nuclear displacements (we neglect the latter, as is usually done); 3) the third and fourth lines constitute the Hamiltonian for the phonon subsystem alone, in the harmonic approximation; 4) the fifth line is a constant with respect to the dynamical coordinates associated with the fixed equilibrium configuration.  

The first line can be written in second quantization as the sum of the terms~\eqref{H_IE} and~\eqref{H_EEI} of the main text, for a general single-electron basis set of wave functions labelled by the index $\alpha = (i_{\alpha}, \sigma_{\alpha})$. 

The displacement operators are quantized as~\cite{Mahan}
\begin{align}
\hat{\boldsymbol{Q}}_{i,n} = i \sum_{\boldsymbol{q}, s} \sqrt{\frac{\hbar}{2 N     \omega_{\boldsymbol{q}, s}} } \frac{ \boldsymbol{\eta}_{\boldsymbol{q}, s, n}}{\sqrt{m_n}} \left( \hat{b}_{\boldsymbol{q},s} + \hat{b}^{\dagger}_{- \boldsymbol{q}, s}\right) e^{i \boldsymbol{q} \cdot \boldsymbol{R}_{ i}}   = \hat{\boldsymbol{Q}}^{\dagger}_{i,n}~,
\label{Q quant}
\end{align}
where $s$ is the branch index, and the polarization vectors satisfy $\boldsymbol{\eta}_{\boldsymbol{q}, s, n} = - \boldsymbol{\eta}_{- \boldsymbol{q}, s, n}$. The polarization vectors and the mass, in general, depend on $n$, as they might be different for different atoms within the same unit cell (so, they depend on the basis vector $\boldsymbol{B}_n$, although not on the lattice vector $\boldsymbol{R}_i$). The quantity $\omega_{\boldsymbol{q}, s} = \omega_{-\boldsymbol{q}, s}$ is, at this stage, introduced as a mere parameter. When Eq.~\eqref{Q quant} is replaced into Eq.~\eqref{Hamiltonian first quantization}, the third and fourth lines of the latter are transformed into the second-quantized IP Hamiltonian given in~Eq.~\eqref{H_IP} of the main text, and $\omega_{\boldsymbol{q}, s}$ acquires the meaning of a phonon (vibrational mode) frequency; it is related to the direct-space matrix $f_{i,j}(s)$ via Eqs.~\eqref{Fourier diagonalization}.

The linear term of the EPI [second line of Eq.~\eqref{Hamiltonian first quantization}] becomes:
\begin{align}
& -   \sum_{\boldsymbol{r}, (j,n)  }  \hat{\boldsymbol{Q}}_{j,n}  \cdot  \left. \frac{ \partial V_{\rm eN}( \boldsymbol{R}   )}{\partial \boldsymbol{R} } \right|_{\boldsymbol{R} = \boldsymbol{r}  - \boldsymbol{R}_{j,n}} \nonumber \\
& =  i \sum_{\boldsymbol{q}, s ,   (j,n)  }  \sqrt{\frac{\hbar}{2 N     \omega_{\boldsymbol{q}, s}} }  \frac{ e^{-i \boldsymbol{q} \cdot \boldsymbol{R}_j } }{\sqrt{m_n}} \, \boldsymbol{\eta}_{\boldsymbol{q}, s, n} \cdot \left[ \sum_{\boldsymbol{r}} \left.   \nabla V_{\rm eN}( \boldsymbol{R}   )  \right|_{\boldsymbol{R} = \boldsymbol{r} - \boldsymbol{R}_{j,n}}   \right] \!  \left( \hat{b}_{-\boldsymbol{q},s} + \hat{b}^{\dagger}_{  \boldsymbol{q}, s}\right)    .
\label{EPI first quantization}
\end{align}

We now write the EPI Hamiltonian in second quantization. Let $\sigma = ( n, \tau )$, where $n$ denotes the orbital within a unit cell, and $\tau$ denotes the electron spin. Let $\psi_{i_{\alpha}, n_{\alpha}}(\boldsymbol{r})$ be the spatial part of the single-electron basis wave function corresponding to the composite index $\alpha$. We introduce the quantity
\begin{align}
M^{(\boldsymbol{q},s) }_{\alpha, \beta} &  \equiv  \delta_{\tau_{\alpha}, \tau_{\beta}} i \sum_{j,n} F^*_{\boldsymbol{q}, j}   \sqrt{ \frac{ \hbar }{m_n  } } \, \boldsymbol{\eta}_{\boldsymbol{q}, s, n} \cdot \int d \boldsymbol{r} \psi_{i_{\alpha}, n_{\alpha}}^*(\boldsymbol{r})  \left[  \left.   \nabla V_{\rm eN}( \boldsymbol{R}   )  \right|_{\boldsymbol{R} = \boldsymbol{r} - \boldsymbol{R}_{j,n}}   \right] \nonumber \\
& \quad \times \psi_{i_{\beta}, n_{\beta}}(\boldsymbol{r})~, 
\end{align}
and the operator
\begin{align}
\hat{Q}_{\boldsymbol{q} , s } =        \frac{1}{\sqrt{2   \omega_{\boldsymbol{q},s}}} \left( \hat{b}^{\dagger}_{\boldsymbol{q}, s} + \hat{b}_{-\boldsymbol{q}, s} \right)~.   
\end{align}
Using these, we turn Eq.~\eqref{EPI first quantization} into its second-quantized expression,
\begin{align}
\hat{\cal H}_{\rm EPI}  =    \sum_{\boldsymbol{q}, s     } \sum_{\alpha, \beta}  \, M^{(\boldsymbol{q},s) }_{\alpha, \beta}         \hat{Q}_{\boldsymbol{q}, s }  \hat{c}^{\dagger}_{\alpha} \hat{c}_{\beta} \,.
\label{EPI for momentum considerations}
\end{align}
We turn to the position representation by applying the Fourier transformation, with the coefficients defined in Eq.~\eqref{Fourier matrix}, and we obtain
\begin{align}
\hat{\cal H}_{\rm EPI}  =    \sum_{(j , s)} \sum_{\alpha, \beta}  \,               M^{(j,s)}_{\alpha, \beta} \hat{Q}_{j, s }  \hat{c}^{\dagger}_{\alpha} \hat{c}_{\beta}~,
\label{EPI second quantization}
\end{align}
where
\begin{align}
M^{(j,s)}_{\alpha, \beta} \equiv \sum_{\boldsymbol{q}} F_{\boldsymbol{q}, j}    M^{(\boldsymbol{q},s) }_{\alpha, \beta} &  = \delta_{\tau_{\alpha}, \tau_{\beta}}  i      \sum_{j',n}  \sqrt{ \frac{ \hbar }{m_n  } } \, \left( \sum_{\boldsymbol{q}} F_{\boldsymbol{q}, j}  F^*_{\boldsymbol{q}, j'}   \boldsymbol{\eta}_{\boldsymbol{q}, s, n} \right) \nonumber \\
& \quad \cdot \int d \boldsymbol{r} \, \psi_{i_{\alpha}, n_{\alpha}}^*(\boldsymbol{r})  \left[    \nabla_{\boldsymbol{r}} V_{\rm eN}( \boldsymbol{r} - \boldsymbol{R}_{j',n}   )    \right] \psi_{i_{\beta}, n_{\beta}}(\boldsymbol{r})~,
\label{microscopic definition M}
\end{align}
and
\begin{align}
\hat{Q}_{j , s } \equiv \sum_{\boldsymbol{q}} F^*_{\boldsymbol{q}, j}  \hat{Q}_{\boldsymbol{q} , s }  =  \sum_{\boldsymbol{q}} F^*_{\boldsymbol{q}, j}    \frac{1}{\sqrt{2   \omega_{\boldsymbol{q},s}}} \left( \hat{b}^{\dagger}_{\boldsymbol{q}, s} + \hat{b}_{-\boldsymbol{q}, s} \right)~.
\label{Q phonon}    
\end{align}
Switching to the Greek-indices notation, we immediately turn Eq.~\eqref{EPI second quantization} into Eq.~\eqref{H_EPI} of the main text.

Let us investigate Eq.~\eqref{microscopic definition M} further. We write the Fourier representation of $V_{\rm eN}( \boldsymbol{r} - \boldsymbol{R}_{j'} -  \boldsymbol{B}_n )$ as
\begin{align}
V_{\rm eN}( \boldsymbol{r} - \boldsymbol{R}_{j'}  -  \boldsymbol{B}_n ) & = \sum_{\boldsymbol{q}'} \sum_{\boldsymbol{K}} \frac{ e^{- i (\boldsymbol{q}' + \boldsymbol{K}) \cdot ( \boldsymbol{r} - \boldsymbol{R}_{j'}  -  \boldsymbol{B}_n )}}{N \sqrt{N}} \, V_{\rm eN}(\boldsymbol{q}'+ \boldsymbol{K})  \nonumber \\
&  =  \sum_{\boldsymbol{q}'} \sum_{\boldsymbol{K}}   F_{\boldsymbol{q}', j'}  \,    e^{- i  ( \boldsymbol{q}' + \boldsymbol{K} ) \cdot  ( \boldsymbol{r} -  \boldsymbol{B}_n ) }    \, \frac{ V_{\rm eN}(\boldsymbol{q}' + \boldsymbol{K}) }{N}~,
\end{align}
where $\boldsymbol{q}'$ belongs to the first Brillouin zone, while $\boldsymbol{K}$ is a reciprocal lattice vector. Substituting this into Eq.~\eqref{microscopic definition M}, we obtain
\begin{align}
M^{(j,s)}_{\alpha, \beta}   &  = \delta_{\tau_{\alpha}, \tau_{\beta}}         \sum_{\boldsymbol{q}} F_{\boldsymbol{q}, j}  \sum_{ n}  \sqrt{ \frac{ \hbar }{m_n  } } \,  \boldsymbol{\eta}_{\boldsymbol{q}, s, n}   \cdot  \sum_{\boldsymbol{K}} ( \boldsymbol{q}  + \boldsymbol{K} ) \frac{ V_{\rm eN}(\boldsymbol{q}  + \boldsymbol{K}) }{N} e^{  i  ( \boldsymbol{q}  + \boldsymbol{K} ) \cdot      \boldsymbol{B}_n   } \nonumber \\
& \quad \times \int d \boldsymbol{r} \, \psi_{i_{\alpha}, n_{\alpha}}^*(\boldsymbol{r})         \,    e^{- i  ( \boldsymbol{q}  + \boldsymbol{K} ) \cdot    \boldsymbol{r}   }        \, \psi_{i_{\beta}, n_{\beta}}(\boldsymbol{r})~.
\label{M before Bloch}
\end{align}
We now represent the single-electron wave functions as
\begin{align}
\psi_{i , n }(\boldsymbol{r}) = \sum_{\boldsymbol{k}} F^*_{i, \boldsymbol{k}} \psi_{\boldsymbol{k} , n }(\boldsymbol{r}) \equiv \sum_{\boldsymbol{k}} F^*_{i, \boldsymbol{k}} e^{i \boldsymbol{k} \cdot \boldsymbol{r}} \sum_{\boldsymbol{G}} e^{i \boldsymbol{G} \cdot \boldsymbol{r}}\psi_{\boldsymbol{k} , n }(\boldsymbol{G})~,
\end{align}
where $\psi_{\boldsymbol{k} , n }(\boldsymbol{r})$ is a Bloch state, whose Fourier decomposition is written explicitly in the last equality ($\boldsymbol{G}$ is a reciprocal lattice vector). This allows us to perform the integration over $\boldsymbol{r}$, which yields a factor 
\begin{align}
\int d \boldsymbol{r} \,  e^{-i \left( \boldsymbol{k}  + \boldsymbol{G} + \boldsymbol{q}  + \boldsymbol{K} - \boldsymbol{k}' - \boldsymbol{G}' \right) \cdot    \boldsymbol{r}   } = N \mathcal{V}_{\rm cell} \delta_{\boldsymbol{k}   + \boldsymbol{q}    - \boldsymbol{k}'    , \boldsymbol{G}' - \boldsymbol{G} - \boldsymbol{K} }~,
\end{align}
where $\mathcal{V}_{\rm cell}$ is the volume of a unit cell. We obtain
\begin{align}
M^{(j,s)}_{\alpha, \beta}   &  = \delta_{\tau_{\alpha}, \tau_{\beta}}         \sum_{\boldsymbol{q}} F_{\boldsymbol{q}, j}  \sum_{ n}  \sqrt{ \frac{ \hbar }{m_n  } } \,  \boldsymbol{\eta}_{\boldsymbol{q}, s, n}   \cdot  \sum_{\boldsymbol{K}} ( \boldsymbol{q}  + \boldsymbol{K} )   V_{\rm eN}(\boldsymbol{q}  + \boldsymbol{K})   e^{  i  ( \boldsymbol{q}  + \boldsymbol{K} ) \cdot      \boldsymbol{B}_n   } \nonumber \\
& \quad \times \sum_{\boldsymbol{k} , \boldsymbol{k}'} \sum_{\boldsymbol{G} , \boldsymbol{G}'}  F_{i_{\alpha}, \boldsymbol{k}} F^*_{i_{\beta}, \boldsymbol{k}'} \psi^*_{\boldsymbol{k} , n_{\alpha} }(\boldsymbol{G}) \psi_{\boldsymbol{k}' , n_{\beta} }(\boldsymbol{G}')    \mathcal{V}_{\rm cell} \delta_{\boldsymbol{k}   + \boldsymbol{q}    - \boldsymbol{k}'    , \boldsymbol{G}' - \boldsymbol{G} - \boldsymbol{K} }~.
\end{align}
The Kronecker delta allows to eliminate $\boldsymbol{k}'$ as $\boldsymbol{k}' = \boldsymbol{k}   + \boldsymbol{q}    -\boldsymbol{G}' + \boldsymbol{G} + \boldsymbol{K}$. Since the vector $-\boldsymbol{G}' + \boldsymbol{G} + \boldsymbol{K}$ belongs to the reciprocal lattice, we have $F^*_{i_{\beta}, \boldsymbol{k}   + \boldsymbol{q}    -\boldsymbol{G}' + \boldsymbol{G} + \boldsymbol{K}} = F^*_{i_{\beta}, \boldsymbol{k}   + \boldsymbol{q}  }$. We end up with the following (exact) expression:
\begin{align}
M^{(j,s)}_{\alpha, \beta}   &  = \delta_{\tau_{\alpha}, \tau_{\beta}}         \mathcal{V}_{\rm cell}    \sum_{\boldsymbol{q}} F_{\boldsymbol{q}, j}  \sum_{ n}  \sqrt{ \frac{ \hbar }{m_n  } } \,  \boldsymbol{\eta}_{\boldsymbol{q}, s, n}   \cdot  \sum_{\boldsymbol{K}} ( \boldsymbol{q}  + \boldsymbol{K} )  \,  V_{\rm eN}(\boldsymbol{q}  + \boldsymbol{K})    \nonumber \\
& \quad \times e^{  i  ( \boldsymbol{q}  + \boldsymbol{K} ) \cdot      \boldsymbol{B}_n   } \sum_{\boldsymbol{k}  } \sum_{\boldsymbol{G} , \boldsymbol{G}'}  F_{i_{\alpha}, \boldsymbol{k}} F^*_{i_{\beta}, \boldsymbol{k}   + \boldsymbol{q}} \psi^*_{\boldsymbol{k} , n_{\alpha} }(\boldsymbol{G}) \psi_{\boldsymbol{k}   + \boldsymbol{q}    -\boldsymbol{G}' + \boldsymbol{G} + \boldsymbol{K} , n_{\beta} }(\boldsymbol{G}')~.
\label{exact expression for M bare}
\end{align}

We briefly mention two approximations which are often applied to the expression for the EPI matrix elements. The first possible approximation consists in neglecting the {\it Umklapp} processes, i.e.~in keeping only the term with $\boldsymbol{K} = \boldsymbol{0}$---an approximation which relies on the fact that $V_{\rm eN}(\boldsymbol{q}  + \boldsymbol{K})$ decays quickly with increasing $| \boldsymbol{K} |$. The resulting expression depends on the scalar product $\boldsymbol{\eta}_{\boldsymbol{q}, s, n}\cdot \boldsymbol{q}$. Therefore, in this approximation, only longitudinal phonon modes couple with the electrons (see also the discussion in Ref.~\cite{Giustino17}). Another possible approximation, which can be dubbed ``{\it strong tight-binding}" assumption, consists in putting
\begin{align}
\psi_{i_{\alpha}, n_{\alpha}}^*(\boldsymbol{r})         \,     \psi_{i_{\beta}, n_{\beta}}(\boldsymbol{r}) \approx \delta_{i_{\alpha}, i_{\beta}} \delta_{n_{\alpha}, n_{\beta}} \left| \psi_{i_{\alpha}, n_{\alpha}}(\boldsymbol{r})  \right|^2 \approx \delta_{i_{\alpha}, i_{\beta}} \delta_{n_{\alpha}, n_{\beta}} \delta(\boldsymbol{r} - \boldsymbol{R}_{i_{\alpha}} - \boldsymbol{B}_{n_{\alpha}}) 
\end{align} 
in Eq.~\eqref{M before Bloch}. This yields
\begin{align}
M^{(j,s)}_{\alpha, \beta}   &  \approx \delta_{ \alpha ,  \beta }         \sum_{\boldsymbol{q}} F_{\boldsymbol{q}, j - i_{\alpha}}  \sum_{ n}  \sqrt{ \frac{ \hbar }{m_n  } } \,  \boldsymbol{\eta}_{\boldsymbol{q}, s, n}   \cdot  \sum_{\boldsymbol{K}} ( \boldsymbol{q}  + \boldsymbol{K} ) \frac{ V_{\rm eN}(\boldsymbol{q}  + \boldsymbol{K}) }{N}  \nonumber \\
& \quad \times      e^{  i  ( \boldsymbol{q}  + \boldsymbol{K} ) \cdot    \left(  \boldsymbol{B}_n - \boldsymbol{B}_{n_{\alpha}}  \right) }          \equiv \delta_{\alpha, \beta} M^{(j,s)}_{(i_{\alpha},\sigma_{ \alpha })}~. 
\label{local approx for M}
\end{align}

For the purposes of our derivation, we do not need to apply these approximations, and we can just use Eq.~\eqref{exact expression for M bare}. We write it as
\begin{align}
M^{(j,s)}_{\alpha, \beta}   &  \equiv \delta_{\tau_{\alpha}, \tau_{\beta}}             \sum_{\boldsymbol{q}} \sum_{\boldsymbol{k}  }  F_{\boldsymbol{q}, j}  F_{i_{\alpha}, \boldsymbol{k}} F^*_{i_{\beta}, \boldsymbol{k}   + \boldsymbol{q}}  M^{(\boldsymbol{q},s)}_{  \boldsymbol{k} ; n_{\alpha}, n_{\beta}}~,
\label{exact expression for M bare - Fourier}
\end{align}
emphasizing its Fourier representation. The definition of $M^{(\boldsymbol{q},s)}_{  \boldsymbol{k} ; n_{\alpha}, n_{\beta}} $ can be determined by comparison with Eq.~\eqref{exact expression for M bare}; in the case of a single orbital at each lattice site, it bears no dependence on $n_{\alpha}$ and $n_{\beta}$, and we write it as $M^{(\boldsymbol{q},s)}_{  \boldsymbol{k}  }$, as we have done in Section~\ref{sec: Eliashberg}. 

\section{Phonon displacement operator in the position representation}
\label{app: Q}

From Eq.~\eqref{Fourier diagonalization}, it follows that
\begin{align}
\frac{1}{\sqrt{ \omega_{\boldsymbol{q}, s} }} =  \sum_{i,j} F^*_{\boldsymbol{q}, i} \, \left[ {\bm f}^{-1/2}(s) \right]_{i, j} \, F_{ \boldsymbol{q}  , j}  \equiv \sum_{i,j} F^*_{\boldsymbol{q}, i} \, f^{-1/2}_{i,j}(s)  \, F_{ \boldsymbol{q}  , j}~.
\end{align}
Also, using
\begin{align}
& \hat{b}^{\dagger}_{\boldsymbol{q}, s} = \sum_{i'} F_{\boldsymbol{q}, i'} \hat{b}^{\dagger}_{i', s}~, \quad \quad \hat{b}_{-\boldsymbol{q}, s} = \sum_{i'} F_{\boldsymbol{q}, i'} \hat{b}_{i', s}~,
\end{align}
we obtain
\begin{align}
  \hat{Q}_{l, s}  
 & =  \frac{1}{\sqrt{2}} \sum_{\boldsymbol{q}} F_{\boldsymbol{q}, l}   \sum_{i,j,i'} F^*_{\boldsymbol{q}, i} \,   f_{i, j}^{-1/2}(s)   \, F_{ \boldsymbol{q}  , j} F_{\boldsymbol{q}, i'}   \left( \hat{b}^{\dagger}_{i', s} + \hat{b}_{i', s} \right) \nonumber \\
& =  \frac{1}{N }   \sum_{i,j,i'}    f_{i, j}^{-1/2}(s)   \,  \delta_{\boldsymbol{R}_i - \boldsymbol{R}_j ,    \boldsymbol{R}_l + \boldsymbol{R}_{i'}}   \frac{1}{\sqrt{2}} \left( \hat{b}^{\dagger}_{i', s} + \hat{b}_{i', s} \right)~. 
\end{align}
It is convenient to abbreviate $\boldsymbol{R}_i - \boldsymbol{R}_j$ as $i - j$. Since $f_{i, j}(s)  =   f_{i-j}(s)  $, we also have $  f_{i, j}^{-1/2}(s)   = f_{i - j}^{-1/2}(s) $ , and we obtain:
\begin{align}
\hat{Q}_{l, s}  
& =      \sum_{ j}   f_{ l + j  }^{-1/2}(s)  \,   \frac{1}{\sqrt{2}} \left( \hat{b}^{\dagger}_{j, s} + \hat{b}_{j, s} \right)~. 
\end{align}
Using Greek indices, this can be rewritten as in Eq.~\eqref{Q operator}, with $f_{ \lambda, -\kappa }^{-1/2} \equiv \delta_{s_{\lambda}, s_{\kappa}} f_{ i_{\lambda} + i_{\kappa} }^{-1/2}$.

\section{Proof - Non-interacting grand potential as a function of the non-interacting GFs}
\label{app: Proof nonint Omega Tr ln}

In the absence of EEIs and EPIs, the action reduces to
\begin{align}
  A^{(c, \overline{c}, b, b^*)}_{{\bm t}, {\bf 0}, {\bm f}, {\bf 0}}  
 & = \frac{1}{T} \sum_n \sum_{\alpha, \beta} \overline{c}_{\alpha}(i \omega_n) G^{-1}_{{\bm t}, {\bm 0}, {\bm f}, {\bm 0}; \alpha, \beta}(i \omega_n) \, c_{\beta}(i \omega_n)  \nonumber \\
  & \quad + \frac{1}{T}\sum_n \sum_{\kappa, \lambda}  b^*_{\kappa}(i \Omega_n) P^{-1}_{{\bm t}, {\bm 0}, {\bm f}, {\bm 0}; \kappa, \lambda}(i \Omega_n) \, b_{\lambda}(i \Omega_n)~.
\end{align}
The partition function can therefore be evaluated analytically:
\begin{align}
{\cal Z}_{{\bm t}, {\bm 0}, {\bm f}, {\bm 0}} & = \int  D(\overline{c}, c) D(b^*, b) \,   e^{A^{(c, \overline{c}, b, b^*)}_{{\bm t}, {\bf 0}, {\bm f}, {\bf 0}}} \nonumber \\
&  =  \int  D(\overline{c}, c)   \,   
\exp\left[ \frac{1}{T} \sum_n \sum_{\alpha, \beta} \overline{c}_{\alpha}(i \omega_n) G^{-1}_{{\bm t}, {\bm 0}, {\bm f}, {\bm 0}; \alpha, \beta}(i \omega_n) \, c_{\beta}(i \omega_n)  \right] \nonumber \\
& \quad \times  \int    D(b^*, b) \,   
\exp\left[ \frac{1}{T}\sum_n \sum_{\kappa, \lambda}  b^*_{\kappa}(i \Omega_n) P^{-1}_{{\bm t}, {\bm 0}, {\bm f}, {\bm 0}; \kappa, \lambda}(i \Omega_n) \, b_{\lambda}(i \Omega_n)    \right] \nonumber \\
& = \frac{ \prod_{ \omega_n} \det\left[ - \frac{1}{T}{\bm G}^{-1}_{{\bm t}, {\bm 0}, {\bm f}, {\bm 0}}(i \omega_n)   \right]   }{\prod_{ \Omega_n} \det\left[ - \frac{1}{T} {\bm P}^{-1}_{{\bm t}, {\bm 0}, {\bm f}, {\bm 0}}(i \Omega_n)   \right]   }~.
\label{nonint Z}
\end{align}
From Eq.~\eqref{nonint Z}, we evaluate the non-interacting grand potential as
\begin{align}
\Omega_{{\bm t}, {\bm 0}, {\bm f}, {\bm 0}} & = - T \ln {\cal Z}_{{\bm t}, {\bm 0}, {\bm f}, {\bm 0}} \nonumber \\
& =  - T \sum_n \ln      \det\left[ - \frac{1}{T} {\bm G}^{-1}_{{\bm t}, {\bm 0}, {\bm f}, {\bm 0}}(i \omega_n)   \right]    + T \sum_n \ln   \det\left[ -  \frac{1}{T} {\bm P}^{-1}_{{\bm t}, {\bm 0}, {\bm f}, {\bm 0}}(i \Omega_n)   \right] \nonumber \\
&   =    T \sum_n \ln      \det\left[ -  T {\bm G}_{{\bm t}, {\bm 0}, {\bm f}, {\bm 0}}(i \omega_n)   \right]    - T \sum_n \ln   \det\left[ -  T {\bm P}_{{\bm t}, {\bm 0}, {\bm f}, {\bm 0}}(i \Omega_n)   \right]      \nonumber \\
& =  T {\rm Tr} \ln    \left( -  T {\bm G}_{{\bm t}, {\bm 0}, {\bm f}, {\bm 0}}   \right)    - T {\rm Tr} \ln   \left( -   T {\bm P}_{{\bm t}, {\bm 0}, {\bm f}, {\bm 0}}  \right)~,
\label{nonint Omega}
\end{align}
which is Eq.~\eqref{noninteracting Omega Tr ln}. This is used in the main text to derive Eq.~\eqref{nonintPhi}.

\section{Proof - GF functionals as functional derivatives}
\label{app: Proof 0}

In this Appendix we prove Eqs.~\eqref{Sigma derivative} and~\eqref{Lambda derivative}. 

Denoting by $\delta \widetilde{F}_{{\bm U},  {\bm M}}[{\bm \Sigma} ; {\bm \Lambda}] $ the variation of the functional $\widetilde{F}_{{\bm U},  {\bm M}}[{\bm \Sigma} ; {\bm \Lambda}]$ with respect to variations of ${\bm \Sigma}$ and ${\bm \Lambda}$ (while keeping fixed the parameters $\bm U$ and $\bm M$), we have
\begin{align}
&   - \frac{1}{T}  \delta \widetilde{F}_{{\bm U},  {\bm M}}[{\bm \Sigma} ; {\bm \Lambda}]  \nonumber \\  
 & =  - \frac{1}{T}  \delta  \widetilde{\Omega}_{{\bm U},  {\bm M}}[ \widetilde{\bm G}_{{\bm U},  {\bm M}}[{\bm \Sigma} ; {\bm \Lambda}]^{-1}  + {\bm \Sigma} ;  \widetilde{\bm P}_{{\bm U},  {\bm M}}[{\bm \Sigma} ; {\bm \Lambda}]^{-1} + {\bm \Lambda}] \nonumber \\
& \quad +   \delta   \sum_n \sum_{\alpha} e^{i \omega_n 0^+} \left\{ \ln  \left( T \widetilde{\bm G}_{{\bm U},  {\bm M}}[{\bm \Sigma} ; {\bm \Lambda}](i \omega_n) \right) \right\}_{\alpha, \alpha}  \nonumber \\
& \quad 
-   \delta   \sum_n \sum_{\alpha} e^{i \Omega_n 0^+} \left\{ \ln  \left( T \widetilde{\bm P}_{{\bm U},  {\bm M}}[{\bm \Sigma} ; {\bm \Lambda}](i \Omega_n) \right) \right\}_{\alpha, \alpha} \nonumber \\
& =     \sum_{n} \sum_{\alpha, \beta}  \widetilde{G}_{{\bm U},  {\bm M}; \alpha, \beta}[  {\bm \Sigma} ; {\bm \Lambda} ](i \omega_n) \,\, \left\{ \delta \widetilde{  G}^{-1}_{{\bm U},  {\bm M}; \beta, \alpha}[{\bm \Sigma} ; {\bm \Lambda}](i \omega_n)   + \delta { \Sigma}_{\beta, \alpha}(i \omega_n) \right\} \nonumber \\
& \quad - \sum_{n} \sum_{\alpha, \beta}  \widetilde{P}_{{\bm U},  {\bm M}; \alpha, \beta}[  {\bm \Sigma} ; {\bm \Lambda} ](i \Omega_n) \,\, \left\{ \delta \widetilde{  P}^{-1}_{{\bm U},  {\bm M}; \beta, \alpha}[{\bm \Sigma} ; {\bm \Lambda}](i \Omega_n)   + \delta { \Lambda}_{\beta, \alpha}(i \Omega_n) \right\} \nonumber \\
& \quad +      \sum_{n } \sum_{\alpha, \beta }    \widetilde{G}^{-1}_{{\bm U},  {\bm M}; \alpha, \beta}[{\bm \Sigma} ; {\bm \Lambda}](i \omega_n)   \left\{   \delta  \widetilde{  G}_{{\bm U},  {\bm M}; \beta, \alpha}[{\bm \Sigma} ; {\bm \Lambda}](i \omega_n)      \right\}  \nonumber \\
& \quad -     \sum_{n } \sum_{\alpha, \beta }    \widetilde{P}^{-1}_{{\bm U},  {\bm M}; \alpha, \beta}[{\bm \Sigma} ; {\bm \Lambda}](i \Omega_n)   \left\{   \delta  \widetilde{  P}_{{\bm U},  {\bm M}; \beta, \alpha}[{\bm \Sigma} ; {\bm \Lambda}](i \Omega_n)      \right\}  \nonumber \\
& =     \sum_{n} \sum_{\alpha, \beta} \left\{ \widetilde{G}_{{\bm U},  {\bm M}; \alpha, \beta}[  {\bm \Sigma} ; {\bm \Lambda} ](i \omega_n) \,\,   \delta { \Sigma}_{\beta, \alpha}(i \omega_n)   \right. \nonumber \\
& \quad \left.  -  \,  \widetilde{P}_{{\bm U},  {\bm M}; \alpha, \beta}[  {\bm \Sigma} ; {\bm \Lambda} ](i \Omega_n) \,\,   \delta { \Lambda}_{\beta, \alpha}(i \Omega_n)      \right\}~,
\label{proof derivatives}
\end{align}
where we have used that (in condensed notation)
\begin{align}
{\rm Tr} \left[ X^{-1} \cdot (\delta X) + ( \delta X^{-1}) \cdot X \right] = {\rm Tr}~\delta( X^{-1} \cdot X) = {\rm Tr}~\delta (1) = 0~.
\end{align} 
The result in the last line of \eqref{proof derivatives} is equivalent to Eqs~.\eqref{Sigma derivative}-\eqref{Lambda derivative}.

\section{Proof - Self-energy functionals as functional derivatives}
\label{app: Proof 1}

In this Appendix, we prove Eqs.~\eqref{G derivative} and \eqref{P derivative} by using Eq.~\eqref{Definition LWF} together with ~\eqref{Sigma derivative} and~\eqref{Lambda derivative}. 

Denoting by $\delta \widetilde{\Phi}_{{\bm U},  {\bm I}}[{\bm G} ; {\bm P}] $ the variation of the functional $\widetilde{\Phi}_{{\bm U},  {\bm I}}[{\bm G} ; {\bm P}] $ with respect to variations of ${\bm G}$ and ${\bm P}$ (while keeping fixed the parameters $\bm U$ and $\bm I$), we have
\begin{align}
&  \frac{1}{T} \delta \widetilde{\Phi}_{{\bm U},  {\bm I}}[{\bm G} ; {\bm P}] \nonumber \\
& = \!  \sum_n \sum_{\alpha, \beta} \Big\{ \! - G_{\beta, \alpha}(i \omega_n) \, \delta \widetilde{  \Sigma}_{{\bm U},  {\bm I}; \alpha, \beta}[{\bm G} ; {\bm P}](i \omega_n)     +    P_{\beta, \alpha}(i \Omega_n) \, \delta \widetilde{  \Lambda}_{{\bm U},  {\bm I}; \alpha, \beta}[{\bm G} ; {\bm P}](i \Omega_n)   \nonumber \\
& \quad  +   \delta  \Big[   \widetilde{ \Sigma}_{{\bm U},  {\bm I} ; \alpha, \beta}[{\bm G} ; {\bm P}](i \omega_n)  \,  G_{\beta, \alpha}(i \omega_n) \Big]  -   \delta \Big[    \widetilde{  \Lambda}_{{\bm U},  {\bm I}; \alpha, \beta}[{\bm G} ; {\bm P}](i \Omega_n) \,   P_{\beta, \alpha}(i \Omega_n) \Big] \Big\}   \nonumber \\
& =   \!      \sum_n \sum_{\alpha, \beta}  \left\{ \widetilde{ \Sigma}_{{\bm U},  {\bm I} ; \alpha, \beta}[{\bm G} ; {\bm P}](i \omega_n)  \,  \delta G_{\beta, \alpha}(i \omega_n)   -      \widetilde{  \Lambda}_{{\bm U},  {\bm I}; \alpha, \beta}[{\bm G} ; {\bm P}](i \Omega_n) \, \delta  P_{\beta, \alpha}(i \Omega_n)  \right\}  ,
\end{align}
which leads directly to Eqs.~\eqref{G derivative} and \eqref{P derivative}.

\section{Simplification of the functional derivative of the four-leg reducible vertex}
\label{app: simplification}

By applying the functional derivative to Eq.~\eqref{Bethe-Salpeter}, we obtain
\begin{align}
&      \widetilde{\Gamma}^{[6]}_{{\bm U}; (\mu  , \nu ; \phi, \theta ; \nu', \mu')}[{\bm G}](i \omega_m, i \omega_n, i \omega_s )    \nonumber \\
& =  \widetilde{U}^{[6]}_{{\bm U}; (\mu  , \nu ; \phi, \theta; \nu', \mu')}[{\bm G}](i \omega_m, i \omega_n, i \omega_s )  
+ \delta_{n,m} \sum_{ \xi',   \eta'}   \widetilde{U}^{[4]}_{{\bm U}; (\mu  , \theta ; \xi', \mu')}[{\bm G}](i \omega_n, i \omega_s )  \nonumber \\
& \quad \times  G_{\eta', \xi'}(i \omega_s) \,  \widetilde{\Gamma}_{{\bm U}; (\phi  , \nu ; \nu', \eta')}[{\bm G}](i \omega_n, i \omega_s )  \nonumber \\
& \quad  + \delta_{n, s} \sum_{\xi,  \eta } \widetilde{U}^{[4]}_{{\bm U}; (\mu  , \xi ; \phi, \mu')}[{\bm G}](i \omega_m, i \omega_n ) \, G_{\xi, \eta}(i \omega_m) \,    \widetilde{\Gamma}_{{\bm U}; (\eta  , \nu ; \nu', \theta)}[{\bm G}](i \omega_m, i \omega_n ) \nonumber \\
& \quad  + \sum_{\xi, \xi', \eta, \eta'}  
 \widetilde{U}^{[6]}_{{\bm U}; (\mu, \xi ; \phi, \theta; \xi', \mu')}[{\bm G}](i \omega_m, i \omega_n, i \omega_s )  \, G_{\xi, \eta}(i \omega_m) \, G_{\eta', \xi'}(i \omega_s) \nonumber \\
 & \quad \times  \widetilde{\Gamma}_{{\bm U}; (\eta  , \nu ; \nu', \eta')}[{\bm G}](i \omega_m, i \omega_s ) \nonumber \\ 
& \quad  + \sum_{\xi, \xi', \eta, \eta'} \widetilde{U}^{[4]}_{{\bm U}; (\mu  , \xi ; \xi', \mu')}[{\bm G}](i \omega_m, i \omega_s ) \, G_{\xi, \eta}(i \omega_m) \, G_{\eta', \xi'}(i \omega_s) \nonumber \\
& \quad \times   \widetilde{\Gamma}^{[6]}_{{\bm U}; (\eta  , \nu ; \phi, \theta; \nu', \eta')}[{\bm G}](i \omega_m, i \omega_n, i \omega_s )~,
\label{derivative of Bethe-Salpeter}
\end{align}
where we have introduced the functional derivative of the irreducible four-leg vertex as in Eq.~\eqref{irreducible hexagon}.

We proceed to simplify Eq.~\eqref{derivative of Bethe-Salpeter}, with the aim of expressing its right-hand side in terms of the irreducible functionals $U^{[4]}$ and $U^{[6]}$, and the reducible functional $\Gamma$, analogously to what was done in Ref.~\cite{Anokhin96} (i.e.~we eliminate the reducible six-leg vertex $\Gamma^{[6]}$). It should be noted that Ref.~\cite{Anokhin96} considered quenched disorder, while here we are considering EEIs in clean systems as the source of the electronic self-energy in the absence of EPI. Because of this difference, although our final expression is similar to their Eq.~(40), some important details are different. Here, in particular, the six-leg vertices depend on three distinct fermionic frequencies, while, in their case, two of the three frequencies coincide, in all vertices. This produces some topologically distinct structures in the final expressions.

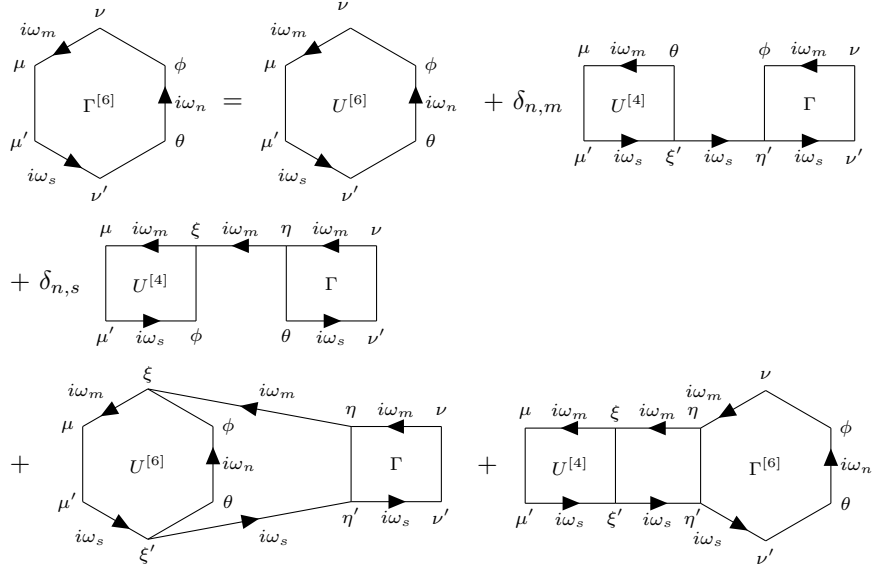
\begin{figure}
\begin{tikzpicture}
\begin{feynman}
\vertex (a) at(0,0);
\vertex (A) at(-0.2,0){\( \scriptstyle \mu' \)};
\vertex (b) at(0,1);
\vertex (B) at(-0.2,1){\( \scriptstyle \mu \)};
\vertex (c) at(0.866,1.5);
\vertex (C) at(0.866,1.7){\( \scriptstyle \nu \)};
\vertex (d) at(1.732,1);
\vertex (D) at(1.932,1){\( \scriptstyle \phi \)};
\vertex (e) at(1.732,0);
\vertex (E) at(1.932,0){\( \scriptstyle \theta \)};
\vertex (f) at(0.866,-0.5);
\vertex (F) at(0.866,-0.7){\( \scriptstyle \nu' \)};
\vertex (name1) at(0.866,0.5){\( \scriptstyle \Gamma^{[6]} \)};
\vertex (eq) at(2.632,0.5){\( = \)};
\vertex (g) at(3.332,0);
\vertex (G) at(3.132,0){\( \scriptstyle \mu' \)};
\vertex (h) at(3.332,1);
\vertex (H) at(3.132,1){\( \scriptstyle \mu \)};
\vertex (i) at(4.198,1.5);
\vertex (I) at(4.098,1.8){\( \scriptstyle \nu \)};
\vertex (j) at(5.064,1);
\vertex (J) at(5.264,1){\( \scriptstyle \phi \)};
\vertex (k) at(5.064,0);
\vertex (K) at(5.264,0){\( \scriptstyle \theta \)};
\vertex (l) at(4.198,-0.5);
\vertex (L) at(4.198,-0.7){\( \scriptstyle \nu' \)};
\vertex (name2) at(4.198,0.5){\( \scriptstyle U^{[6]} \)};
\vertex (plus) at(6.5,0.5){\( + \,\, \delta_{n,m} \)};
\vertex (m) at(7.3,0);
\vertex (M) at(7.3,-0.2){\( \scriptstyle \mu' \)};
\vertex (n) at(8.5,0);
\vertex (N) at(8.5,-0.2){\( \scriptstyle \xi' \)};
\vertex (o) at(8.5,1);
\vertex (O) at(8.5,1.2){\( \scriptstyle \theta \)};
\vertex (p) at(7.3,1);
\vertex (P) at(7.3,1.2){\( \scriptstyle \mu \)};
\vertex (name3) at(7.9,0.5){\( \scriptstyle U^{[4]} \)};
\vertex (q) at(9.7,0);
\vertex (Q) at(9.7,-0.2){\( \scriptstyle \eta' \)};
\vertex (r) at(10.9,0);
\vertex (R) at(10.9,-0.2){\( \scriptstyle \nu' \)};
\vertex (s) at(10.9,1);
\vertex (S) at(10.9,1.2){\( \scriptstyle \nu \)};
\vertex (t) at(9.7,1);
\vertex (T) at(9.7,1.2){\( \scriptstyle \phi \)};
\vertex (name4) at(10.3,0.5){\( \scriptstyle \Gamma \)};
\diagram* {
(a) -- (b),
(b) -- [anti fermion, edge label=\(\scriptstyle i\omega_m\)] (c),
(c) -- (d),
(d) -- [anti fermion, edge label=\(\scriptstyle i\omega_n\)] (e),
(e) -- (f),
(f) -- [anti fermion, edge label=\(\scriptstyle i\omega_s\)] (a),
(g) -- (h),
(h) -- [anti fermion, edge label=\(\scriptstyle i\omega_m\)] (i),
(i) -- (j),
(j) -- [anti fermion, edge label=\(\scriptstyle i\omega_n\)] (k),
(k) -- (l),
(l) -- [anti fermion, edge label=\(\scriptstyle i\omega_s\)] (g),
(m) -- [fermion, edge label'=\(\scriptstyle i\omega_s\)] (n),
(n) -- (o),
(o) -- [fermion, edge label'=\(\scriptstyle i\omega_m\)] (p),
(p) -- (m),
(q) -- [fermion, edge label'=\(\scriptstyle i\omega_s\)] (r),
(r) -- (s),
(s) -- [fermion, edge label'=\(\scriptstyle i\omega_m\)] (t),
(t) -- (q),
(n) -- [fermion, edge label'=\(\scriptstyle i\omega_s\)] (q),
};
\end{feynman}
\end{tikzpicture}

\begin{tikzpicture}
\begin{feynman}
\vertex (plus) at(4,0.5){\( + \,\, \delta_{n,s} \)};
\vertex (m) at(4.8,0);
\vertex (M) at(4.8,-0.2){\( \scriptstyle \mu' \)};
\vertex (n) at(6,0);
\vertex (N) at(6,-0.2){\( \scriptstyle \phi \)};
\vertex (o) at(6,1);
\vertex (O) at(6,1.2){\( \scriptstyle \xi \)};
\vertex (p) at(4.8,1);
\vertex (P) at(4.8,1.2){\( \scriptstyle \mu \)};
\vertex (name3) at(5.4,0.5){\( \scriptstyle U^{[4]} \)};
\vertex (q) at(7.2,0);
\vertex (Q) at(7.2,-0.2){\( \scriptstyle \theta \)};
\vertex (r) at(8.4,0);
\vertex (R) at(8.4,-0.2){\( \scriptstyle \nu' \)};
\vertex (s) at(8.4,1);
\vertex (S) at(8.4,1.2){\( \scriptstyle \nu \)};
\vertex (t) at(7.2,1);
\vertex (T) at(7.2,1.2){\( \scriptstyle \eta \)};
\vertex (name4) at(7.8,0.5){\( \scriptstyle \Gamma \)};
\diagram* {
(m) -- [fermion, edge label'=\( \scriptstyle i\omega_s\)] (n),
(n) -- (o),
(o) -- [fermion, edge label'=\(\scriptstyle i\omega_m\)] (p),
(p) -- (m),
(q) -- [fermion, edge label'=\(\scriptstyle i\omega_s\)] (r),
(r) -- (s),
(s) -- [fermion, edge label'=\(\scriptstyle i\omega_m\)] (t),
(t) -- (q),
(t) -- [fermion, edge label'=\(\scriptstyle i\omega_m\)] (o),
};
\end{feynman}
\end{tikzpicture}

\begin{tikzpicture}
\begin{feynman}
\vertex (plus) at(4,0.5){\( +  \)};
\vertex (g) at(4.832,0);
\vertex (G) at(4.632,0){\( \scriptstyle \mu' \)};
\vertex (h) at(4.832,1);
\vertex (H) at(4.632,1){\( \scriptstyle \mu \)};
\vertex (i) at(5.698,1.5);
\vertex (I) at(5.698,1.7){\( \scriptstyle \xi \)};
\vertex (j) at(6.564,1);
\vertex (J) at(6.764,1){\( \scriptstyle \phi \)};
\vertex (k) at(6.564,0);
\vertex (K) at(6.764,0){\( \scriptstyle \theta \)};
\vertex (l) at(5.698,-0.5);
\vertex (L) at(5.698,-0.7){\( \scriptstyle \xi' \)};
\vertex (name2) at(5.698,0.5){\( \scriptstyle U^{[6]} \)};
\vertex (q) at(8.4,0);
\vertex (Q) at(8.4,-0.2){\( \scriptstyle \eta' \)};
\vertex (r) at(9.6,0);
\vertex (R) at(9.6,-0.2){\( \scriptstyle \nu' \)};
\vertex (s) at(9.6,1);
\vertex (S) at(9.6,1.2){\( \scriptstyle \nu \)};
\vertex (t) at(8.4,1);
\vertex (T) at(8.4,1.2){\( \scriptstyle \eta \)};
\vertex (name4) at(9,0.5){\( \scriptstyle \Gamma \)};
\diagram* {
(g) -- (h)  -- [anti fermion, edge label=\(\scriptstyle i\omega_m\)] (i)  -- (j),
(j) -- [anti fermion, edge label=\(\scriptstyle i\omega_n\)] (k)  -- (l)  -- [anti fermion, edge label=\(\scriptstyle i\omega_s\)] (g),
(q) -- [fermion, edge label'=\(\scriptstyle i\omega_s\)] (r)  -- (s)  -- [fermion, edge label'=\(\scriptstyle i\omega_m\)] (t)  -- (q),
(l) -- [fermion, edge label'=\(\scriptstyle i\omega_s\)] (q),
(t) -- [fermion, edge label'=\(\scriptstyle i\omega_m\)] (i),
};
\end{feynman}
\end{tikzpicture}
\begin{tikzpicture}
\begin{feynman}
\vertex (plus) at(4,0.5){\( +  \)};
\vertex (g) at(6.832,0);
\vertex (G) at(6.732,-0.2){\( \scriptstyle \eta' \)};
\vertex (h) at(6.832,1);
\vertex (H) at(6.732,1.2){\( \scriptstyle \eta \)};
\vertex (i) at(7.698,1.5);
\vertex (I) at(7.698,1.7){\( \scriptstyle \nu \)};
\vertex (j) at(8.564,1);
\vertex (J) at(8.764,1){\( \scriptstyle \phi \)};
\vertex (k) at(8.564,0);
\vertex (K) at(8.764,0){\( \scriptstyle \theta \)};
\vertex (l) at(7.698,-0.5);
\vertex (L) at(7.698,-0.7){\( \scriptstyle \nu' \)};
\vertex (name2) at(7.698,0.5){\( \scriptstyle \Gamma^{[6]} \)};
\vertex (q) at(4.5,0);
\vertex (Q) at(4.5,-0.2){\( \scriptstyle \mu' \)};
\vertex (r) at(5.7,0);
\vertex (R) at(5.7,-0.2){\( \scriptstyle \xi' \)};
\vertex (s) at(5.7,1);
\vertex (S) at(5.7,1.2){\( \scriptstyle \xi \)};
\vertex (t) at(4.5,1);
\vertex (T) at(4.5,1.2){\( \scriptstyle \mu \)};
\vertex (name4) at(5.1,0.5){\( \scriptstyle U^{[4]} \)};
\diagram* {
(g) -- (h)  -- [anti fermion, edge label=\(\scriptstyle i\omega_m\)] (i)  -- (j),
(j) -- [anti fermion, edge label=\(\scriptstyle i\omega_n\)] (k)  -- (l)  -- [anti fermion, edge label=\(\scriptstyle i\omega_s\)] (g),
(q) -- [fermion, edge label'=\(\scriptstyle i\omega_s\)] (r),
(r) -- (s),
(s) -- [fermion, edge label'=\(\scriptstyle i\omega_m\)] (t),
(t) -- (q),
(r) -- [fermion, edge label'=\(\scriptstyle i\omega_s\)] (g),
(h) -- [fermion, edge label'=\(\scriptstyle i\omega_m\)] (s),
};
\end{feynman}
\end{tikzpicture}
\caption{Diagrammatic representation of Eq.~\eqref{derivative of Bethe-Salpeter}.}
\label{diagram passage 0}
\end{figure}

We now go through the derivation by employing Feynman diagrams. First, we represent Eq.~\eqref{derivative of Bethe-Salpeter} by means of Feynman diagrams, obtaining Fig.~\ref{diagram passage 0}, where we explicitly label the internal vertices as well as the external vertices (it is intended that internal indices are summed over). In the following steps, we will omit the dummy labels of the internal vertices. Moreover, to avoid cluttering the figures with redundant information, we will indicate only one frequency argument for each fermionic oriented line, with the understanding that it applies to all the connected segments that form that line. 

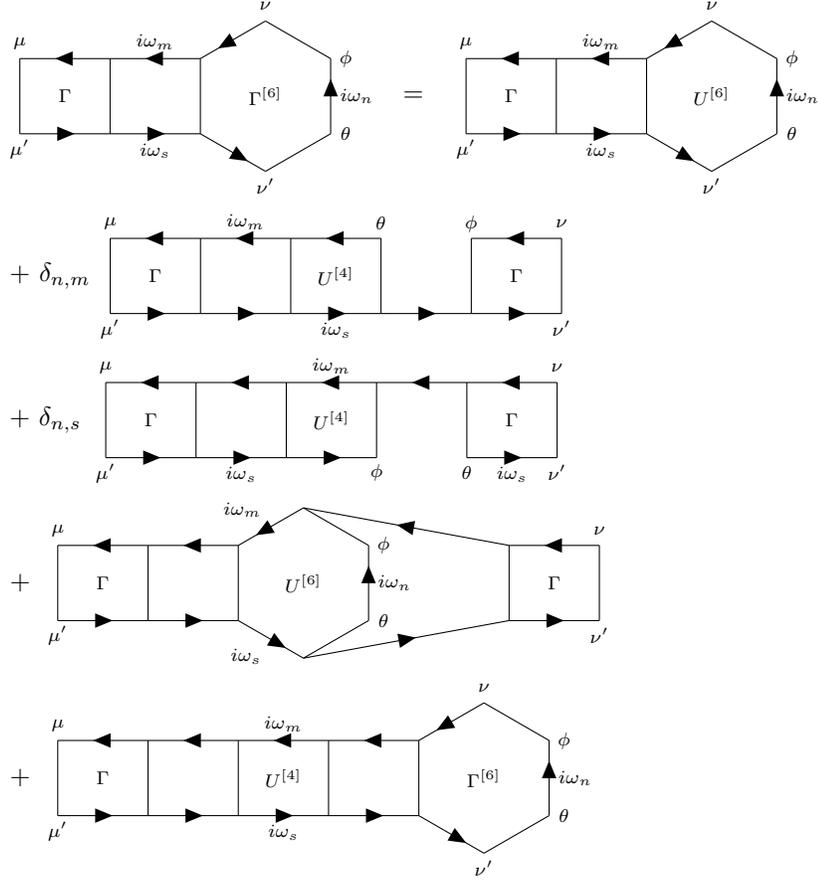
\begin{figure}
\begin{tikzpicture}
\begin{feynman}
\vertex (q) at(0,0);
\vertex (Q) at(0,-0.2){\( \scriptstyle \mu' \)};
\vertex (r) at(1.2,0); 
\vertex (s) at(1.2,1); 
\vertex (t) at(0,1);
\vertex (T) at(0,1.2){\( \scriptstyle \mu \)};
\vertex (name4) at(0.6,0.5){\( \scriptstyle \Gamma \)};
\vertex (a) at(2.4,0); 
\vertex (b) at(2.4,1); 
\vertex (c) at(3.266,1.5);
\vertex (C) at(3.266,1.7){\( \scriptstyle \nu \)};
\vertex (d) at(4.132,1);
\vertex (D) at(4.332,1){\( \scriptstyle \phi \)};
\vertex (e) at(4.132,0);
\vertex (E) at(4.332,0){\( \scriptstyle \theta \)};
\vertex (f) at(3.266,-0.5);
\vertex (F) at(3.266,-0.7){\( \scriptstyle \nu' \)};
\vertex (name1) at(3.266,0.5){\( \scriptstyle \Gamma^{[6]} \)};
\vertex (eq) at(5.232,0.5){\( = \)};
\vertex (g) at(8.332,0); 
\vertex (h) at(8.332,1); 
\vertex (i) at(9.198,1.5);
\vertex (I) at(9.198,1.7){\( \scriptstyle \nu \)};
\vertex (j) at(10.064,1);
\vertex (J) at(10.264,1){\( \scriptstyle \phi \)};
\vertex (k) at(10.064,0);
\vertex (K) at(10.264,0){\( \scriptstyle \theta \)};
\vertex (l) at(9.198,-0.5);
\vertex (L) at(9.198,-0.7){\( \scriptstyle \nu' \)};
\vertex (name2) at(9.198,0.5){\( \scriptstyle U^{[6]} \)}; 
\vertex (m) at(5.932,0);
\vertex (M) at(5.932,-0.2){\( \scriptstyle \mu' \)};
\vertex (n) at(7.132,0); 
\vertex (o) at(7.132,1); 
\vertex (p) at(5.932,1);
\vertex (P) at(5.932,1.2){\( \scriptstyle \mu \)};
\vertex (name3) at(6.532,0.5){\( \scriptstyle \Gamma \)};
\diagram* {
(a) -- (b),
(b) -- [anti fermion] (c),
(c) -- (d),
(d) -- [anti fermion, edge label=\(\scriptstyle i\omega_n\)] (e),
(e) -- (f),
(f) -- [anti fermion] (a),
(g) -- (h),
(h) -- [anti fermion] (i),
(i) -- (j),
(j) -- [anti fermion, edge label=\(\scriptstyle i\omega_n\)] (k),
(k) -- (l),
(l) -- [anti fermion] (g),
(m) -- [fermion] (n)  -- (o)  -- [fermion] (p)  -- (m),
(q) -- [fermion] (r)  -- (s)  -- [fermion] (t)  -- (q),
(r) -- [fermion, edge label'=\(\scriptstyle i\omega_s\)] (a),
(b) -- [fermion, edge label'=\(\scriptstyle i\omega_m\)] (s),
(n) -- [fermion, edge label'=\(\scriptstyle i\omega_s\)] (g),
(h) -- [fermion, edge label'=\(\scriptstyle i\omega_m\)] (o),
};
\end{feynman}
\end{tikzpicture} 

\begin{tikzpicture}
\begin{feynman}
\vertex (plus) at(4,0.5){\( + \,\, \delta_{n,m} \)};
\vertex (g) at(4.8,0);
\vertex (G) at(4.8,-0.2){\( \scriptstyle \mu' \)};
\vertex (h) at(6,0); 
\vertex (i) at(6,1); 
\vertex (j) at(4.8,1);
\vertex (J) at(4.8,1.2){\( \scriptstyle \mu \)};
\vertex (name1) at(5.4,0.5){\( \scriptstyle \Gamma \)};
\vertex (m) at(7.2,0); 
\vertex (n) at(8.4,0); 
\vertex (o) at(8.4,1); 
\vertex (O) at(8.4,1.2){\( \scriptstyle \theta \)};
\vertex (p) at(7.2,1); 
\vertex (name2) at(7.8,0.5){\( \scriptstyle U^{[4]} \)};
\vertex (q) at(9.6,0);
\vertex (r) at(10.8,0);
\vertex (R) at(10.8,-0.2){\( \scriptstyle \nu' \)};
\vertex (s) at(10.8,1);
\vertex (S) at(10.8,1.2){\( \scriptstyle \nu \)};
\vertex (t) at(9.6,1); 
\vertex (T) at(9.6,1.2){\( \scriptstyle \phi \)};
\vertex (name3) at(10.2,0.5){\( \scriptstyle \Gamma \)};
\diagram* { 
(g) -- [fermion] (h)  -- (i)  -- [fermion] (j)  -- (g),
(m) -- [fermion, edge label'=\(\scriptstyle i\omega_s\)] (n)  -- (o)  -- [fermion] (p)  -- (m),
(q) -- [fermion] (r)  -- (s)  -- [fermion] (t)  -- (q),
(n) -- [fermion] (q),
(h) -- [fermion] (m), 
(p) -- [fermion, edge label'=\(\scriptstyle i\omega_m\)] (i),
};
\end{feynman}
\end{tikzpicture}

\begin{tikzpicture}
\begin{feynman}
\vertex (plus) at(4,0.5){\( + \,\, \delta_{n,s} \)};
\vertex (g) at(4.8,0);
\vertex (G) at(4.8,-0.2){\( \scriptstyle \mu' \)};
\vertex (h) at(6,0); 
\vertex (i) at(6,1); 
\vertex (j) at(4.8,1);
\vertex (J) at(4.8,1.2){\( \scriptstyle \mu \)};
\vertex (name1) at(5.4,0.5){\( \scriptstyle \Gamma \)};
\vertex (m) at(7.2,0); 
\vertex (n) at(8.4,0);
\vertex (N) at(8.4,-0.2){\( \scriptstyle \phi \)};
\vertex (o) at(8.4,1); 
\vertex (p) at(7.2,1); 
\vertex (name2) at(7.8,0.5){\( \scriptstyle U^{[4]} \)};
\vertex (q) at(9.6,0);
\vertex (Q) at(9.6,-0.2){\( \scriptstyle \theta \)};
\vertex (r) at(10.8,0);
\vertex (R) at(10.8,-0.2){\( \scriptstyle \nu' \)};
\vertex (s) at(10.8,1);
\vertex (S) at(10.8,1.2){\( \scriptstyle \nu \)};
\vertex (t) at(9.6,1); 
\vertex (name3) at(10.2,0.5){\( \scriptstyle \Gamma \)};
\diagram* { 
(g) -- [fermion] (h)  -- (i)  -- [fermion] (j)  -- (g),
(m) -- [fermion] (n)  -- (o)  -- [fermion, edge label'=\(\scriptstyle i\omega_m\)] (p)  -- (m),
(q) -- [fermion, edge label'=\(\scriptstyle i\omega_s\)] (r)  -- (s)  -- [fermion] (t)  -- (q),
(t) -- [fermion] (o),
(h) -- [fermion, edge label'=\(\scriptstyle i\omega_s\)] (m), 
(p) -- [fermion] (i),
};
\end{feynman}
\end{tikzpicture}

\begin{tikzpicture}
\begin{feynman}
\vertex (plus) at(0,0.5){\( +  \)};
\vertex (a) at(0.5,0);
\vertex (A) at(0.5,-0.2){\( \scriptstyle \mu' \)};
\vertex (b) at(1.7,0); 
\vertex (c) at(1.7,1); 
\vertex (d) at(0.5,1);
\vertex (D) at(0.5,1.2){\( \scriptstyle \mu \)};
\vertex (name1) at(1.1,0.5){\( \scriptstyle \Gamma \)};
\vertex (g) at(2.9,0); 
\vertex (h) at(2.9,1); 
\vertex (i) at(3.766,1.5); 
\vertex (j) at(4.632,1);
\vertex (J) at(4.832,1){\( \scriptstyle \phi \)};
\vertex (k) at(4.632,0);
\vertex (K) at(4.832,0){\( \scriptstyle \theta \)};
\vertex (l) at(3.766,-0.5); 
\vertex (name2) at(3.766,0.5){\( \scriptstyle U^{[6]} \)};
\vertex (q) at(6.5,0); 
\vertex (r) at(7.7,0);
\vertex (R) at(7.7,-0.2){\( \scriptstyle \nu' \)};
\vertex (s) at(7.7,1);
\vertex (S) at(7.7,1.2){\( \scriptstyle \nu \)};
\vertex (t) at(6.5,1); 
\vertex (name3) at(7.1,0.5){\( \scriptstyle \Gamma \)};
\diagram* { 
(a) -- [fermion] (b)  -- (c)  -- [fermion] (d)  -- (a),
(g) -- (h)  -- [anti fermion, edge label=\(\scriptstyle i\omega_m\)] (i)  -- (j),
(j) -- [anti fermion, edge label=\(\scriptstyle i\omega_n\)] (k)  -- (l)  -- [anti fermion, edge label=\(\scriptstyle i\omega_s\)] (g),
(q) -- [fermion] (r)  -- (s)  -- [fermion] (t)  -- (q),
(l) -- [fermion] (q),
(t) -- [fermion] (i),
(b) -- [fermion] (g),
(h) -- [fermion] (c),
};
\end{feynman}
\end{tikzpicture}

\begin{tikzpicture}
\begin{feynman}
\vertex (plus) at(0,0.5){\( +  \)};
\vertex (g) at(5.3,0); 
\vertex (h) at(5.3,1); 
\vertex (i) at(6.166,1.5);
\vertex (I) at(6.166,1.7){\( \scriptstyle \nu \)};
\vertex (j) at(7.032,1);
\vertex (J) at(7.232,1){\( \scriptstyle \phi \)};
\vertex (k) at(7.032,0);
\vertex (K) at(7.232,0){\( \scriptstyle \theta \)};
\vertex (l) at(6.166,-0.5);
\vertex (L) at(6.166,-0.7){\( \scriptstyle \nu' \)};
\vertex (name3) at(6.166,0.5){\( \scriptstyle \Gamma^{[6]} \)};
\vertex (q) at(0.5,0);
\vertex (Q) at(0.5,-0.2){\( \scriptstyle \mu' \)};
\vertex (r) at(1.7,0); 
\vertex (s) at(1.7,1); 
\vertex (t) at(0.5,1);
\vertex (T) at(0.5,1.2){\( \scriptstyle \mu \)};
\vertex (name1) at(1.1,0.5){\( \scriptstyle \Gamma \)};
\vertex (a) at(2.9,0); 
\vertex (b) at(4.1,0); 
\vertex (c) at(4.1,1); 
\vertex (d) at(2.9,1); 
\vertex (name2) at(3.5,0.5){\( \scriptstyle U^{[4]} \)};
\diagram*{
(g) -- (h),
(h) -- [anti fermion] (i),
(i) -- (j),
(j) -- [anti fermion, edge label=\(\scriptstyle i\omega_n\)] (k),
(k) -- (l),
(l) -- [anti fermion] (g),
(q) -- [fermion] (r)  -- (s)  -- [fermion] (t)  -- (q),
(a) -- [fermion, edge label'=\(\scriptstyle i\omega_s\)] (b)  -- (c)  -- [fermion, edge label'=\(\scriptstyle i\omega_m\)] (d)  -- (a),
(r) -- [fermion] (a),
(d) -- [fermion] (s),
(b) -- [fermion] (g),
(h) -- [fermion] (c),
}; 
\end{feynman}
\end{tikzpicture}
\caption{First step of the derivation that leads from Eq.~\eqref{derivative of Bethe-Salpeter} to Eq.~\eqref{derivative of Bethe-Salpeter simplified}.
\label{diagram passage 1}}
\end{figure}

The first step of the derivation consists in applying a $\Gamma$ diagram from the left, by attaching it (via two GFs) to the $\mu$ and $\mu'$ vertices in Fig.~\ref{diagram passage 0}. After renaming the vertices so that $\mu$ and $\mu'$ remain the labels of the external vertices on the left of the diagrams, we obtain Fig.~\ref{diagram passage 1}.

\begin{figure}
\begin{tikzpicture}
\begin{feynman}
\vertex (a) at(0,0);
\vertex (A) at(0,-0.2){\( \scriptstyle \mu' \)};
\vertex (b) at(1.2,0);
\vertex (B) at(1.2,-0.2){\( \scriptstyle \nu' \)};
\vertex (c) at(1.2,1);
\vertex (C) at(1.2,1.2){\( \scriptstyle \nu \)};
\vertex (d) at(0,1);
\vertex (D) at(0,1.2){\( \scriptstyle \mu \)};
\vertex (name1) at(0.6,0.5){\( \scriptstyle \Gamma \)};
\vertex (equal) at(1.8,0.5){\( = \)};
\vertex (e) at(2.4,0);
\vertex (E) at(2.4,-0.2){\( \scriptstyle \mu' \)};
\vertex (f) at(3.6,0);
\vertex (F) at(3.6,-0.2){\( \scriptstyle \nu' \)};
\vertex (g) at(3.6,1);
\vertex (G) at(3.6,1.2){\( \scriptstyle \nu \)};
\vertex (h) at(2.4,1);
\vertex (H) at(2.4,1.2){\( \scriptstyle \mu \)};
\vertex (name2) at(3,0.5){\( \scriptstyle U^{[4]} \)};
\vertex (plus) at(4.2,0.5){\( + \)};
\vertex (i) at(4.8,0);
\vertex (I) at(4.8,-0.2){\( \scriptstyle \mu' \)};
\vertex (j) at(6,0);
\vertex (J) at(6,-0.3);
\vertex (k) at(6,1);
\vertex (K) at(6,1.3);
\vertex (l) at(4.8,1);
\vertex (L) at(4.8,1.2){\( \scriptstyle \mu \)};
\vertex (name3) at(5.4,0.5){\( \scriptstyle \Gamma \)};
\vertex (m) at(7.2,0);
\vertex (M) at(7.2,-0.3);
\vertex (n) at(8.4,0);
\vertex (N) at(8.4,-0.2){\( \scriptstyle \nu' \)};
\vertex (o) at(8.4,1);
\vertex (O) at(8.4,1.2){\( \scriptstyle \nu \)};
\vertex (p) at(7.2,1);
\vertex (P) at(7.2,1.3);
\vertex (name4) at(7.8,0.5){\( \scriptstyle U^{[4]} \)};
\diagram* {
(a)[dot] -- [fermion, edge label'=\(\scriptstyle i\omega_s\)] (b)[dot]  -- (c)  -- [fermion, edge label'=\(\scriptstyle i\omega_m\)] (d)  -- (a),
(e) -- [fermion, edge label'=\(\scriptstyle i\omega_s\)] (f)  -- (g)  -- [fermion, edge label'=\(\scriptstyle i\omega_m\)] (h)  -- (e),
(i) -- [fermion] (j)  -- (k)  -- [fermion] (l)  -- (i),
(m) -- [fermion] (n)  -- (o)  -- [fermion] (p)  -- (m),
(p) -- [fermion, edge label'=\(\scriptstyle i\omega_m\)] (k),
(j) -- [fermion, edge label'=\(\scriptstyle i\omega_s\)] (m),
};
\end{feynman}
\end{tikzpicture}
\caption{Diagrammatic representation of the Bethe-Salpeter equation, Eq.~\eqref{Bethe-Salpeter}. The connected blocks $\Gamma$ and $U^{[4]}$ on the right-hand side can be interchanged.}
\label{BS figure}
\end{figure}
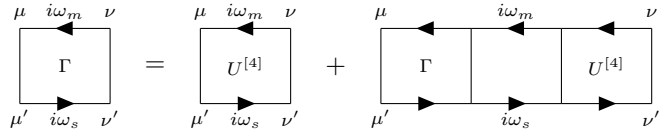

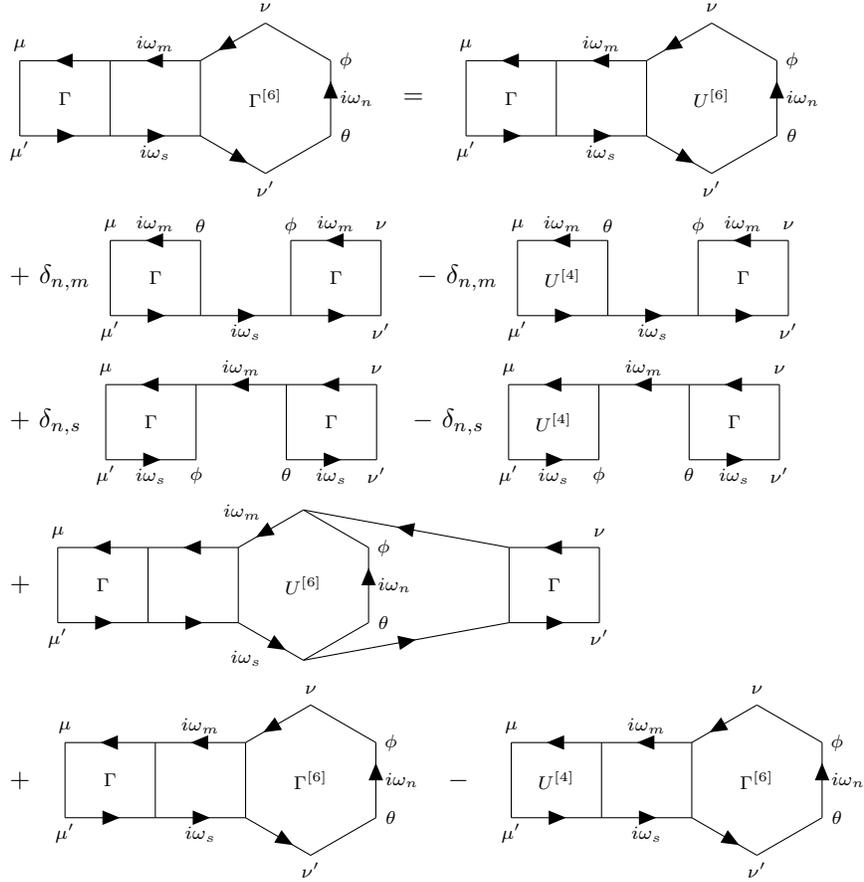
\begin{figure}
\begin{tikzpicture}
\begin{feynman}
\vertex (q) at(0,0);
\vertex (Q) at(0,-0.2){\( \scriptstyle \mu' \)};
\vertex (r) at(1.2,0); 
\vertex (s) at(1.2,1); 
\vertex (t) at(0,1);
\vertex (T) at(0,1.2){\( \scriptstyle \mu \)};
\vertex (name4) at(0.6,0.5){\( \scriptstyle \Gamma \)};
\vertex (a) at(2.4,0); 
\vertex (b) at(2.4,1); 
\vertex (c) at(3.266,1.5);
\vertex (C) at(3.266,1.7){\( \scriptstyle \nu \)};
\vertex (d) at(4.132,1);
\vertex (D) at(4.332,1){\( \scriptstyle \phi \)};
\vertex (e) at(4.132,0);
\vertex (E) at(4.332,0){\( \scriptstyle \theta \)};
\vertex (f) at(3.266,-0.5);
\vertex (F) at(3.266,-0.7){\( \scriptstyle \nu' \)};
\vertex (name1) at(3.266,0.5){\( \scriptstyle \Gamma^{[6]} \)};
\vertex (eq) at(5.232,0.5){\( = \)};
\vertex (g) at(8.332,0); 
\vertex (h) at(8.332,1); 
\vertex (i) at(9.198,1.5);
\vertex (I) at(9.198,1.7){\( \scriptstyle \nu \)};
\vertex (j) at(10.064,1);
\vertex (J) at(10.264,1){\( \scriptstyle \phi \)};
\vertex (k) at(10.064,0);
\vertex (K) at(10.264,0){\( \scriptstyle \theta \)};
\vertex (l) at(9.198,-0.5);
\vertex (L) at(9.198,-0.7){\( \scriptstyle \nu' \)};
\vertex (name2) at(9.198,0.5){\( \scriptstyle U^{[6]} \)}; 
\vertex (m) at(5.932,0);
\vertex (M) at(5.932,-0.2){\( \scriptstyle \mu' \)};
\vertex (n) at(7.132,0); 
\vertex (o) at(7.132,1); 
\vertex (p) at(5.932,1);
\vertex (P) at(5.932,1.2){\( \scriptstyle \mu \)};
\vertex (name3) at(6.532,0.5){\( \scriptstyle \Gamma \)};
\diagram* {
(a) -- (b),
(b) -- [anti fermion] (c),
(c) -- (d),
(d) -- [anti fermion, edge label=\(\scriptstyle i\omega_n\)] (e),
(e) -- (f),
(f) -- [anti fermion] (a),
(g) -- (h),
(h) -- [anti fermion] (i),
(i) -- (j),
(j) -- [anti fermion, edge label=\(\scriptstyle i\omega_n\)] (k),
(k) -- (l),
(l) -- [anti fermion] (g),
(m) -- [fermion] (n)  -- (o)  -- [fermion] (p)  -- (m),
(q) -- [fermion] (r)  -- (s)  -- [fermion] (t)  -- (q),
(r) -- [fermion, edge label'=\(\scriptstyle i\omega_s\)] (a),
(b) -- [fermion, edge label'=\(\scriptstyle i\omega_m\)] (s),
(n) -- [fermion, edge label'=\(\scriptstyle i\omega_s\)] (g),
(h) -- [fermion, edge label'=\(\scriptstyle i\omega_m\)] (o),
};
\end{feynman}
\end{tikzpicture} 

\begin{tikzpicture}
\begin{feynman}
\vertex (plus) at(0,0.5){\( + \,\, \delta_{n,m} \)};
\vertex (g) at(0.8,0);
\vertex (G) at(0.8,-0.2){\( \scriptstyle \mu' \)};
\vertex (h) at(2,0); 
\vertex (i) at(2,1); 
\vertex (I) at(2,1.2){\( \scriptstyle \theta \)};
\vertex (j) at(0.8,1);
\vertex (J) at(0.8,1.2){\( \scriptstyle \mu \)};
\vertex (name1) at(1.4,0.5){\( \scriptstyle \Gamma \)};
\vertex (q) at(3.2,0);
\vertex (r) at(4.4,0);
\vertex (R) at(4.4,-0.2){\( \scriptstyle \nu' \)};
\vertex (s) at(4.4,1);
\vertex (S) at(4.4,1.2){\( \scriptstyle \nu \)};
\vertex (t) at(3.2,1); 
\vertex (T) at(3.2,1.2){\( \scriptstyle \phi \)};
\vertex (name3) at(3.8,0.5){\( \scriptstyle \Gamma \)};
\diagram* { 
(g) -- [fermion] (h)  -- (i)  -- [fermion, edge label'=\(\scriptstyle i\omega_m\)] (j)  -- (g),
(q) -- [fermion] (r)  -- (s)  -- [fermion, edge label'=\(\scriptstyle i\omega_m\)] (t)  -- (q),
(h) -- [fermion, edge label'=\(\scriptstyle i\omega_s\)] (q),
};
\end{feynman}
\end{tikzpicture}
\begin{tikzpicture}
\begin{feynman}
\vertex (minus) at(0,0.5){\( - \,\, \delta_{n,m} \)};
\vertex (g) at(0.8,0);
\vertex (G) at(0.8,-0.2){\( \scriptstyle \mu' \)};
\vertex (h) at(2,0); 
\vertex (i) at(2,1); 
\vertex (I) at(2,1.2){\( \scriptstyle \theta \)};
\vertex (j) at(0.8,1);
\vertex (J) at(0.8,1.2){\( \scriptstyle \mu \)};
\vertex (name1) at(1.4,0.5){\( \scriptstyle U^{[4]} \)};
\vertex (q) at(3.2,0);
\vertex (r) at(4.4,0);
\vertex (R) at(4.4,-0.2){\( \scriptstyle \nu' \)};
\vertex (s) at(4.4,1);
\vertex (S) at(4.4,1.2){\( \scriptstyle \nu \)};
\vertex (t) at(3.2,1); 
\vertex (T) at(3.2,1.2){\( \scriptstyle \phi \)};
\vertex (name3) at(3.8,0.5){\( \scriptstyle \Gamma \)};
\diagram* { 
(g) -- [fermion] (h)  -- (i)  -- [fermion, edge label'=\(\scriptstyle i\omega_m\)] (j)  -- (g),
(q) -- [fermion] (r)  -- (s)  -- [fermion, edge label'=\(\scriptstyle i\omega_m\)] (t)  -- (q),
(h) -- [fermion, edge label'=\(\scriptstyle i\omega_s\)] (q),
};
\end{feynman}
\end{tikzpicture}

\begin{tikzpicture}
\begin{feynman}
\vertex (plus) at(0,0.5){\( + \,\, \delta_{n,s} \)};
\vertex (g) at(0.8,0);
\vertex (G) at(0.8,-0.2){\( \scriptstyle \mu' \)};
\vertex (h) at(2,0); 
\vertex (H) at(2,-0.2){\( \scriptstyle \phi \)};
\vertex (i) at(2,1); 
\vertex (j) at(0.8,1);
\vertex (J) at(0.8,1.2){\( \scriptstyle \mu \)};
\vertex (name1) at(1.4,0.5){\( \scriptstyle \Gamma \)};
\vertex (q) at(3.2,0);
\vertex (Q) at(3.2,-0.2){\( \scriptstyle \theta \)};
\vertex (r) at(4.4,0);
\vertex (R) at(4.4,-0.2){\( \scriptstyle \nu' \)};
\vertex (s) at(4.4,1);
\vertex (S) at(4.4,1.2){\( \scriptstyle \nu \)};
\vertex (t) at(3.2,1); 
\vertex (name3) at(3.8,0.5){\( \scriptstyle \Gamma \)};
\diagram* { 
(g) -- [fermion, edge label'=\(\scriptstyle i\omega_s\)] (h)  -- (i)  -- [fermion] (j)  -- (g),
(q) -- [fermion, edge label'=\(\scriptstyle i\omega_s\)] (r)  -- (s)  -- [fermion] (t)  -- (q),
(t) -- [fermion, edge label'=\(\scriptstyle i\omega_m\)] (i),
};
\end{feynman}
\end{tikzpicture}
\begin{tikzpicture}
\begin{feynman}
\vertex (plus) at(0,0.5){\( - \,\, \delta_{n,s} \)};
\vertex (g) at(0.8,0);
\vertex (G) at(0.8,-0.2){\( \scriptstyle \mu' \)};
\vertex (h) at(2,0); 
\vertex (H) at(2,-0.2){\( \scriptstyle \phi \)};
\vertex (i) at(2,1); 
\vertex (j) at(0.8,1);
\vertex (J) at(0.8,1.2){\( \scriptstyle \mu \)};
\vertex (name1) at(1.4,0.5){\( \scriptstyle U^{[4]} \)};
\vertex (q) at(3.2,0);
\vertex (Q) at(3.2,-0.2){\( \scriptstyle \theta \)};
\vertex (r) at(4.4,0);
\vertex (R) at(4.4,-0.2){\( \scriptstyle \nu' \)};
\vertex (s) at(4.4,1);
\vertex (S) at(4.4,1.2){\( \scriptstyle \nu \)};
\vertex (t) at(3.2,1); 
\vertex (name3) at(3.8,0.5){\( \scriptstyle \Gamma \)};
\diagram* { 
(g) -- [fermion, edge label'=\(\scriptstyle i\omega_s\)] (h)  -- (i)  -- [fermion] (j)  -- (g),
(q) -- [fermion, edge label'=\(\scriptstyle i\omega_s\)] (r)  -- (s)  -- [fermion] (t)  -- (q),
(t) -- [fermion, edge label'=\(\scriptstyle i\omega_m\)] (i),
};
\end{feynman}
\end{tikzpicture}

\begin{tikzpicture}
\begin{feynman}
\vertex (plus) at(0,0.5){\( +  \)};
\vertex (a) at(0.5,0);
\vertex (A) at(0.5,-0.2){\( \scriptstyle \mu' \)};
\vertex (b) at(1.7,0); 
\vertex (c) at(1.7,1); 
\vertex (d) at(0.5,1);
\vertex (D) at(0.5,1.2){\( \scriptstyle \mu \)};
\vertex (name1) at(1.1,0.5){\( \scriptstyle \Gamma \)};
\vertex (g) at(2.9,0); 
\vertex (h) at(2.9,1); 
\vertex (i) at(3.766,1.5); 
\vertex (j) at(4.632,1);
\vertex (J) at(4.832,1){\( \scriptstyle \phi \)};
\vertex (k) at(4.632,0);
\vertex (K) at(4.832,0){\( \scriptstyle \theta \)};
\vertex (l) at(3.766,-0.5); 
\vertex (name2) at(3.766,0.5){\( \scriptstyle U^{[6]} \)};
\vertex (q) at(6.5,0); 
\vertex (r) at(7.7,0);
\vertex (R) at(7.7,-0.2){\( \scriptstyle \nu' \)};
\vertex (s) at(7.7,1);
\vertex (S) at(7.7,1.2){\( \scriptstyle \nu \)};
\vertex (t) at(6.5,1); 
\vertex (name3) at(7.1,0.5){\( \scriptstyle \Gamma \)};
\diagram* { 
(a) -- [fermion] (b)  -- (c)  -- [fermion] (d)  -- (a),
(g) -- (h)  -- [anti fermion, edge label=\(\scriptstyle i\omega_m\)] (i)  -- (j),
(j) -- [anti fermion, edge label=\(\scriptstyle i\omega_n\)] (k)  -- (l)  -- [anti fermion, edge label=\(\scriptstyle i\omega_s\)] (g),
(q) -- [fermion] (r)  -- (s)  -- [fermion] (t)  -- (q),
(l) -- [fermion] (q),
(t) -- [fermion] (i),
(b) -- [fermion] (g),
(h) -- [fermion] (c),
};
\end{feynman}
\end{tikzpicture}

\begin{tikzpicture}
\begin{feynman}
\vertex (plus) at(-0.6,0.5){\( +  \)};
\vertex (q) at(0,0);
\vertex (Q) at(0,-0.2){\( \scriptstyle \mu' \)};
\vertex (r) at(1.2,0); 
\vertex (s) at(1.2,1); 
\vertex (t) at(0,1);
\vertex (T) at(0,1.2){\( \scriptstyle \mu \)};
\vertex (name4) at(0.6,0.5){\( \scriptstyle \Gamma \)};
\vertex (a) at(2.4,0); 
\vertex (b) at(2.4,1); 
\vertex (c) at(3.266,1.5);
\vertex (C) at(3.266,1.7){\( \scriptstyle \nu \)};
\vertex (d) at(4.132,1);
\vertex (D) at(4.332,1){\( \scriptstyle \phi \)};
\vertex (e) at(4.132,0);
\vertex (E) at(4.332,0){\( \scriptstyle \theta \)};
\vertex (f) at(3.266,-0.5);
\vertex (F) at(3.266,-0.7){\( \scriptstyle \nu' \)};
\vertex (name1) at(3.266,0.5){\( \scriptstyle \Gamma^{[6]} \)};
\vertex (minus) at(5.232,0.5){\( - \)};
\vertex (g) at(8.332,0); 
\vertex (h) at(8.332,1); 
\vertex (i) at(9.198,1.5);
\vertex (I) at(9.198,1.7){\( \scriptstyle \nu \)};
\vertex (j) at(10.064,1);
\vertex (J) at(10.264,1){\( \scriptstyle \phi \)};
\vertex (k) at(10.064,0);
\vertex (K) at(10.264,0){\( \scriptstyle \theta \)};
\vertex (l) at(9.198,-0.5);
\vertex (L) at(9.198,-0.7){\( \scriptstyle \nu' \)};
\vertex (name2) at(9.198,0.5){\( \scriptstyle \Gamma^{[6]} \)}; 
\vertex (m) at(5.932,0);
\vertex (M) at(5.932,-0.2){\( \scriptstyle \mu' \)};
\vertex (n) at(7.132,0); 
\vertex (o) at(7.132,1); 
\vertex (p) at(5.932,1);
\vertex (P) at(5.932,1.2){\( \scriptstyle \mu \)};
\vertex (name3) at(6.532,0.5){\( \scriptstyle U^{[4]} \)};
\diagram* {
(a) -- (b),
(b) -- [anti fermion] (c),
(c) -- (d),
(d) -- [anti fermion, edge label=\(\scriptstyle i\omega_n\)] (e),
(e) -- (f),
(f) -- [anti fermion] (a),
(g) -- (h),
(h) -- [anti fermion] (i),
(i) -- (j),
(j) -- [anti fermion, edge label=\(\scriptstyle i\omega_n\)] (k),
(k) -- (l),
(l) -- [anti fermion] (g),
(m) -- [fermion] (n)  -- (o)  -- [fermion] (p)  -- (m),
(q) -- [fermion] (r)  -- (s)  -- [fermion] (t)  -- (q),
(r) -- [fermion, edge label'=\(\scriptstyle i\omega_s\)] (a),
(b) -- [fermion, edge label'=\(\scriptstyle i\omega_m\)] (s),
(n) -- [fermion, edge label'=\(\scriptstyle i\omega_s\)] (g),
(h) -- [fermion, edge label'=\(\scriptstyle i\omega_m\)] (o),
};
\end{feynman}
\end{tikzpicture} 
\caption{Second step of the derivation that leads from Eq.~\eqref{derivative of Bethe-Salpeter} to Eq.~\eqref{derivative of Bethe-Salpeter simplified}.
\label{diagram passage 2}}
\end{figure}

We then use the Bethe-Salpeter equation, Eq.~\eqref{Bethe-Salpeter}, which relates the reducible four-leg vertex $\Gamma$ to the irreducible four-leg vertex $U^{[4]}$. In our diagrammatic notation, Eq.~\eqref{Bethe-Salpeter} is translated into Fig.~\ref{BS figure}. By substituting it into Fig.~\ref{diagram passage 1}, we obtain Fig.~\ref{diagram passage 2}. 

\begin{figure}
\begin{tikzpicture}
\begin{feynman}
\vertex (a) at(0,0);
\vertex (A) at(-0.2,0){\( \scriptstyle \mu' \)};
\vertex (b) at(0,1);
\vertex (B) at(-0.2,1){\( \scriptstyle \mu \)};
\vertex (c) at(0.866,1.5);
\vertex (C) at(0.866,1.7){\( \scriptstyle \nu \)};
\vertex (d) at(1.732,1);
\vertex (D) at(1.932,1){\( \scriptstyle \phi \)};
\vertex (e) at(1.732,0);
\vertex (E) at(1.932,0){\( \scriptstyle \theta \)};
\vertex (f) at(0.866,-0.5);
\vertex (F) at(0.866,-0.7){\( \scriptstyle \nu' \)};
\vertex (name1) at(0.866,0.5){\( \scriptstyle \Gamma^{[6]} \)};
\vertex (eq) at(2.632,0.5){\( = \)};
\vertex (g) at(3.332,0);
\vertex (G) at(3.132,0){\( \scriptstyle \mu' \)};
\vertex (h) at(3.332,1);
\vertex (H) at(3.132,1){\( \scriptstyle \mu \)};
\vertex (i) at(4.198,1.5);
\vertex (I) at(4.198,1.7){\( \scriptstyle \nu \)};
\vertex (j) at(5.064,1);
\vertex (J) at(5.264,1){\( \scriptstyle \phi \)};
\vertex (k) at(5.064,0);
\vertex (K) at(5.264,0){\( \scriptstyle \theta \)};
\vertex (l) at(4.198,-0.5);
\vertex (L) at(4.198,-0.7){\(\scriptstyle  \nu' \)};
\vertex (name2) at(4.198,0.5){\( \scriptstyle U^{[6]} \)};
\vertex (plus) at(6.5,0.5){\( + \,\, \delta_{n,m} \)};
\vertex (m) at(7.3,0);
\vertex (M) at(7.3,-0.2){\( \scriptstyle \mu' \)};
\vertex (n) at(8.5,0);
\vertex (N) at(8.5,-0.2){\( \scriptstyle \xi' \)};
\vertex (o) at(8.5,1);
\vertex (O) at(8.5,1.2){\( \scriptstyle \theta \)};
\vertex (p) at(7.3,1);
\vertex (P) at(7.3,1.2){\( \scriptstyle \mu \)};
\vertex (name3) at(7.9,0.5){\( \scriptstyle \Gamma \)};
\vertex (q) at(9.7,0);
\vertex (Q) at(9.7,-0.2){\( \scriptstyle \eta' \)};
\vertex (r) at(10.9,0);
\vertex (R) at(10.9,-0.2){\( \scriptstyle \nu' \)};
\vertex (s) at(10.9,1);
\vertex (S) at(10.9,1.2){\( \scriptstyle \nu \)};
\vertex (t) at(9.7,1);
\vertex (T) at(9.7,1.2){\( \scriptstyle \phi \)};
\vertex (name4) at(10.3,0.5){\( \scriptstyle \Gamma \)};
\diagram* {
(a) -- (b),
(b) -- [anti fermion, edge label=\(\scriptstyle i\omega_m\)] (c),
(c) -- (d),
(d) -- [anti fermion, edge label=\(\scriptstyle i\omega_n\)] (e),
(e) -- (f),
(f) -- [anti fermion, edge label=\(\scriptstyle i\omega_s\)] (a),
(g) -- (h),
(h) -- [anti fermion, edge label=\(\scriptstyle i\omega_m\)] (i),
(i) -- (j),
(j) -- [anti fermion, edge label=\(\scriptstyle i\omega_n\)] (k),
(k) -- (l),
(l) -- [anti fermion, edge label=\(\scriptstyle i\omega_s\)] (g),
(m) -- [fermion, edge label'=\(\scriptstyle i\omega_s\)] (n),
(n) -- (o),
(o) -- [fermion, edge label'=\(\scriptstyle i\omega_m\)] (p),
(p) -- (m),
(q) -- [fermion, edge label'=\(\scriptstyle i\omega_s\)] (r),
(r) -- (s),
(s) -- [fermion, edge label'=\(\scriptstyle i\omega_m\)] (t),
(t) -- (q),
(n) -- [fermion, edge label'=\(\scriptstyle i\omega_s\)] (q),
};
\end{feynman}
\end{tikzpicture}

\begin{tikzpicture}
\begin{feynman}
\vertex (plus) at(4,0.5){\( + \,\, \delta_{n,s} \)};
\vertex (m) at(4.8,0);
\vertex (M) at(4.8,-0.2){\( \scriptstyle \mu' \)};
\vertex (n) at(6,0);
\vertex (N) at(6,-0.2){\( \scriptstyle \phi \)};
\vertex (o) at(6,1);
\vertex (O) at(6,1.2){\( \scriptstyle \xi \)};
\vertex (p) at(4.8,1);
\vertex (P) at(4.8,1.2){\( \scriptstyle \mu \)};
\vertex (name3) at(5.4,0.5){\( \scriptstyle \Gamma \)};
\vertex (q) at(7.2,0);
\vertex (Q) at(7.2,-0.2){\( \scriptstyle \theta \)};
\vertex (r) at(8.4,0);
\vertex (R) at(8.4,-0.2){\( \scriptstyle \nu' \)};
\vertex (s) at(8.4,1);
\vertex (S) at(8.4,1.2){\( \scriptstyle \nu \)};
\vertex (t) at(7.2,1);
\vertex (T) at(7.2,1.2){\( \scriptstyle \eta \)};
\vertex (name4) at(7.8,0.5){\( \scriptstyle \Gamma \)};
\diagram* {
(m) -- [fermion, edge label'=\(\scriptstyle i\omega_s\)] (n),
(n) -- (o),
(o) -- [fermion, edge label'=\(\scriptstyle i\omega_m\)] (p),
(p) -- (m),
(q) -- [fermion, edge label'=\(\scriptstyle i\omega_s\)] (r),
(r) -- (s),
(s) -- [fermion, edge label'=\(\scriptstyle i\omega_m\)] (t),
(t) -- (q),
(t) -- [fermion, edge label'=\(\scriptstyle i\omega_m\)] (o),
};
\end{feynman}
\end{tikzpicture}

\begin{tikzpicture}
\begin{feynman}
\vertex (plus) at(4,0.5){\( +  \)};
\vertex (g) at(4.832,0);
\vertex (G) at(4.632,0){\( \scriptstyle \mu' \)};
\vertex (h) at(4.832,1);
\vertex (H) at(4.632,1){\( \scriptstyle \mu \)};
\vertex (i) at(5.698,1.5);
\vertex (I) at(5.698,1.7){\( \scriptstyle \xi \)};
\vertex (j) at(6.564,1);
\vertex (J) at(6.764,1){\( \scriptstyle \phi \)};
\vertex (k) at(6.564,0);
\vertex (K) at(6.764,0){\( \scriptstyle \theta \)};
\vertex (l) at(5.698,-0.5);
\vertex (L) at(5.698,-0.7){\( \scriptstyle \xi' \)};
\vertex (name2) at(5.698,0.5){\( \scriptstyle U^{[6]} \)};
\vertex (q) at(8.4,0);
\vertex (Q) at(8.4,-0.2){\( \scriptstyle \eta' \)};
\vertex (r) at(9.6,0);
\vertex (R) at(9.6,-0.2){\( \scriptstyle \nu' \)};
\vertex (s) at(9.6,1);
\vertex (S) at(9.6,1.2){\( \scriptstyle \nu \)};
\vertex (t) at(8.4,1);
\vertex (T) at(8.4,1.2){\( \scriptstyle \eta \)};
\vertex (name4) at(9,0.5){\( \scriptstyle \Gamma \)};
\diagram* {
(g) -- (h)  -- [anti fermion, edge label=\(\scriptstyle i\omega_m\)] (i)  -- (j),
(j) -- [anti fermion, edge label=\(\scriptstyle i\omega_n\)] (k)  -- (l)  -- [anti fermion, edge label=\(\scriptstyle i\omega_s\)] (g),
(q) -- [fermion, edge label'=\(\scriptstyle i\omega_s\)] (r)  -- (s)  -- [fermion, edge label'=\(\scriptstyle i\omega_m\)] (t)  -- (q),
(l) -- [fermion, edge label'=\(\scriptstyle i\omega_s\)] (q),
(t) -- [fermion, edge label'=\(\scriptstyle i\omega_m\)] (i),
};
\end{feynman}
\end{tikzpicture}
\begin{tikzpicture}
\begin{feynman}
\vertex (plus) at(4,0.5){\( +  \)};
\vertex (g) at(6.832,0);
\vertex (G) at(6.732,-0.2){\( \scriptstyle \eta' \)};
\vertex (h) at(6.832,1);
\vertex (H) at(6.732,1.2){\( \scriptstyle \eta \)};
\vertex (i) at(7.698,1.5);
\vertex (I) at(7.698,1.7){\( \scriptstyle \nu \)};
\vertex (j) at(8.564,1);
\vertex (J) at(8.764,1){\( \scriptstyle \phi \)};
\vertex (k) at(8.564,0);
\vertex (K) at(8.764,0){\( \scriptstyle \theta \)};
\vertex (l) at(7.698,-0.5);
\vertex (L) at(7.698,-0.7){\( \scriptstyle \nu' \)};
\vertex (name2) at(7.698,0.5){\( \scriptstyle U^{[6]} \)};
\vertex (q) at(4.5,0);
\vertex (Q) at(4.5,-0.2){\( \scriptstyle \mu' \)};
\vertex (r) at(5.7,0);
\vertex (R) at(5.7,-0.2){\( \scriptstyle \xi' \)};
\vertex (s) at(5.7,1);
\vertex (S) at(5.7,1.2){\( \scriptstyle \xi \)};
\vertex (t) at(4.5,1);
\vertex (T) at(4.5,1.2){\( \scriptstyle \mu \)};
\vertex (name4) at(5.1,0.5){\( \scriptstyle \Gamma \)};
\diagram* {
(g) -- (h)  -- [anti fermion, edge label=\(\scriptstyle i\omega_m\)] (i)  -- (j),
(j) -- [anti fermion, edge label=\(\scriptstyle i\omega_n\)] (k)  -- (l)  -- [anti fermion, edge label=\(\scriptstyle i\omega_s\)] (g),
(q) -- [fermion, edge label'=\(\scriptstyle i\omega_s\)] (r),
(r) -- (s),
(s) -- [fermion, edge label'=\(\scriptstyle i\omega_m\)] (t),
(t) -- (q),
(r) -- [fermion, edge label'=\(\scriptstyle i\omega_s\)] (g),
(h) -- [fermion, edge label'=\(\scriptstyle i\omega_m\)] (s),
};
\end{feynman}
\end{tikzpicture}

\begin{tikzpicture}
\begin{feynman}
\vertex (plus) at(0,0.5){\( +  \)};
\vertex (a) at(0.5,0);
\vertex (A) at(0.5,-0.2){\( \scriptstyle \mu' \)};
\vertex (b) at(1.7,0); 
\vertex (B) at(1.7,-0.2){\( \scriptstyle \xi' \)};
\vertex (c) at(1.7,1); 
\vertex (C) at(1.7,1.2){\( \scriptstyle \xi \)};
\vertex (d) at(0.5,1);
\vertex (D) at(0.5,1.2){\( \scriptstyle \mu \)};
\vertex (name1) at(1.1,0.5){\( \scriptstyle \Gamma \)};
\vertex (g) at(2.9,0);
\vertex (G) at(2.8,-0.2){\( \scriptstyle \eta' \)}; 
\vertex (h) at(2.9,1); 
\vertex (H) at(2.8,1.2){\( \scriptstyle \eta \)}; 
\vertex (i) at(3.766,1.5); 
\vertex (I) at(3.766,1.7){\( \scriptstyle \zeta \)}; 
\vertex (j) at(4.632,1);
\vertex (J) at(4.832,1){\( \scriptstyle \phi \)};
\vertex (k) at(4.632,0);
\vertex (K) at(4.832,0){\( \scriptstyle \theta \)};
\vertex (l) at(3.766,-0.5); 
\vertex (L) at(3.766,-0.7){\( \scriptstyle \zeta' \)}; 
\vertex (name2) at(3.766,0.5){\( \scriptstyle U^{[6]} \)};
\vertex (q) at(6.5,0); 
\vertex (Q) at(6.5,-0.2){\( \scriptstyle \chi' \)}; 
\vertex (r) at(7.7,0);
\vertex (R) at(7.7,-0.2){\( \scriptstyle \nu' \)};
\vertex (s) at(7.7,1);
\vertex (S) at(7.7,1.2){\( \scriptstyle \nu \)};
\vertex (t) at(6.5,1); 
\vertex (T) at(6.5,1.2){\( \scriptstyle \chi \)};
\vertex (name3) at(7.1,0.5){\( \scriptstyle \Gamma \)};
\diagram* { 
(a) -- [fermion, edge label'=\(\scriptstyle i\omega_s\)] (b)  -- (c)  -- [fermion, edge label'=\(\scriptstyle i\omega_m\)] (d)  -- (a),
(g) -- (h)  -- [anti fermion, edge label=\(\scriptstyle i\omega_m\)] (i)  -- (j),
(j) -- [anti fermion, edge label=\(\scriptstyle i\omega_n\)] (k)  -- (l)  -- [anti fermion, edge label=\(\scriptstyle i\omega_s\)] (g),
(q) -- [fermion, edge label'=\(\scriptstyle i\omega_s\)] (r)  -- (s)  -- [fermion, edge label'=\(\scriptstyle i\omega_m\)] (t)  -- (q),
(l) -- [fermion, edge label'=\(\scriptstyle i\omega_s\)] (q),
(t) -- [fermion, edge label'=\(\scriptstyle i\omega_m\)] (i),
(b) -- [fermion, edge label'=\(\scriptstyle i\omega_s\)] (g),
(h) -- [fermion, edge label'=\(\scriptstyle i\omega_m\)] (c),
};
\end{feynman}
\end{tikzpicture}

\caption{Result of the derivation. This is the diagrammatic representation of Eq.~\eqref{derivative of Bethe-Salpeter simplified}.}
\label{diagram passage 3}
\end{figure}
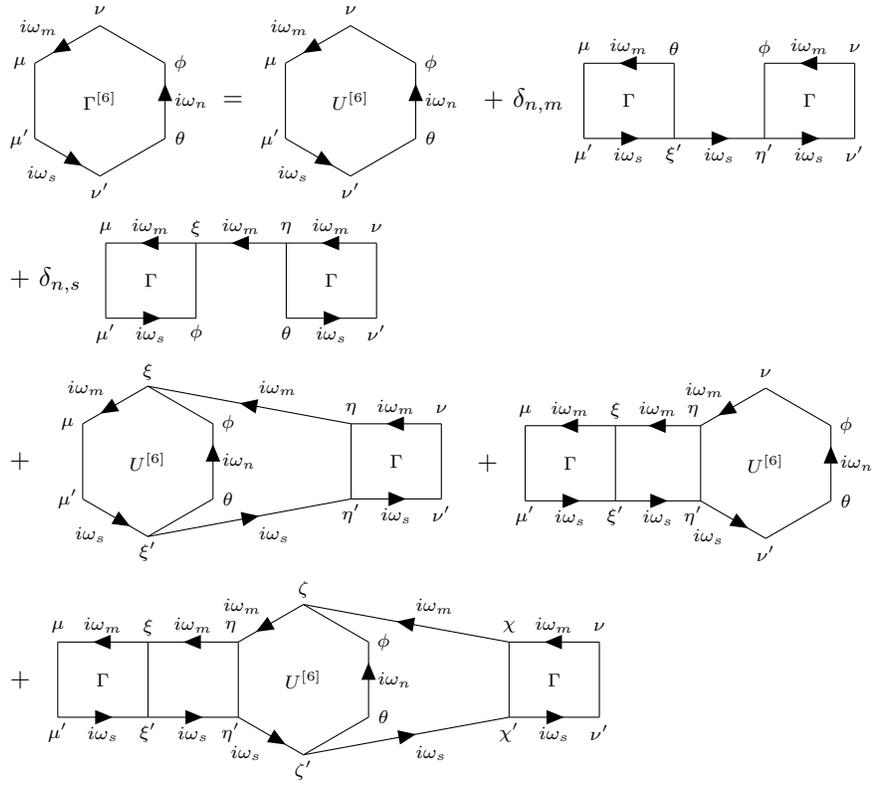

We now observe that the last line of Fig.~\ref{diagram passage 2} contains a term which is identical to the quantity on the left-hand side of the same diagrammatic equation (so they cancel out), and another term, which depends on $U^{[4]}$ and $\Gamma^{[6]}$, which appears in the last line of Fig.~\ref{diagram passage 0}. We then solve Fig.~\ref{diagram passage 2} for this quantity, substitute in to Fig.~\ref{diagram passage 0}, and we obtain the result, Fig.~\ref{diagram passage 3}.

Finally, we convert the result displayed in Fig.~\ref{diagram passage 3} into the following algebraic expression:
\begin{align}
&      \widetilde{\Gamma}^{[6]}_{{\bm U}; (\mu  , \nu ; \phi, \theta ; \nu', \mu')}[{\bm G}](i \omega_m, i \omega_n, i \omega_s )    \nonumber \\
& =  \widetilde{U}^{[6]}_{{\bm U}; (\mu  , \nu ; \phi, \theta; \nu', \mu')}[{\bm G}](i \omega_m, i \omega_n, i \omega_s )  
+ \delta_{n,m} \sum_{ \xi',   \eta'}   \widetilde{\Gamma}_{{\bm U}; (\mu  , \theta ; \xi', \mu')}[{\bm G}](i \omega_m, i \omega_s )  \nonumber \\
& \quad \times  G_{\eta', \xi'}(i \omega_s) \,  \widetilde{\Gamma}_{{\bm U}; (\phi  , \nu ; \nu', \eta')}[{\bm G}](i \omega_m, i \omega_s )  \nonumber \\
& \quad  + \delta_{n, s} \sum_{\xi,  \eta } \widetilde{\Gamma}_{{\bm U}; (\mu  , \xi ; \phi, \mu')}[{\bm G}](i \omega_m, i \omega_s ) \, G_{\xi, \eta}(i \omega_m) \,    \widetilde{\Gamma}_{{\bm U}; (\eta  , \nu ; \nu', \theta)}[{\bm G}](i \omega_m, i \omega_s ) \nonumber \\
& \quad  + \sum_{\xi, \xi', \eta, \eta'}   \widetilde{U}^{[6]}_{{\bm U}; (\mu  , \xi ; \phi, \theta ; \xi', \mu' )}[{\bm G}](i \omega_m, i \omega_n, i \omega_s )  \, G_{\xi, \eta}(i \omega_m) \, G_{\eta', \xi'}(i \omega_s) \nonumber \\
& \quad \times  \widetilde{\Gamma}_{{\bm U}; (\eta  , \nu ; \nu', \eta')}[{\bm G}](i \omega_m, i \omega_s ) \nonumber \\ 
& \quad  + \sum_{\xi, \xi', \eta, \eta'} \widetilde{\Gamma}_{{\bm U}; (\mu  , \xi ; \xi', \mu')}[{\bm G}](i \omega_m, i \omega_s ) \, G_{\xi, \eta}(i \omega_m) \, G_{\eta', \xi'}(i \omega_s) \nonumber \\
& \quad \times    \widetilde{U}^{[6]}_{{\bm U}; (\eta  , \nu ; \phi, \theta; \nu', \eta')}[{\bm G}](i \omega_m, i \omega_n, i \omega_s )   \nonumber \\ 
& \quad + \sum_{\xi, \xi', \eta, \eta', \zeta, \zeta', \chi, \chi'} \widetilde{\Gamma}_{{\bm U}; (\mu  , \xi ; \xi', \mu')}[{\bm G}](i \omega_m, i \omega_s ) \, G_{\xi, \eta}(i \omega_m) \, G_{\eta', \xi'}(i \omega_s) \nonumber \\
& \quad \times    \widetilde{U}^{[6]}_{{\bm U}; (\eta  , \zeta; \phi, \theta ; \zeta', \eta' )}[{\bm G}](i \omega_m , i \omega_n , i \omega_s ) \, G_{\zeta, \chi}(i \omega_m) \, G_{\chi', \zeta' }(i \omega_s)   \nonumber \\
&   \quad \times   \widetilde{\Gamma}_{{\bm U}; (\chi  , \nu ; \nu', \chi')}[{\bm G}](i \omega_m, i \omega_s )~.
\label{derivative of Bethe-Salpeter simplified}
\end{align}
\section{G-N transformation}
\label{app:Nambu}

Under the assumptions stated at the beginning of Section~\ref{sec: Eliashberg}, the total Hamiltonian becomes
\begin{align}
\hat{\cal H} & = \sum_{i,j} t_{i,j} \sum_{\sigma} \hat{c}^{\dagger}_{i, \sigma} \hat{c}_{j, \sigma} + \frac{1}{2} \sum_{i,j,k,l}  U_{i,j,l,k}  \sum_{\sigma, \sigma'} \hat{c}^{\dagger}_{i, \sigma} \hat{c}^{\dagger}_{j, \sigma'} \hat{c}_{k, \sigma'} \hat{c}_{l, \sigma} \nonumber \\
& \quad + \sum_{i,j,k} \sum_s I^{(k)}_{i,j}(s) \, \hat{q}_{k,s} \sum_{\sigma} \hat{c}^{\dagger}_{i, \sigma} \hat{c}_{j, \sigma} + \sum_{i, j} \sum_s f_{i, j}(s) \, \hat{b}^{\dagger}_{i, s} \hat{b}_{j, s}~.
\end{align}
We perform the G-N transformation on the fermions, 
\begin{align}
\hat{c}_{i, \uparrow} \equiv \hat{d}_{i, \uparrow}~, \quad \hat{c}_{i, \downarrow} \equiv \hat{d}^{\dagger}_{i, \downarrow}~,
\end{align}
and we normal-order the result. Neglecting constant terms, we obtain 
\begin{align}
\hat{\cal H} & = \sum_{i,j} t_{i,j} \sum_{\sigma} \sigma \hat{d}^{\dagger}_{i, \sigma} \hat{d}_{j, \sigma}  +   \sum_{i, j} \sum_s f_{i, j}(s) \, \hat{b}^{\dagger}_{i, s} \hat{b}_{j, s} \nonumber \\
& \quad + \frac{1}{2} \sum_{i,j,k,l}  U_{i,j,l,k}  \sum_{\sigma, \sigma'} \sigma \sigma' \, \hat{d}^{\dagger}_{i, \sigma} \hat{d}^{\dagger}_{j, \sigma'} \hat{d}_{k, \sigma'} \hat{d}_{l, \sigma} + \sum_{i, j} \left( \sum_k U_{i, k, j, k} \right) \sum_{\sigma} \sigma \hat{d}^{\dagger}_{i, \sigma} \hat{d}_{j, \sigma} \nonumber \\
& \quad + \sum_{i,j,k} \sum_s I^{(k)}_{i,j}(s) \, \hat{q}_{k,s} \sum_{\sigma} \sigma \hat{d}^{\dagger}_{i, \sigma} \hat{d}_{j, \sigma} + \sum_{k} \sum_s \hat{q}_{k,s} \sum_i I^{(k)}_{i,i}(s)~.
\label{Nambu being done}
\end{align}
Let us examine Eq.~\eqref{Nambu being done}. Notice, in particular, the last term in the second line, which results from the normal-ordering of the transformed EEI Hamiltonian, and the last term in the third line, which results from the normal-ordering of the transformed EPI Hamiltonian. The first of these terms (purely fermionic) can be directly incorporated into a renormalization of the electronic hopping parameters. The second term, instead, which is purely phononic, apparently differs from the other phononic terms. In order to recover the form of the Hamiltonian, we make the bosonic transformation
\begin{align}
\hat{b}_{i, s} \equiv \hat{a}_{i, s} + K_{i, s}~, \quad \hat{b}^{\dagger}_{i, s} \equiv \hat{a}^{\dagger}_{i, s} + K^*_{i, s}~,
\label{Nambu boson}
\end{align}
where the quantities $K_{i,s}$ are constants that will be determined shortly, and the fields $\hat{a}_{i, s}$ are bosons. When this transformation is done, the first term in the third line of Eq.~\eqref{Nambu being done} generates an additional purely electronic term, while the second term in the first line generates both quadratic and linear terms in the new bosonic fields. By imposing that the terms which are linear in the bosons (and do not involve fermionic fields) cancel out, we determine the values of the parameters $K_{i,s}$:
\begin{align}
K_{i, s} = K^*_{i, s} \equiv - \frac{1}{\sqrt{2}} \sum_j f^{-1}_{i, j}(s) \, \sum_l I^{(j)}_{l,l}(s)~,
\end{align}
where we have exploited that $f_{k, i}(s) = f_{i, k}(s)$. Neglecting further constants, the Hamiltonian becomes
\begin{align}
\hat{\cal H} & = \sum_{i,j} \left[ t_{i,j} +  \sum_k U_{i, k, j, k} -  \sum_{k, s} I^{(k)}_{i,j}(s) \,    \sum_m f^{-1}_{k, m}(s) \, \sum_l I^{(m)}_{l,l}(s) \right] \sum_{\sigma} \sigma \hat{d}^{\dagger}_{i, \sigma} \hat{d}_{j, \sigma}  \nonumber \\
& \quad +   \sum_{i, j} f_{i, j}(s) \,   \hat{a}^{\dagger}_{i, s} \hat{a}_{j, s}  + \sum_{i,j } \sum_{k, s} I^{(k)}_{i,j}(s) \, \frac{1}{\sqrt{2}} \left( \hat{a}_{k,s} + \hat{a}^{\dagger}_{k,s} \right) \sum_{\sigma} \sigma \hat{d}^{\dagger}_{i, \sigma} \hat{d}_{j, \sigma}   \nonumber \\
& \quad + \frac{1}{2} \sum_{i,j,k,l}  U_{i,j,l,k}  \sum_{\sigma, \sigma'} \sigma \sigma' \, \hat{d}^{\dagger}_{i, \sigma} \hat{d}^{\dagger}_{j, \sigma'} \hat{d}_{k, \sigma'} \hat{d}_{l, \sigma}~. 
\label{Nambu done}
\end{align}
This is mathematically equivalent to the initial Hamiltonian \eqref{H} of the main text, provided that the fields are renamed and the parameters are specified as in Eqs.~\eqref{t Nambu}, \eqref{M Nambu}, and \eqref{U Nambu}.

The transformation \eqref{Nambu boson} is equivalent to a fixed displacement of the lattice ions, dependent on the lattice position and on the phonon mode. It can be interpreted as an effect of the interaction between the phonon system and the vacuum of the Nambu fields: when the state of the system contains zero Nambu fermions, the equilibrium positions of the ions are different with respect to those characterizing the state with zero electrons (i.e., the lattice sites). An analogous transformation was discussed in Ref.~\cite{Secchi18}.


\begin{thebibliography}{50}
\bibitem{note1} 
We use a different prefactor with respect to Ref.~\cite{Potthoff06}, so that all fields $\overline{c}, c, b, b^*$ are dimensionless in all representations. Consequently, some of our formulas   have additional prefactors $T$ with respect to the corresponding ones in Ref.~\cite{Potthoff06}.
\bibitem{note2} 
In Ref.~\cite{Potthoff06}, the matrix transposition is not indicated explicitly.
\end{thebibliography}
\end{document}